\documentclass[12pt]{article}
\pdfoutput=1
\usepackage{jheppub}
\usepackage{epsfig}
\usepackage{amsmath}
\usepackage{amssymb}
\usepackage{amsfonts}
\usepackage{amsxtra}
\usepackage{amsthm}
\usepackage{mathrsfs}
\usepackage{makeidx}
\usepackage{graphics}
\usepackage{dsfont}
\usepackage{mathtools}
\usepackage{graphicx}
\usepackage{placeins}
\usepackage{bm}
\usepackage[capitalise]{cleveref}
\usepackage{empheq}
\usepackage{colortbl}
\usepackage{xcolor}
\usepackage{enumerate}
\usepackage{titlesec}
\usepackage{longtable}
\usepackage{hhline}
\usepackage{float}
\usepackage{color}
\usepackage{tikz}
\usepackage{xfrac}
\usepackage{footnote}
\usepackage{rotating}
\usepackage{lscape}
\usepackage{makecell}
\usepackage{environ}
\usepackage{tabularx}
\usepackage{subfiles}
\usepackage[export]{adjustbox}
\usepackage{ytableau}
\usepackage{tikz-3dplot}
\usepackage{slashed}
\usepackage{pifont}
\usepackage{multirow}
\usepackage{mdframed}
\usepackage{bbm}
\usepackage[normalem]{ulem}
\usepackage{arydshln}
\usepackage{enumitem} 
\usepackage{fancybox}
\usetikzlibrary{positioning,trees,decorations.pathmorphing,decorations.markings,decorations.pathreplacing,calc,shapes,patterns,arrows,chains,arrows.meta,fit,fadings,decorations.markings,graphs,graphs.standard,quotes,plotmarks,calligraphy}

\DeclareMathOperator{\U}{U}
\DeclareMathOperator{\SU}{SU}
\DeclareMathOperator{\SO}{SO}
\DeclareMathOperator{\SL}{SL}

\DeclareMathOperator{\USp}{USp}
\DeclareMathOperator{\tr}{Tr}

\newcommand{\coma}{\, , \quad}
\newcommand{\fstop}{\, .}

\newcommand{\ID}{\mathds{1}}

\def\ZZ{{\mathbb{Z}}}

\renewcommand{\epsilon}{\varepsilon}

\theoremstyle{plain}% default
\theoremstyle{definition}
\let\oldFootnote\footnote
\newcommand\nextToken\relax

\renewcommand\footnote[1]{%
    \oldFootnote{#1}\futurelet\nextToken\isFootnote}

\newcommand\isFootnote{%
    \ifx\footnote\nextToken\textsuperscript{,}\fi}

\tikzset{cross/.style={cross out, draw=black, fill=none, minimum size=2*(#1-\pgflinewidth), inner sep=0pt, outer sep=0pt}, cross/.default={2pt}}
\usetikzlibrary{positioning}
\usetikzlibrary{chains}
\usetikzlibrary{arrows, arrows.meta ,fit,decorations.pathreplacing}
\tikzstyle{every picture}+=[remember picture]
\tikzstyle{na} = [baseline]
\tikzstyle{ligne}=[draw, thick]

\usetikzlibrary{arrows, decorations.markings, calc, fadings, decorations.pathreplacing, patterns, decorations.pathmorphing, positioning}
\tikzset{>={Latex[width=1.5mm,length=1.5mm]}}
\tikzset{bd/.style={circle, draw=black, inner sep=0pt, fill=black, minimum size=1.2mm}}
\tikzset{bld/.style={circle, draw=blue, inner sep=0pt, fill=blue, minimum size=1.2mm}}
\tikzset{wd/.style={circle, draw=black, inner sep=0pt, fill=white, minimum size=1.2mm}}
\tikzset{rd/.style={circle, draw=red, inner sep=0pt, fill=red, minimum size=.9mm}}
\tikzset{wrd/.style={circle, draw=red, inner sep=0pt, fill=white, minimum size=.9mm}}
\usetikzlibrary{graphs,graphs.standard,quotes}

\def\node#1#2{\overset{#1}{\underset{#2}{{\color{gray} \bullet}}}}

\def\node#1#2{\overset{#1}{\underset{#2}{\circ}}}

\tikzstyle{every picture}+=[remember picture]
\tikzstyle{na} = [baseline=-.5ex]

\newcommand{\eg}{e.g. }

\newcommand{\ie}{i.e. }

\numberwithin{equation}{section}
\newcommand{\bes}[1]{\begin{equation} \begin{split} #1\end{split} \end{equation}}

\newcommand{\be}{\begin{equation}} \newcommand{\ee}{\end{equation}}
\newcommand{\bea}{\begin{equation} \begin{aligned}} \newcommand{\eea}{\end{aligned} \end{equation}}

\def\tilde{\widetilde}

\def\hat{\widehat}

\def\bar{\overline}

\def\rt2{\sqrt{2}}

\def\mod{{\rm mod}}
\def\det{\mathop{\rm det}}

\def\tr{\mathop{\rm tr}}

\def\CF{{\cal F}}

\def\CI{{\cal I}}

\def\CN{{\cal N}}

\def\CR{{\cal R}}
\def\CS{{\cal S}}
\def\CT{{\cal T}}

\def\CX{{\cal X}}

\def\CZ{{\cal Z}}

\def\1{{\ds 1}}

\newcommand{\fm}{\mathfrak{m}}
\newcommand{\fn}{\mathfrak{n}}

\newcommand{\fz}{\mathfrak{z}}

\def\SO{\mathrm{SO}}
\def\O{\mathrm{O}}
\def\SU{\mathrm{SU}}
\def\SL{\mathrm{SL}}

\def\Spin{\mathrm{Spin}}

\def\su{\mathfrak{su}}

\def\so{\mathfrak{so}}

\def\usp{\mathfrak{usp}}

\def\repa{\raise4pt\hbox{$\square$}\mkern-14mu\raise-4pt\hbox{$\square$}}
\def\repab{\overline{\raise4pt\hbox{$\square$}\mkern-14mu\raise-4pt\hbox{$\square$}\mkern-1mu}}

\def\smileface{\ensuremath{\hbox{\large$\bigcirc$}\mkern-15mu\raise-1pt\hbox{\scriptsize$\smallsmile$}%
\mkern-10mu\raise4pt\hbox{..}\mkern4mu}}
\def\frownface{\ensuremath{\hbox{\large$\bigcirc$}\mkern-15mu\raise-1pt\hbox{\scriptsize$\smallfrown$}%
\mkern-10mu\raise4pt\hbox{..}\mkern4mu}}

\newcommand{\ba}{\begin{array}}
\newcommand{\ea}{\end{array}}
\newcommand{\bi}{\begin{itemize}}
\newcommand{\ei}{\end{itemize}}
\def\vec#1{\bm{#1}}
\def\bea#1\eea{\allowdisplaybreaks \begin{align}#1\end{align}}
 \newcommand{\ben}{\begin{enumerate}}
\newcommand{\een}{\end{enumerate}}
\newcommand{\bean}{\begin{eqnarray*}}
\newcommand{\eean}{\end{eqnarray*}}
\newcommand{\eref}[1]{(\ref{#1})}

\newcommand{\PE}{\mathop{\rm PE}}

\newcommand{\BC}{\mathbb{C}}
\newcommand{\BR}{\mathbb{R}}

\newcommand{\BZ}{\mathbb{Z}}

\newcommand{\diag}{\mathrm{diag}}

\newcommand{\Sym}{\mathrm{Sym}}

\definecolor{light-gray}{gray}{0.5}
\definecolor{new-green}{rgb}{0,0.7,0.3}

\newcommand{\blue}{\color{blue}}

\newcommand{\red}{\color{red}}

\def\aup#1 {\overset{#1}{\uparrow} \, \overset{\tilde{#1}}{\downarrow}}

\tikzset{snake it/.style={decorate, decoration={snake, amplitude=.4mm, segment length=2mm,
                       post length=0mm,pre length=0mm}}}

\newcommand{\MA}{\mathds{A}}
\newcommand{\BM}{\mathds{B}}
\newcommand{\SM}{\mathds{S}}
\newcommand{\TM}{\mathds{T}}
\newcommand{\Bt}{\mathbf{t}}
\newcommand{\PM}{\mathds{P}}
\newcommand{\QM}{\mathds{Q}}
\newcommand{\ZM}{\mymathds{0}}
\newcommand{\BD}{\mathbb{D}}
\def\u{\mathfrak{u}}
\DeclareMathAlphabet{\mymathds}{U}{BOONDOX-ds}{m}{n}
\tikzstyle{double_border} = [draw, double, double distance=1pt]

\hypersetup{
	pdftitle={Discrete Global Symmetries: Gauging and Twisted Compactification},    % title
	pdfauthor={\textcopyright\ Simone Giacomelli, William Harding, Noppadol Mekareeya, Alessandro Mininno},     % author
	pdfsubject={HEP},   % subject of the document
	pdfcreator={pdfLaTex},   % creator of the document
	pdfproducer={LaTex}, % producer of the document
	pdfkeywords={},
	colorlinks=true,
}

\makeatletter
\newsavebox{\measure@tikzpicture}
\NewEnviron{scaletikzpicturetowidth}[1]{%
  \def\tikz@width{#1}%
  \begin{lrbox}{\measure@tikzpicture}%
  \BODY
  \end{lrbox}%
  \pgfmathparse{#1/\wd\measure@tikzpicture}%
  \BODY
}
\makeatother

\makeatletter
\def\squarecorner#1{
    \pgf@x=\the\wd\pgfnodeparttextbox%
    \pgfmathsetlength\pgf@xc{\pgfkeysvalueof{/pgf/inner xsep}}%
    \advance\pgf@x by 2\pgf@xc%
    \pgfmathsetlength\pgf@xb{\pgfkeysvalueof{/pgf/minimum width}}%
    \ifdim\pgf@x<\pgf@xb%
        \pgf@x=\pgf@xb%
    \fi%
    \pgf@y=\ht\pgfnodeparttextbox%
    \advance\pgf@y by\dp\pgfnodeparttextbox%
    \pgfmathsetlength\pgf@yc{\pgfkeysvalueof{/pgf/inner ysep}}%
    \advance\pgf@y by 2\pgf@yc%
    \pgfmathsetlength\pgf@yb{\pgfkeysvalueof{/pgf/minimum height}}%
    \ifdim\pgf@y<\pgf@yb%
        \pgf@y=\pgf@yb%
    \fi%
    \ifdim\pgf@x<\pgf@y%
        \pgf@x=\pgf@y%
    \else
        \pgf@y=\pgf@x%
    \fi
    \pgf@x=#1.5\pgf@x%
    \advance\pgf@x by.5\wd\pgfnodeparttextbox%
    \pgfmathsetlength\pgf@xa{\pgfkeysvalueof{/pgf/outer xsep}}%
    \advance\pgf@x by#1\pgf@xa%
    \pgf@y=#1.5\pgf@y%
    \advance\pgf@y by-.5\dp\pgfnodeparttextbox%
    \advance\pgf@y by.5\ht\pgfnodeparttextbox%
    \pgfmathsetlength\pgf@ya{\pgfkeysvalueof{/pgf/outer ysep}}%
    \advance\pgf@y by#1\pgf@ya%
}
\makeatother

\pgfdeclareshape{square}{
    \savedanchor\northeast{\squarecorner{}}
    \savedanchor\southwest{\squarecorner{-}}

    \foreach \x in {east,west} \foreach \y in {north,mid,base,south} {
        \inheritanchor[from=rectangle]{\y\space\x}
    }
    \foreach \x in {east,west,north,mid,base,south,center,text} {
        \inheritanchor[from=rectangle]{\x}
    }
    \inheritanchorborder[from=rectangle]
    \inheritbackgroundpath[from=rectangle]
}

\tikzset{stretch/.initial=1}
\newcommand\drawloop[4][]%
   {\draw[shorten <=0pt, shorten >=0pt,#1]
      ($(#2)!\pgfkeysvalueof{/tikz/stretch}!(#2.#3)$)
      let \p1=($(#2.center)!\pgfkeysvalueof{/tikz/stretch}!(#2.north)-(#2)$),
          \n1= {veclen(\x1,\y1)*sin(0.5*(#4-#3))/sin(0.5*(180-#4+#3))}
      in arc [start angle={#3-90}, end angle={#4+90}, radius=\n1]%
   }

\preprint{ZMP-HH/24-2}
\title{Discrete Global Symmetries: Gauging and Twisted Compactification}
\author[a]{Simone Giacomelli,}
\author[a,b]{William Harding,}
\author[b,c]{Noppadol Mekareeya}
\author[d]{and Alessandro Mininno}
\affiliation[a]{Dipartimento di Fisica, Universit\`a di Milano-Bicocca, \\ Piazza della Scienza 3, I-20126 Milano, Italy}
\affiliation[b]{INFN, sezione di Milano-Bicocca, \\Piazza della Scienza 3,  I-20126 Milano, Italy}
\affiliation[c]{Department of Physics, Faculty of Science, Chulalongkorn University, \\ Phayathai Road,
Pathumwan, Bangkok 10330, Thailand}
\affiliation[d]{II. Institut f\"ur Theoretische Physik, Universit\"at Hamburg,\\
Luruper Chaussee 149, 22607 Hamburg, Germany}
\emailAdd{simone.giacomelli@unimib.it}
\emailAdd{w.harding@campus.unimib.it}
\emailAdd{n.mekareeya@gmail.com}
\emailAdd{alessandro.mininno@desy.de}
\abstract{Discrete global symmetries of 4d $\mathcal{N}=2$ SCFTs are studied via two operations: gauging and twisted compactification. We consider gauging of discrete symmetries in several well-known 4d $\mathcal{N}=2$ SCFTs, including $\mathrm{SU}(n)$ SQCD with $2n$ flavors, theories of class $\mathcal{S}$ of type $A_{2n-1}$, and Argyres--Douglas theories of type $(A_N, A_N)$, as well as propose new 4d SCFTs as a result. The wreathing technique, which involves gauging a subgroup of the automorphism group of the quiver diagram of the corresponding 3d mirror theory, is exploited. This allows us to understand several properties of discretely gauged theories, including moduli spaces and how discrete gauging affects the mixed 't Hooft anomaly between the 1-form symmetry and the 0-form flavor symmetry. Many examples are viewed through the lens of the Argyres--Seiberg duality and its generalization. We also examine discrete gauging of $\mathrm{SU}(2)$ SQCD with 4 flavors by various $\mathbb{Z}_2$ and $\mathbb{Z}_2 \times \mathbb{Z}_2$ subgroups of the permutation group $S_4$ using the superconformal index. Regarding compactification, we propose a magnetic quiver for 4d $\mathcal{N}=2$ $\SU(n)$ SQCD with $2n$ flavors compactified on a circle with a $\mathbb{Z}_2$ twist. The twisted compactification by non-invertible symmetries of the 4d $\mathcal{N}=4$ SYM theory with gauge group $\mathrm{SU}(N)$ is revisited. The non-invertible symmetry naturally gives rise to a $\mathbb{Z}_k$ action on the scalar fields parametrizing  the moduli space.  Upon examining the $\mathbb{Z}_k$ invariant chiral ring of the Higgs branch, we find that, in addition to the largest branch of the moduli space that is expected to be captured by the ABJ(M) theory, there exist in general nilpotent operators that lead to a branch of the moduli space which is a radical ideal.   
}

\allowdisplaybreaks[1]

\begin{document} 

\maketitle

\section{Introduction}
\label{sec:introduction}
The central theme of this paper involves discrete global symmetries of superconformal field theories (SCFTs) in three and four dimensions. The two operations that will be studied are gauging of discrete symmetries and compactification of SCFTs on a circle with an outer-automorphism twist. Both are classic methods that can be used to obtain new theories, some of which are dual to each other in an intricate manner. 

In gauging a discrete symmetry, the background gauge field associated with the symmetry in question is summed over in the path integral. A notable class of 4d SCFTs that can be obtained from discrete gauging is that with $\CN=3$ supersymmetry \cite{Garcia-Etxebarria:2015wns, Aharony:2016kai}. A large family of them can be obtained by gauging an appropriate discrete subgroup of the $\mathrm{SL}(2,\BZ)$ duality group of 4d $\CN=4$ super Yang--Mills (SYM) theories at a certain value of the coupling. This idea can be generalized to theories with 4d $\CN=2$ supersymmetry in many ways, see \eg  \cite{Argyres:2016yzz, Bourget:2017tmt, Bourget:2018ond, Arias-Tamargo:2019jyh, Apruzzi:2020pmv, Giacomelli:2020gee, Giacomelli:2020jel, Arias-Tamargo:2023duo}. Moreover, discrete gauging can also involve discrete higher-form symmetries \cite{Kapustin:2014gua, Gaiotto:2014kfa}. In particular, we will investigate theories whose 1-form symmetry has a mixed 't Hooft anomaly with a continuous flavor symmetry and gauging thereof. 
%Another type of symmetry that will play an important role in this paper is discrete higher-form symmetry \cite{Kapustin:2014gua, Gaiotto:2014kfa}. We will investigate theories whose 1-form symmetry has a mixed 't Hooft anomaly with continuous flavor symmetry and gauging thereof. 

On the other hand, for a circle compactification, we may twist the theory around the circle by introducing a holonomy for the background gauge fields associated to discrete global symmetries of the theory, giving rise to a twisted compactified theory on a circle.  Upon shrinking the size of the circle to zero, we obtain a new theory in one dimension lower. For example, many new 5d theories can be obtained from 6d SCFTs in this way, see \eg  \cite{Bhardwaj:2018yhy, Bhardwaj:2018vuu, Kim:2019dqn, Bhardwaj:2019fzv, Bhardwaj:2019xeg, Bhardwaj:2020kim, Bhardwaj:2020gyu}.  Another prominent example is S-duality of 5d maximally SYM theories on $S^1$, whereby the tower of Kaluza--Klein W-bosons and the tower of instantonic monopoles are interchanged. This maps a simply-laced gauge theory on $S^1$ to itself and maps a non-simply-laced gauge theory on $S^1$ to a simply-laced gauge theory twisted by an outer-automorphism around $S^1$, see \eg  \cite{Lee:1998vu, Davies:2000nw, Hanany:2001iy, Witten:2009at, Tachikawa:2011ch}.

This paper contains two major parts. In the first part, we consider discrete gauging of certain well-known 4d $\CN=2$ SCFTs, including the $\SU(n)$ gauge theory with $2n$ flavors of hypermultiplets, certain theories of class $\CS$ of type $A_{2n-1}$ \cite{Gaiotto:2009we, Gaiotto:2009hg}, and the Argyres--Douglas theories of type $(A_N, A_N)$ \cite{Cecotti:2010fi}. Importantly, new 4d $\CN=2$ SCFTs are proposed as a result of this study.  Various properties of the discretely gauged theories, such as their moduli spaces, 1-form global symmetries and the mixed 't Hooft anomalies with the 0-form flavor symmetries, are investigated. The main tools that we exploit are wreathing \cite{Bourget:2020bxh} the 3d mirror theories \cite{Intriligator:1996ex} of the 4d SCFTs in question, as well as their Hilbert series (see also \cite{Hanany:2018vph, Hanany:2018cgo, Hanany:2018dvd, Bourget:2018ond, Arias-Tamargo:2019jyh, Arias-Tamargo:2021ppf, Hanany:2023uzn}). Roughly speaking, wreathing is a method of discretely gauging the automorphism group, or its subgroup, of a given quiver. The Coulomb branch of the wreathed quiver can be used to study the Higgs branch of the 4d $\CN=2$ SCFTs at issue. We also examine how wreathing affects the mixed 't Hooft anomalies between the $1$-form and the 0-form flavor symmetries in a theory of class $\CS$. For the $\SU(2)$ gauge theory with $4$ hypermultiplets in the fundamental representation, we also consider the wreathing by various $\BZ_2$ and $\BZ_2 \times \BZ_2$ subgroups of the permutation group $S_4$ from the perspective of the 3d superconformal index \cite{Bhattacharya:2008zy,Bhattacharya:2008bja, Kim:2009wb,Imamura:2011su, Kapustin:2011jm, Dimofte:2011py, Aharony:2013dha, Aharony:2013kma}. With this approach, the physical origin of the discrete groups involved in wreathing becomes clear: they consist of a $0$-form symmetry that acts on a subset of the hypermultiplets and various charge conjugation symmetries associated with the flavor symmetries. 

The second part of the paper deals with compactification of 4d SCFTs on a circle with a twist. We consider the twisted compactification of the 4d $\CN=2$ $\SU(n)$ gauge theory with $2n$ flavors.  Along the line of \cite{Bourget:2020asf, Bourget:2020mez}, we propose the 3d mirror (also known as magnetic quiver) containing a non-simply-lace edge associated with the 3d theory arising from the $\BZ_2$ twisted compactification.  This non-simply-laced quiver can be obtained by ``folding'' a certain part of the 3d mirror theory of the $\SU(n)$ gauge theory with $2n$ flavors. As before, the Coulomb branch Hilbert series of such a non-simply-laced quiver can be computed using the prescription of \cite{Cremonesi:2014xha} and this should be identified with that of the Higgs branch of the compactified theory. We study how the Higgs branch operators are projected out as a result of the $\BZ_2$ twist. The final part of the paper revisits the twisted compactification by non-invertible symmetries of the 4d $\CN=4$ SYM theory with $\SU(N)$ gauge group studied in \cite{Kaidi:2022uux}. It was pointed out that the $\BZ_k$ twist projects out operators in such a way that the moduli space of the resulting 3d theory is a symmetric product of a $\BZ_k$ orbifold of flat space. The latter leads to the identification of the 3d theory as the ABJ(M) theory \cite{Aharony:2008ug, Aharony:2008gk}, which indeed possesses the aforementioned moduli space. In this work, we view the $\CN=4$ SYM theory as an $\CN=2$ gauge theory and focus on its Higgs branch. This is parametrized by the gauge invariant quantities constructed from two commuting chiral fields in the adjoint representation of $\SU(N)$. Upon considering the $\BZ_k$ invariant sector of the said Higgs branch chiral ring, we find that the component of the largest dimension is always described by a symmetric product of a $\BZ_k$ orbifold of flat space. However, to our surprise, there is generally a component of the quotient ring that is parametrized by nilpotent operators, leading to the so-called {\it radical ideal}. Since the moduli space of the ABJ(M) theory is expected to be a symmetric product of a $\BZ_k$ orbifold of flat space, such radical ideals are not captured by the ABJ(M) theory. The origin of such radical ideals and their physical relation to the twisted compactification are not clear to us at present, and these deserve further investigations in the future.

\subsubsection*{Structure of the Paper}

The paper is structured as follows. The first part of the paper consists of Sections \ref{sec:newSCFTdiscretegauging} to \ref{app:mirrordualsD4}. In Section \ref{sec:newSCFTdiscretegauging}, we propose new 4d $\mathcal{N}=2$ SCFTs obtained by a $\BZ_2$ discrete gauging of various 4d parent theories. In Sections \ref{sec:newrank2SCFT} and \ref{sec:NewRankn-1SCFT}, the parent theory is $\SU(n)$ SQCD with $2n$ flavors, while in Section \ref{sec:NewRank2SCFT-2}, the parent theory is the class $\mathcal{S}$ theory of type $A_1$ on a Riemann surface of genus 1 and two punctures. The discrete gauging is explored by wreathing the corresponding 3d mirror theories, as well as through the Argyres--Seiberg duality \cite{Argyres:2007cn} and its generalization. The computation of the Coulomb Branch Hilbert series of the wreathed quiver is reviewed in Section \ref{sec:reviewWreathing}. In Section \ref{sec:disgauging-hooft}, we investigate the effects of discrete gauging on the mixed 't Hooft anomalies between the $\BZ_n$ 1-form symmetry and the 0-form flavor symmetry in the theory of class $\CS$ of type $A_{2n-1}$ with four twisted trivial punctures and four untwisted minimal punctures. The method that we adopt to examine such 't Hooft anomalies is the one discussed in \cite{Mekareeya:2022spm, Sacchi:2023omn} applied to the 3d mirror theory wreathed by various subgroups of the permutation group $S_4$. The case of $n=1$ of this class $\CS$ theory is a close cousin of $\SU(2)$ SQCD with $4$ flavors, whereby if the 1-form symmetry is gauged, the resulting theory is precisely the aforementioned SQCD. We analyze discrete gauging of this theory, as well as the SQCD theory, by the $\BZ_2$ and $\BZ_2 \times \BZ_2$ subgroups of $S_4$ from the perspective of the 3d superconformal index in Section \ref{app:mirrordualsD4}. The analysis of the $n=2$ case is delegated to Appendix \ref{sec:Casen2USp4}.  The second part of the paper consists of Sections \ref{sec:twistedcompactificationsSUN} and \ref{sec:twistedcompactification-SYM}. In Section \ref{sec:twistedcompactificationsSUN}, we examine the $\BZ_2$ twisted compactification of 4d $\CN=2$ $\SU(n)$ SQCD with $2n$ flavors on a circle and provide a description of the magnetic quiver in terms of that with a non-simply-laced edge.  In Section \ref{sec:twistedcompactification-SYM}, we analyze the $\BZ_k$ invariant chiral ring of the Higgs branch of the 4d $\mathcal{N}=4$ $\SU(N)$ SYM theory viewed as an $\CN=2$ gauge theory. We point out the presence of nilpotent operators and radical ideals in the cases of $N=3$ and $N=4$.  In Appendix \ref{sec:discretegaugingAA}, we consider discrete gauging of the $(A_N, A_N)$ Argyres--Douglas theories via wreathing of the 3d mirror theories, which admit the description in terms of complete graphs. Finally, in Appendix \ref{app:complexreflectiongroup}, we provide a detailed computation of the Hilbert series of the quotient of flat space by complex reflection groups.  This is useful for studying the moduli space of $4d$ $\CN=3$ SCFTs, as well as that of the ABJ(M) theory.

\section{New SCFTs from Discrete Gauging}
\label{sec:newSCFTdiscretegauging}

In this section, we obtain new SCFTs as a result of gauging a discrete $\BZ_2$ symmetry of well-known 4d $\CN = 2$ theories, including $\SU(n)$ SQCD with $2 n$ flavors and the $A_1$ class $\mathcal{S}$ theory associated with a genus-1 Riemann surface with two punctures. As a result of discrete gauging of SQCD, we obtain novel $\CN = 2$ SCFTs which, to the best of our knowledge, have never been explored before in literature. Specifically, the case of $\SU(3)$ SQCD with $6$ flavors is analyzed by exploiting the Argyres--Seiberg duality \cite{Argyres:2007cn}, and a generalization thereof is implemented in order to construct new SCFTs arising from $\SU(n)$ SQCD with $n > 3$. Furthermore, the $\BZ_2$ transformation of our interest is carefully analyzed in the case of $\SU(2)$ SQCD with 4 flavors. We subsequently examine the theory of class $\CS$ and the new SCFT with $\so(4)$ global symmetry.
The Higgs branch of the said theories can be studied via wreathing of the corresponding 3d $\CN = 4$ magnetic quivers.

\subsection{Review of Wreathed Quivers}
\label{sec:reviewWreathing}

In this section, we review the basic concepts of wreathing, which is a vital tool for studying discrete gauging of 3d $\CN=4$ quiver gauge theories. We refer the reader to \cite{Bourget:2020bxh} for a more detailed discussion. Given a finite group $A$, the wreathing of $A$ by a subgroup $\Gamma$ of the permutation group $S_n$ is denoted by $A \wr \Gamma$, or $A_{\text{wr}\, \Gamma}$, and is defined by letting $\Gamma$ act on the direct product of $n$ copies of $A$. For example, $\BZ_m \wr S_n$ can be realized as follows: we start from the $n$-fold direct product $\BZ_m^n = \BZ_m \times \ldots \times \BZ_m$ ($n$ times) and consider the action $f:S_n \rightarrow \mathrm{Aut}(\BZ_m^n)$ such that $f(\sigma)(x_1, x_2, \ldots, x_n) = (x_{\sigma(1)}, x_{\sigma(2)}, \ldots, x_{\sigma(n)})$, where $\sigma \in S_n$ and $x_j \in \BZ_m$ for all $j=1, \ldots, n$.  Note that the order of $\BZ_m \wr S_n$ is $m^n|S_n| = m^n n!$.

Given a 3d $\CN=4$ quiver gauge theory $\CX$ containing a family of unitary gauge groups $G=\prod_{i=1}^M \U(n_i)$ connected by a set of edges associated with bifundamental hypermultiplets, the Coulomb branch Hilbert series of theory $\CX$ can be written as
\bes{ \label{CBHST}
\mathrm{HS}\left[\text{CB of $\CX$}\right](t)= \frac{1}{|W|} \sum_{\vec m \in \BZ^r} \sum_{\gamma \in W(\vec m)} \frac{t^{2\Delta(\vec m)}}{\det(\ID-t^2 \gamma)} = \sum_{\vec \fm \in \BZ^r/W} t^{2\Delta(\vec \fm)} P_G(\vec \fm;t)~,
}
where $W$ is the Weyl group $W=\prod_{i = 1}^M S_{n_i}$, $r$ is the total rank of the gauge groups $r= \sum_{i = 1}^M n_i$, and $\Delta(\vec m)$ is the dimension of the monopole operator with magnetic flux $\vec m$.  In the second equality, we provide an equivalent way of writing the formula, as discussed in \cite{Cremonesi:2013lqa}, where
\bes{ \label{dressinggen}
P_G(\vec \fm;t) = \frac{1}{|W_{H(\vec \fm)}|} \sum_{g \in W_{H(\vec \fm)}} \frac{1}{\det(\ID- t^2g)}
}
is known as the dressing factor. Here, $H(\vec \fm)$ is the residual gauge symmetry (depending on the magnetic flux $\vec \fm$), which is a subgroup of $G$, and $W_{H(\vec \fm)}$ is the associated Weyl group. Note that \eref{dressinggen} is indeed the discrete Molien formula applied to $H(\vec \fm)$.

Suppose that the quiver diagram of $\CX$ has a symmetry $\Gamma$, and let us say that $\Gamma$ leaves $\Delta(\vec m)$ invariant. We can let $\Gamma$ act on the gauge factors of $\CX$ and the resulting quiver is known as a wreathed quiver, which is denoted by $\CX \wr \Gamma$, or $\CX_{\text{wr}\, \Gamma}$. The action of $\Gamma$ on the Coulomb branch coordinates of $\CX \wr \Gamma$ was discussed in \cite{Bourget:2020bxh}. The Coulomb branch Hilbert series of the wreathed quiver is
\begin{equation} \label{CBHSTwr}
    \mathrm{HS}\left[\text{CB of $\CX\wr \Gamma$}\right](t) = \frac{1}{|W_\Gamma|}\sum_{\vec n \in \BZ^r}\sum_{\gamma\in W_\Gamma(\vec{n})}\frac{t^{2\Delta(\vec {n})}}{\det(\ID-t^2\gamma)} = \sum_{\fn \in \BZ^r/W_\Gamma} t^{2\Delta(\fn)} P_G(\vec \fn;t)~,
\end{equation}
where $W_\Gamma = W \wr \Gamma$, i.e. the wreathing of the Weyl group $W$ by $\Gamma$, as discussed above. We emphasize that, although the dressing factor is given by the discrete Molien formula of the residual gauge symmetry, the explicit forms of the dressing factors in \eref{CBHST} and \eref{CBHSTwr} may be different due to the different summations over $\fm$ and $\fn$ which may result in the difference in $H(\fm)$ and $H(\fn)$.

Let us also comment on the effect of wreathing on the topological symmetry. First, we consider the theory $\CX$. Let $z_i$ be the topological fugacity associated to the gauge group $\U(n_i)$. The Coulomb branch Hilbert series can be refined with respect to this symmetry by inserting in the summation of \eref{CBHST} the factor $\prod_{i=1}^M z_i^{\sum_{j=1}^{n_i} \fm^{(j)}_i}$, where $\fm = (\fm_1, \fm_2, \ldots, \fm_M)$ and $\fm^{(j)}_i$ denotes the $j$-th component of $\fm_i$.  Now, let $\Gamma$ act {\it non-trivially} on a subset of gauge nodes of $\CX$, say $\U(n_1), \ldots, \U(n_p)$. The topological symmetry $\U(1)^p$ associated with such gauge groups is broken to its $\U(1)$ diagonal subgroup, which means that we must set $z_1=z_2=\ldots=z_p$ when computing the Hilbert series.  For this reason, the Coulomb branch symmetry of the wreathed quiver $\CX \wr \Gamma$ is a subset of that of the original theory $\CX$.

If a 4d $\CN=2$ SCFT $\CT$ is known to have a 3d mirror theory described by $\CX$, we can study the Higgs branch of $\CT$ by examining the Coulomb branch of $\CX$. It is shown by a collection of results in \cite{Argyres:2016yzz, Hanany:2018vph, Hanany:2018cgo, Hanany:2018dvd, Bourget:2020bxh, Hanany:2023uzn} that the Coulomb branch of the wreathed quiver $\CX \wr \Gamma$ can be identified with the Higgs branch of $\CT$ discretely gauged by $\Gamma$. In this sense, the wreathed quiver serves as a magnetic quiver of the 4d $\CN=2$ discretely gauged SCFT in question.

\subsection{New Rank-2 SCFT with \texorpdfstring{$\USp(6) \times \U(1)$}{USp(6)xU(1)} Global Symmetry}
\label{sec:newrank2SCFT}

In this section, we discuss a new 4d $\CN=2$ rank-2 SCFT, which we shall denote by $\CT^{r=2}_{C_3U_1}$, obtained by a $\BZ_2$ discrete gauging of the 4d $\CN=2$ $\SU(3)$ gauge theory with 6 flavors.  The two Coulomb branch operators of $\CT^{r=2}_{C_3U_1}$ have scaling dimensions $2$ and $6$.  We will investigate the Higgs branch of $\CT^{r=2}_{C_3U_1}$ in two ways. The first way is by studying the Coulomb branch of a wreathed quiver, and the second way is by exploiting the Argyres--Seiberg duality.

\subsubsection{Wreathing the Mirror of \texorpdfstring{$\SU(3)$}{SU(3)} SQCD with 6 Flavors}
\label{sec:wreathingSU3SQCD}

Let us discuss the first approach. We start from the mirror theory of 3d $\CN=4$ $\SU(3)$ SQCD with 6 flavors and gauge a $\BZ_2$ symmetry of the quiver in the following way: 
\bes{ \label{mirrSU3w6}
\begin{tikzpicture}[baseline,font=\footnotesize,scale=1]
\node (U3) at (0,0) {$\U(3)$}; 
\node (U1a) at (-0.5,1.1) {$\U(1)$}; 
\node (U1b) at (0.5,1.1) {$\U(1)$}; 
\node (U2l) at (-1.6,0) {$\U(2)$};
\node (U2r) at (1.6,0) {$\U(2)$};
\node[draw=none] (U1r) at (3.1,0) {$\U(1)$};
\node[draw=none] (U1l) at (-3.1,0) {$\U(1)$};
\draw (U1l)--(U2l)--(U3)--(U2r)--(U1r);
\draw (U1a)--(U3)--(U1b);
\draw [<->,red] (U2l) to [out=-140,in=-40,looseness=0.7] (U2r);
\draw [<->,red] (U1l) to [out=-140,in=-40,looseness=0.6] (U1r);
\node at (5,0) {$/\U(1)$};
\end{tikzpicture}
}
where the {\red red} arrows show the action on the gauge groups that are involved in the $\BZ_2$ discrete gauging. The resulting theory is the required wreathed quiver, whose Coulomb branch is identified with the Higgs branch of $\CT^{r=2}_{C_3U_1}$.

It should be noted that this is different from the wreathing considered in \cite[Section 4.2]{Arias-Tamargo:2021ppf} and in \cite[Section 3.3]{Arias-Tamargo:2023duo}. Here, the two $\U(1)$ nodes above the $\U(3)$ node are left untouched, whereas in \cite{Arias-Tamargo:2021ppf} they are identified in the process of discrete gauging, which corresponds to gauging the charge conjugation symmetry of 4d $\CN = 2$ $\SU(3)$ SQCD with 6 flavors. The gauging we are considering here is indeed slightly different, and it does not correspond to gauging the charge conjugation symmetry. Using the Hilbert series, we will show that this wreathed quiver possesses $\usp(6) \oplus \u(1)$ global symmetry, not $\usp(6)$ as in \cite{Arias-Tamargo:2021ppf, Arias-Tamargo:2023duo}.  The $\U(1)$ global symmetry in our discussion can indeed be identified with the baryonic symmetry of the SQCD theory, and this is not projected out by discrete gauging.\footnote{Since the baryonic symmetry survives the projection, the discrete gauging in question cannot be just exchanging quarks and antiquarks, as acted by the charge conjugation symmetry.}  

We now discuss the computation of the Coulomb branch Hilbert series.  First, we recall that the Coulomb branch Hilbert series of \eref{mirrSU3w6} {\it without} wreathing is
\bes{ \label{HSCBmirrSU3w6}
&(1-t^2) \sum_{u_R \in \BZ}\,\,\sum_{u_L \in \BZ}\,\,\sum_{u_{T_R} \in \BZ}\,\, \sum_{u_{T_L} \in  \BZ} \,\, \sum_{v^{(1)}_R \geq v^{(2)}_R > -\infty} \,\, \sum_{v^{(1)}_L \geq v^{(2)}_L > -\infty} \,\, \sum_{m^{(1)} \geq m^{(2)} \geq m^{(3)} =0} t^{2\Delta} \\
& \qquad \quad \times \frac{P_{{\U(2)}}(v^{(1)}_R, v^{(2)}_R; t) P_{{\U(2)}}(v^{(1)}_L, v^{(2)}_L; t) P_{{\U(3)}}(m^{(1)}, m^{(2)}, m^{(3)}; t)}{(1-t^2)^4} \\
& \qquad \quad \times z_{1}^{u_R} z_{2}^{u_L}z_3^{u_{T_R}}z_4^{u_{T_L}}z_5^{v^{(1)}_R+v^{(2)}_R}z_6^{v^{(1)}_L+v^{(2)}_L}z_7^{m^{(1)}+m^{(2)}+m^{(3)}}~,
}
where the factor $(1-t^2)$ and the restriction $m^{(3)}=0$ appearing in the first line are there for modding out an overall $\U(1)$, moreover
\bes{
2\Delta =& - 2|v^{(1)}_R -v^{(2)}_R| + \sum_{i=1}^2 |u_R-v^{(i)}_R| + \sum_{i=1}^2 \sum_{j=1}^3 |v^{(i)}_R-m^{(j)}|  + \left( R \leftrightarrow L\right) \\
& - 2 \sum_{1 \leq k < l \leq 3}| m^{(k)}- m^{(l)}|+\sum_{j=1}^3 \left( |u_{T_R}-m^{(j)}|+ |u_{T_L}-m^{(j)}| \right)~,
}
the dressing factors appearing in the second line are
\bes{ \label{PU2U3}
P_{\U(2)}(n_1, n_2;t) &= \begin{cases}  (1-t^2)^{-2} &\quad n_1\neq n_2 \\
(1 - t^2)^{-1} (1 - t^4)^{-1} &\quad n_1=n_2\end{cases}~, \\
P_{\U(3)}(n_1, n_2, n_3;t) &= \begin{cases}  (1-t^2)^{-3} &\quad n_1\neq n_2 \neq n_3 \\
(1-t^2)^{-2} \left(1-t^4\right)^{-1} &\quad \text{if precisely two $n_i$ are equal} \\
(1 - t^2)^{-1} (1 - t^4)^{-1} (1 - t^6)^{-1} &\quad n_1=n_2=n_3\end{cases} ~,
}
and each $\U(1)$ gauge group contributes $(1-t^2)^{-1}$ so that the factor $(1-t^2)^{-4}$ in the second line comes from the fact that we have four $\U(1)$ gauge groups.  The third line represents the topological fugacities satisfying $z_1 z_2 z_3 z_4 z_5^2 z_6^2 z_7^3=1$.  Evaluating the summations in \eref{HSCBmirrSU3w6}, we indeed obtain the Higgs branch Hilbert series of $\SU(3)$ SQCD with 6 flavors.

To compute the Coulomb branch Hilbert series of the wreathed quiver, we follow the procedure described in \cite[Section 4.2]{Arias-Tamargo:2021ppf}. In particular, the range of summations over the magnetic fluxes in the first line of \eref{HSCBmirrSU3w6} needs to be restricted, the dressing factors in the second line of \eref{HSCBmirrSU3w6} are modified, and finally we make the identification $z_1=z_2$ and $z_5=z_6$.  Explicitly, the range of summations is restricted, and the dressing factors are modified as follows:
\bes{ 
\scalebox{0.94}{$
\renewcommand*{\arraystretch}{1.4}
    \begin{array}{c|c}
        \text{Restriction} & \text{Dressing factor} \\
        \hline
        u_R<u_L & {(1-t^2)^{-2}} P_{{\U(2)}}(\vec v_R; t) P_{{\U(2)}}(\vec v_L; t) P_{{\U(3)}}(\vec m; t)  \\
        u_R=u_L, v^{(2)}_R < v^{(2)}_L & (1-t^2)^{-2} P_{{\U(2)}}(\vec v_R; t) P_{{\U(2)}}(\vec v_L; t) P_{{\U(3)}}(\vec m; t) \\
        u_R=u_L, v^{(2)}_R= v^{(2)}_L , v^{(1)}_R< v^{(1)}_L & (1-t^2)^{-2} P_{{\U(2)}}(\vec v_R; t) P_{{\U(2)}}(\vec v_L; t) P_{{\U(3)}}(\vec m; t) \\
        u_R=u_L, v^{(2)}_R= v^{(2)}_L , v^{(1)}_R= v^{(1)}_L, v^{(2)}_L < v^{(1)}_L  & (1+3t^4)(1-t^2)^{-3}(1-t^4)^{-3} P_{{\U(3)}}(\vec m; t) \\
        u_R=u_L, v^{(2)}_R= v^{(2)}_L , v^{(1)}_R= v^{(1)}_L, v^{(2)}_L = v^{(1)}_L  & (1-t^2+2t^4)(1-t^2)^{-3}(1-t^4)^{-2}(1-t^8) P_{{\U(3)}}(\vec m; t) 
    \end{array}
    $}
    \label{eq:dressTable}
}
The dressing factors, apart from $P_{{\U(3)}}(\vec m; t)$, in the last two lines of \eref{eq:dressTable} are computed from the discrete Molien formula for the appropriate finite groups $G$ as described in \eref{dressinggen}, namely
\bes{ \label{Molien}
\frac{1}{|G|} \sum_{g \in G} \frac{1}{\det(\ID - g t^2)}~,
}
with $G= \langle (12)(35)(46) \rangle$ and $G=\langle (12)(35)(46), (56) \rangle$ respectively.\footnote{Here and in the rest of the paper, we use the notation $\langle c_1, c_2, \ldots, c_\ell \rangle$ to denote the group generated by the permutation cycles $c_1, \ldots, c_\ell$. In this particular case, the former contains two elements, while the latter contains eight elements.} In this case, $\ID$ can be taken as a $6 \times 6$ identity matrix and $g$ can be taken as $6\times 6$ matrix representations of the elements of $G$.  For reference, we list a few terms (with all $z_i$ set to unity) for each restriction as follows:
\bes{
\begin{array}{rcrcrcrcrcrcrcrcl}
  &   & 8 t^2 & + & 8 t^3 & + & 188 t^4 & + & 332 t^5 & + & 2532 t^6 & + & 5300 t^7 & + & 24768 t^8 & + & \mathcal{O}(t^9)~, \\
  &   & 2 t^2 & + & 2 t^3 & + & 41 t^4  & + & 74 t^5  & + & 508 t^6  & + & 1076 t^7 & + & 4629 t^8  & + & \mathcal{O}(t^9)~, \\
  &   & 2 t^2 & + & 2 t^3 & + & 33 t^4  & + & 58 t^5  & + & 336 t^6  & + & 684 t^7  & + & 2599 t^8  & + & \mathcal{O}(t^9)~, \\
  &   & 4 t^2 & + & 8 t^3 & + & 49 t^4  & + & 104 t^5 & + & 405 t^6  & + & 840 t^7  & + & 2489 t^8  & + & \mathcal{O}(t^9)~, \\
1 & + & 6 t^2 & + & 8 t^3 & + & 32 t^4  & + & 48 t^5  & + & 153 t^6  & + & 248 t^7  & + & 606 t^8   & + & \mathcal{O}(t^9)~.
\end{array}
}
Summing all of these contributions, we have the unrefined Hilbert series:
\bes{
\scalebox{0.97}{$
\begin{split}
\mathrm{HS}\left[\text{CB of } \eref{mirrSU3w6}\right] (t) &=
    1+ 22t^2+28 t^3 +343 t^4 +616 t^5 + 3934 t^6 + 8148 t^7 + 35091 t^8+\mathcal{O}(t^9) \\
&= \PE\left[22 t^2 + 28 t^3 +90 t^4 - 476 t^6 -1456 t^7 - 2884 t^8+\mathcal{O}(t^9)\right]~.
\end{split}$}
}
We can turn on $z_i$ and obtain the Higgs branch Hilbert series of $\CT^{r=2}_{C_3U_1}$ in terms of $\usp(6) \oplus \u(1)$ representations.  This can be represented in terms of the highest weight generating function (HWG) as follows:
\bes{ \label{HWGHBTC3U1}
&\mathrm{HWG}\left[\text{HB of} \, \CT^{r=2}_{C_3U_1}\right] (\mu_1, \mu_2, \mu_3; t) = \mathrm{HWG}\left[\text{CB of } \eref{mirrSU3w6}\right] (\mu_1, \mu_2, \mu_3; t) \\
&=\PE \left[\left(\mu_1^2+1\right) t^2 + \mu_3 \left(b^3+b^{-3}\right) t^3 +\left(2 \mu_2^2+\mu_2+1\right) t^4 + \left(\mu_1 \mu_2+\mu_3\right) \left(b^3+b^{-3}\right) t^5 \right.\\
& \left.\qquad \qquad \qquad \quad \, \, \, \, \, +  \left\{\mu_1^2 \left(b^6+1+b^{-6}\right)+\mu_1 \mu_2 \mu_3+\mu_1 \mu_3+\mu_2^2+\mu_3^2\right\} t^6 +\mathcal{O}(t^7)  \right]~.
}
Upon expanding this in powers of $t$, 
the term $\mu_1^{p_1} \mu_2^{p_2} \mu_3^{p_3}$ gives the character of the representation $[p_1,p_2,p_3]_{\usp(6)}$ of $\usp(6)$.\footnote{Throughout this paper, we will denote the character $\chi_{[a_1, \ldots, a_n]}^{\mathfrak{g}}\left(\vec z\right)$ of the representation $[a_1, \ldots, a_n]$ of $\mathfrak{g}$ as $[a_1, \ldots, a_n]_{\mathfrak{g}}$. For simplicity, we omit the explicit dependence from the fugacities $\vec z = \left(z_1, \ldots, z_n\right)$ of $\mathfrak{g}$.}
The Hilbert series is given by
\bes{ \label{HSHBTC3U1}
\scalebox{0.91}{$
\begin{split}
&\mathrm{HS}\left[\text{HB of} \, \CT^{r=2}_{C_3U_1}\right] (\vec{z}; t) = \mathrm{HS}\left[\text{CB of} \eref{mirrSU3w6}\right] (\vec{z}; t) \\
& = \PE \left[ \left([2,0,0]_{\usp(6)} + 1\right) t^2 +\left(b^3+b^{-3}\right)[0,0,1]_{\usp(6)} t^3 + [0,2,0]_{\usp(6)} t^4\right. \\
& \left. \quad \, \, \, \, \, \, - \left([2, 1, 0]_{\usp(6)}+ 2 [2, 0, 0]_{\usp(6)}+ [1, 0, 1]_{\usp(6)} +[0, 2, 0]_{\usp(6)} + 
   [0, 0, 2]_{\usp(6)} +1\right) t^6  + \mathcal{O}(t^7) \right]~.
\end{split}$}
}
The term at order $t^2$ confirms that the global symmetry of $\CT^{r=2}_{C_3U_1}$ is indeed $\usp(6) \oplus \u(1)$. Note that we have chosen the normalization of the $\u(1)$ fugacity $b$ to be as that of the baryonic symmetry of $\SU(3)$ SQCD with 6 flavors.

It is instructive to compare this to the Higgs branch of $\SU(3)$ SQCD with 6 flavors. In terms of the $\usp(6) \oplus \u(1)$ representations, the generators of the latter are
\begin{itemize}
    \item \textbf{Mesons} at order $t^2$: ${\red [0,1,0]_{\usp(6)}(0)}+[2,0,0]_{\usp(6)}(0)+[0,0,0]_{\usp(6)}(0)$~.
    \item \textbf{Baryons} at order $t^3$: ${\red [1,0,0]_{\usp(6)}(3) + [1,0,0]_{\usp(6)}(-3)} +[0,0,1]_{\usp(6)}(3) + [0,0,1]_{\usp(6)}(-3)$~.
\end{itemize}
In particular, we highlighted in {\red red} the generators that are projected out by discrete gauging, \ie those that are {\it absent} in the $\CT^{r=2}_{C_3U_1}$ theory.  We emphasize that the $\CT^{r=2}_{C_3U_1}$ theory also has $90$ generators in $[0,2,0]_{\usp(6)}(0)$ at order $t^4$.  These are not present in $\SU(3)$ SQCD. The reason is that the generators in ${\red [0,1,0]_{\usp(6)}(0)}$ at order $t^2$ are not invariant under the $\BZ_2$ symmetry that we gauge, and so upon gauging they combine with themselves to form gauge invariant objects in $[0,2,0]_{\usp(6)}(0)$ at order $t^4$.

\subsubsection{Exploiting the Argyres--Seiberg Duality}
\label{sec:ASduality}

The 4d $\CN=2$ $\SU(3)$ gauge theory with 6 flavors can be realized using the Argyres--Seiberg duality \cite{Argyres:2007cn} by starting from the rank-1 Minahan--Nemeschansky $E_6$ SCFT, coupling one hypermultiplet in the fundamental representation of an $\SU(2)$ subgroup of $E_6$, and then gauging the $\SU(2)$ symmetry. This can be written schematically as
\bes{ \label{ArgyresSeiberg}
\text{SU(3) with 6 flavors} = \text{(rank-1 $E_6$ SCFT + $[\SU(2)]-[1]$)}///\SU(2)~.
}
The Higgs branch of the $E_6$ theory is the closure of the minimal nilpotent orbit of $E_6$, denoted by $\bar{\min\,E_6}$ \cite{Gaiotto:2009jjh}.  The fact that the flavor central charge of the $E_6$ theory is 6 and that the embedding index of $\su(2)$ into $\mathfrak{e}_6$ is $1$ ensures the vanishing of the beta function of the $\SU(2)$ gauge group on the right side \cite{Argyres:2007cn}.

Theory $\CT^{r=2}_{C_3U_1}$ can be obtained by performing a $\BZ_2$ discrete gauging on the theory on the left-hand side of \eref{ArgyresSeiberg}. We will demonstrate that, on the right-hand side of \eref{ArgyresSeiberg}, such discrete gauging affects only the strong-coupled sector, namely the $E_6$ theory, but not the weakly coupled sector, namely the weakly gauged $\SU(2)$ symmetry and hypermultiplet. The $\BZ_2$ discrete gauging of the $E_6$ theory has been studied in \cite{Argyres:2016yzz}. The resulting theory is a rank-1 SCFT with the Coulomb branch operator of scaling dimension $6$ and with the Higgs branch isomorphic to the next-to-minimal nilpotent orbit of $F_4$, denoted by $\bar{\mathrm{n.min} \, {F_4}}$ (see \cite{BK}, \cite[(7.7)]{Hanany:2020jzl} and \cite{Bourget:2020bxh}). For simplicity, we will refer to the latter as $\CT^{r=1}_{\bar{\mathrm{n.min} \, {F_4}}}$. In fact, we propose that
\bes{ \label{couplednminF4}
\CT^{r=2}_{C_3U_1} = (\CT^{r=1}_{\bar{\mathrm{n.min} \, {F_4}}}  + [\SU(2)]-[1])///\SU(2)~.
}
The two Coulomb branch operators of $\CT^{r=2}_{C_3U_1}$ thus have dimensions $2$ and $6$. We remark that the flavor central charge of the $F_4$ symmetry in the $\CT^{r=1}_{\bar{\mathrm{n.min} \, {F_4}}}$ theory is 6, which remains unchanged from the $E_6$ theory upon discrete gauging \cite{Argyres:2016yzz}, and that the embedding index of $\su(2)$ in $\mathfrak{f}_4$ is 1.\footnote{This can be computed as 
\bes{\nonumber
I_{\su(2) \hookrightarrow \mathfrak{f}_4} = \frac{6T([1]_{\su(2)})+14 T([0]_{\su(2)})}{T([0,0,0,1]_{\mathfrak{f}_4})} =  1~,
}
where $T(\vec r)$ is the quadratic index of the representation $\vec r_{\mathfrak{g}}$ of $\mathfrak{g}$.}
%$\vec r_G$ of group $G$.} 
The beta-function of the $\SU(2)$ gauge group on the right-hand side of \eref{couplednminF4} is therefore zero.

A non-trivial test of the proposal \eref{couplednminF4}
is to show that the Hilbert series of the right-hand side is given by \eref{HWGHBTC3U1} and \eref{HSHBTC3U1}. Let us discuss this in detail. The HWG function of $\bar{\mathrm{n.min} \, {F_4}}$ is \cite{Hanany:2017ooe}
\bes{
\mathrm{HWG}\left[\bar{\mathrm{n.min} \, {F_4}}\right] = \PE \left[\nu_1 t^2 + \nu_4^2 t^4 \right]~,
}
where $\nu_1$ and $\nu_4$ denote the highest weight fugacities for the adjoint and fundamental representations of $F_4$ respectively.  
Now, we decompose representations of $F_4$ to those of $\su(2) \oplus \usp(6)$; for example, we have the following branching rules:
\bes{
\nu_1 ~ &\rightarrow ~  \sigma^2+  \mu_3 \sigma + \mu_1^2~, \\
\nu_4 ~ &\rightarrow ~ \mu _1 \sigma +\mu _2~, \\
\nu_4^2 ~ &\rightarrow ~  \mu _1^2 \sigma ^2+\mu _1 \mu _2 \sigma +\mu _3 \sigma +\mu _2^2+\mu _2+1~,
}
where $\sigma^r$ denotes the highest weight fugacity for the $\su(2)$ representation $[r]_{\su(2)}$, and $\mu_{1,2,3}$ denote the highest weight fugacities for $\usp(6)$, as before. Explicitly, the Hilbert series of $\bar{\mathrm{n.min} \, {F_4}}$ can be written in terms of $\su(2) \oplus \usp(6)$ characters as
\bes{
\mathrm{HS}\left[\bar{\mathrm{n.min} \, {F_4}}\right](z, \vec x; t) = \PE \left[\left({[2]_{\su(2)}} + {[1]_{\su(2)}} {[0,0,1]_{\usp(6)}} +{[2,0,0]_{\usp(6)}} \right) t^2 -t^4 +\ldots  \right]~,
}
where we denote by $z$ and $\vec{x}$ the $\su(2)$ and $\usp(6)$ fugacities respectively.
The Hilbert series of \eref{couplednminF4} is then given by \cite{Benvenuti:2010pq, Hanany:2010qu}
\bes{
\oint_{|z|=1} \frac{dz }{2 \pi i z} (1-z^2) ~\frac{\mathrm{HS}\left[\bar{\mathrm{n.min} \, {F_4}}\right](z, \vec x; t) \, \PE\left[t (z+z^{-1})(b^3+b^{-3})\right]}{\PE\left[(z^2+1+z^{-2}) t^2\right]}~.
}
Upon evaluating the integral, we obtain \eref{HWGHBTC3U1} and \eref{HSHBTC3U1}, as required.

\subsection{New Rank-\texorpdfstring{$(n-1)$}{(n-1)} SCFT with \texorpdfstring{$\USp(2n)\times \U(1)$}{USp(2n)xU(1)} Global Symmetry}
\label{sec:NewRankn-1SCFT}

A generalization of the precedent discussion is to obtain a new rank-$(n-1)$ SCFT with $\usp(2n)\oplus \u(1)$ symmetry, denoted by $\CT^{r=n-1}_{C_nU_1}$, by performing a $\BZ_2$ discrete gauging of 4d $\CN=2$ $\SU(n)$ SQCD with $2n$ flavors.  The Higgs branch of $\CT^{r=n-1}_{C_nU_1}$ can be realized in two ways, as before.  The first way is through the Coulomb branch of the wreathed quiver
\bes{ \label{mirrSUnw2n}
\begin{tikzpicture}[baseline,font=\footnotesize,scale=1]
\node (U3) at (0,0) {$\U(n)$}; 
\node (U1a) at (-0.5,1.1) {$\U(1)$}; 
\node (U1b) at (0.5,1.1) {$\U(1)$}; 
\node (Upl) at (-1.6,0) {$\U(n-1)$};
\node[draw=none] (Udl) at (-3.1,0) {$\cdots$};
\node (U2l) at (-4.2,0) {$\U(2)$};
\node (U1l) at (-5.5,0) {$\U(1)$};
\node (Upr) at (1.6,0) {$\U(n-1)$};
\node[draw=none] (Udr) at (3.1,0) {$\cdots$};
\node (U2r) at (4.2,0) {$\U(2)$};
\node (U1r) at (5.5,0) {$\U(1)$};
\draw (U1l)--(U2l)--(Udl)--(Upl)--(U3)--(Upr)--(Udr)--(U2r)--(U1r);
\draw (U1a)--(U3)--(U1b);
\draw [<->,red] (Upl) to [out=-140,in=-40,looseness=0.7] (Upr);
\draw [<->,red] (U2l) to [out=-140,in=-40,looseness=0.5] (U2r);
\draw [<->,red] (U1l) to [out=-140,in=-40,looseness=0.5] (U1r);
\node at (7,0) {$/\U(1)$};
\end{tikzpicture}
}
The second way is by exploiting a generalization of the Argyres--Seiberg duality, namely \cite{Chacaltana:2010ks}
\bes{ \label{genASduality}
\text{$\SU(n)$ with $2n$ flavors}
= (\text{$R_{0,n}$ SCFT} + [\SU(2)]-[1])///\SU(2)~,
}
where the $R_{0,n}$ SCFT is a theory of class $\mathcal{S}$ of the $A_{n-1}$-type associated with a sphere with two maximal punctures $[1^{n}]$ and a puncture $[n-2,1^2]$.

The two ways discussed above can be reconciled as follows. The 3d mirror theory of the $R_{0,n}$ SCFT is described by the following star-shaped quiver \cite{Benini:2010uu}:
\bes{ \label{mirrR0n}
\begin{tikzpicture}[baseline,font=\footnotesize,scale=1]
\node(U3) at (0,0) {$\U(n)$}; 
\node(U1a) at (0,0.9) {$\U(2)$}; 
\node(U1b) at (0,1.8) {$\U(1)$}; 
\node(Upl) at (-1.6,0) {$\U(n-1)$};
\node[draw=none] (Udl) at (-3.1,0) {$\cdots$};
\node(U2l) at (-4.2,0) {$\U(2)$};
\node(U1l) at (-5.5,0) {$\U(1)$};
\node(Upr) at (1.6,0) {$\U(n-1)$};
\node[draw=none] (Udr) at (3.1,0) {$\cdots$};
\node(U2r) at (4.2,0) {$\U(2)$};
\node(U1r) at (5.5,0) {$\U(1)$};
\draw (U1l)--(U2l)--(Udl)--(Upl)--(U3)--(Upr)--(Udr)--(U2r)--(U1r);
\draw (U3)--(U1a)--(U1b);
\node at (7,0) {$/\U(1)$};
\end{tikzpicture}
}
The duality \eref{genASduality} can be realized in terms of the mirror theory as follows. Coupling two flavors of hypermultiplets to the $\SU(2)$ symmetry of the $R_{0,n}$ theory amounts to attaching a $\U(1)$ gauge node to node $\U(2)$ on top of $\U(n)$ in the previous diagram:
\bes{ \label{mirrR0na}
\begin{tikzpicture}[baseline,font=\footnotesize,scale=1]
\node(U3) at (0,0) {$\U(n)$}; 
\node(U1a) at (0,0.9) {$\U(2)$}; 
\node(U1b) at (-0.5,1.8) {$\U(1)$}; 
\node(U1c) at (0.5,1.8) {$\U(1)$}; 
\node(Upl) at (-1.6,0) {$\U(n-1)$};
\node[draw=none] (Udl) at (-3.1,0) {$\cdots$};
\node(U2l) at (-4.2,0) {$\U(2)$};
\node(U1l) at (-5.5,0) {$\U(1)$};
\node(Upr) at (1.6,0) {$\U(n-1)$};
\node[draw=none] (Udr) at (3.1,0) {$\cdots$};
\node(U2r) at (4.2,0) {$\U(2)$};
\node(U1r) at (5.5,0) {$\U(1)$};
\draw (U1l)--(U2l)--(Udl)--(Upl)--(U3)--(Upr)--(Udr)--(U2r)--(U1r);
\draw (U3)--(U1a)--(U1b);
\draw (U1a)--(U1c);
\node at (7,0) {$/\U(1)$};
\end{tikzpicture}
}
In order to reduce the number of flavors to 1, we give mass to one of the flavors of hypermultiplets and integrate it out.  In terms of the mirror theory, this amounts to giving an FI term to one of the $\U(1)$ gauge nodes above the $\U(2)$ gauge node.  Flowing to the IR, we obtain the 3d mirror theory of $\SU(n)$ SQCD with $2n$ flavors (see \cite{Bourget:2020mez, vanBeest:2021xyt, Bourget:2023uhe}):
\bes{ \label{mirrSUnw2nx}
\begin{tikzpicture}[baseline,font=\footnotesize,scale=1]
\node(U3) at (0,0) {$\U(n)$}; 
\node(U1a) at (-0.5,1.1) {$\U(1)$}; 
\node(U1b) at (0.5,1.1) {$\U(1)$}; 
\node(Upl) at (-1.6,0) {$\U(n-1)$};
\node[draw=none] (Udl) at (-3.1,0) {$\cdots$};
\node(U2l) at (-4.2,0) {$\U(2)$};
\node(U1l) at (-5.5,0) {$\U(1)$};
\node(Upr) at (1.6,0) {$\U(n-1)$};
\node[draw=none] (Udr) at (3.1,0) {$\cdots$};
\node(U2r) at (4.2,0) {$\U(2)$};
\node(U1r) at (5.5,0) {$\U(1)$};
\draw (U1l)--(U2l)--(Udl)--(Upl)--(U3)--(Upr)--(Udr)--(U2r)--(U1r);
\draw (U1a)--(U3)--(U1b);
\node at (7,0) {$/\U(1)$};
\end{tikzpicture}
}
Now let us perform the $\BZ_2$ discrete gauging on the left-hand side of \eref{genASduality}.  We have previously justified, in the case of $n=3$, that discrete gauging only acts on the strongly coupled sector, namely the $R_{0,n}$ theory, but not on the weakly coupled sector. In terms of the mirror theory, this amounts to applying wreathing to each pair of nodes of the same rank on the left and on the right of $\U(n)$ in \eref{mirrR0n}, \eref{mirrR0na} and \eref{mirrSUnw2nx}.  As a result, we obtain \eref{mirrSUnw2n} as required.

We remark that this technique provides information about the Higgs branch of the 4d $\CN=2$ $\CT^{r=n-1}_{C_nU_1}$ theory for general $n$.  However, unlike the $n=3$ case, it does not provide us with information about the Coulomb branch for higher $n$.  This is due to the lack of an explicit description of the discrete gauging of the $R_{0,n}$ theory for $n\geq 4$. We leave the problem to determine the properties of the Coulomb branch of the $\CT^{r=n-1}_{C_nU_1}$ theory, with $n\geq 4$, for future work.

\subsubsection*{The Case of \texorpdfstring{$n=2$}{n=2}}

Let us discuss the special case of $n=2$, corresponding to a $\BZ_2$ discrete gauging of 4d $\CN=2$ $\SU(2)$ SQCD with 4 flavors. In this case, it turns out that the effect of the $\BZ_2$ transformation is to flip the sign of one of the four mass parameters $m_i$, say $m_4$.
In this case, the discrete gauging can be described at the level of the elementary fields as follows.
Using the standard $\SU(n)$ notation, we can denote the adjoint chiral in the vector multiplet as $\Phi_a^b$, with $a,b=1,2$ color indices, and the matter chiral fields as $\widetilde{Q}^{a,i}$ and $Q_{a,i}$, with $i=1,2,3,4$ a flavor index. In terms of these fields, we can write the superpotential term as 
\be\label{sup1} \mathcal{W}=\widetilde{Q}^{a,i}\Phi_a^bQ_{b,i}=\epsilon^{ac}\epsilon^{bd}\Phi_{ad}Q_{b,i}\widetilde{Q}_{c}^{i}\,;\quad \widetilde{Q}_{a}^{i}\equiv\epsilon_{ab}\widetilde{Q}^{b,i}\,,\ee
where we exploit the fact that the fundamental representation of $\SU(2)$ is pseudoreal to raise and lower color indices with the epsilon tensor. Notice that the field $\Phi_{ab}$ is symmetric. The $\BZ_2$ transformation we are interested in can then be written e.g. as 
\be\label{transf1}Q_{a}^{i}\rightarrow Q_{a}^{i}\,,\; \widetilde{Q}_{a}^{i}\rightarrow\widetilde{Q}_{a}^{i}\;\; (i=1,2,3);\quad Q_{a}^{4}\leftrightarrow\widetilde{Q}_{a}^{4}\,.\ee 
The transformation \eqref{transf1} leaves the superpotential invariant and the action on gauge invariant operators is as follows: 
\begin{itemize} 
\item The meson components $\widetilde{Q}^{a,i}Q_{a}^j$ with $i,j=1,2,3$ are invariant. 
\item The meson components $\widetilde{Q}^{a,4}Q_{a}^{j}$ with $j=1,2,3$ map to the baryons $\epsilon^{ab}Q_{a}^4Q_{b}^{j}$ and similarly $\widetilde{Q}^{a,i}Q_{a}^{4}$ with $i=1,2,3$ map to the antibaryons $\epsilon_{ab}\widetilde{Q}^{a,i}\widetilde{Q}^{b,4}$, while $\widetilde{Q}^{a,4}Q_{a}^{4}$ maps to minus itself. 
\item The remaining baryon and antibaryon components are invariant.
\end{itemize} 
Overall, out of the 28 gauge invariant operators, 7 are projected out and only 21 of them survive after gauging. If we now trade the $Q$'s and $\widetilde{Q}$'s for the multiplets $q_a^k$ ($k=1,\dots,8$) transforming in the vector of the $\so(8)$ global symmetry: 
\be\label{transf2} q_a^{2i-1}=Q_a^i+\widetilde{Q}_a^i\,,\quad q_a^{2i}=-i(Q_a^i-\widetilde{Q}_a^i);\quad i=1,\dots,4\,.\ee 
In these variables, the superpotential can be written as 
$$\mathcal{W}=\epsilon^{ab}\epsilon^{cd}\Phi_{ac}q_b^iq_d^j\delta_{ij}$$
and the transformation \eqref{transf1} reads $q_a^8\rightarrow -q_a^8$, while all other flavor components are invariant. This is exactly the $\BZ_2$ transformation discussed in \cite[(4.6)]{Argyres:2016yzz}, 
 where it was noticed that the resulting Higgs branch is isomorphic to the next-to-minimal orbit of $\so(7)$, denoted by $\bar{\mathrm{n.min}\, B_3}$. For convenience, we denote the theory we get after the discrete gauging by $\CT^{r=1}_{\bar{\mathrm{n.min}\, B_3}}$.
In the variables $q_a^i$, it is easy to see that the action on the mass parameters is precisely the expected one. The superpotential mass term indeed reads \be\label{masses}\sum_{i=1}^4\epsilon^{ab}m_iq_a^{2i-1}q_b^{2i}\ee 
and, since only $q_a^8$ changes sign under the $\BZ_2$ action, we find, as expected, that 
\be\label{transf3} m_1\rightarrow m_1\,;\quad m_2\rightarrow m_2\,;\quad m_3\rightarrow m_3\,;\quad m_4\rightarrow -m_4\,.\ee 
As was discussed in detail in \cite[(10.3)]{Seiberg:1994aj}, the action \eqref{transf3} on the mass parameters corresponds to interchanging the two spinor representations of $\so(8)$ while leaving the vector representation invariant. These are the representations in which monopoles and dyons of the SQCD theory transform, and accordingly this transformation is accompanied by the action of the $\mathsf{T}$ generator of $\mathrm{SL}(2,\BZ)$ which shifts $\theta \rightarrow \theta+\pi$, or equivalently $\tau \rightarrow \tau +1$, where $\tau = \frac{\theta}{\pi}+\frac{8\pi i}{g^2}$ is defined as in \cite{Seiberg:1994aj} (see also \cite{Ferrari:1997gu}).\footnote{In the notation of \cite[(16.35), (16.36)]{Seiberg:1994aj}, this corresponds to interchanging $e_2$ and $e_3$, which amounts to shifting $\tau \rightarrow \tau +1$ in \cite[(16.14)]{Seiberg:1994aj}. This is indeed the action of $\mathsf{T}$ generator of $\mathrm{SL}(2,\BZ)$.} Indeed, this shift of the theta angle  does not affect the supercharges, hence supersymmetry is automatically preserved, and we do not need to combine this with a $\U(1)_r$ transformation.\footnote{Under the $\mathrm{SL}(2,\BZ)$ transformation that sends $\tau \rightarrow \frac{a\tau+b}{c\tau+d}$, the supercharges $Q^i_\alpha$ (with $i=1,2$ the $\SU(2)_R$ index and $\alpha=1,2$)  transform as $Q^i_\alpha \rightarrow \left( \frac{|c\tau+d|}{c\tau+d}\right)^{1/2} Q^i_\alpha$. For the  $\mathsf{T}=\begin{psmallmatrix} 1 & 1 \\ 0 & 1\end{psmallmatrix}$ and $\mathsf{S}=\begin{psmallmatrix} 0 & -1 \\ 1 & 0 \end{psmallmatrix}$ generators of $\mathrm{SL}(2,\BZ)$, we see that $\mathsf{T}$ leaves the supercharges invariant, whereas $\mathsf{S}$ transforms the supercharges by a non-trivial phase. } As a result, our $\BZ_2$ transformation acts trivially on the Coulomb branch and the marginal deformation is not removed from the theory.
 From \eqref{transf2} and \eqref{transf3} we are therefore led to propose that upon gauging we simply obtain the following 4d $\CN=2$ gauge theory:\footnote{We stress that, as pointed out in \cite[(3.2) and Page 15]{Seiberg:1994aj}, the $\BZ_2$ symmetry discussed in \eref{transf1} is anomalous unless it is accompanied by the shift $\theta \rightarrow \theta+\pi$ of the theta angle.}
\bes{\label{B3C1O1}
\begin{tikzpicture}[baseline,font=\footnotesize]
\node (SO7) at (3.8,0) {$[\SO(7)]$};
\node (USp2) at (1.9,0) {$\USp(2)$};
\node (O1) at (0,0) {$\O(1)$};
\draw (SO7) -- (USp2) -- (O1);
\end{tikzpicture}
}
Upon compactification on a circle, we therefore obtain the same theory in 3d.
As claimed in \cite[Section 3.7]{Bourget:2020bxh}, \eref{B3C1O1} is actually the mirror theory of the $\BZ_2$ wreathing of the affine $D_4$ quiver depicted in \eref{affineD4}. This represents a nontrivial consistency check of our analysis. We will analyze this wreathing in Sections \ref{sec:disgauging-hooft} and \ref{app:mirrordualsD4}.  In particular, in the latter section, we compute the superconformal index of theory \eref{B3C1O1} and show that its Higgs and Coulomb branch limits give the Hilbert series of $\bar{\mathrm{n.min}\, B_3}$ and $\BC^2/\hat{D}_6$, in perfect agreement with \cite[Figure 9 and Figure 11]{Bourget:2020bxh}. 

As discussed in \cite{Argyres:2016yzz}, there is already a known rank-1 SCFT with Higgs branch $\bar{\mathrm{n.min}\, B_3}$. This model has a Coulomb branch operator of dimension $4$ and therefore a trivial conformal manifold.
We would like to stress that this theory does not coincide with our model $\CT^{r=1}_{\bar{\mathrm{n.min}\, B_3}}$, even though the two theories have the same Higgs branch: our theory has a nontrivial conformal manifold and a Coulomb branch operator of dimension 2. 
This is in contrast with the model discussed in \cite{Argyres:2016yzz}, in which the discrete symmetry being gauged involves the $\mathsf{S}$ generator of $\mathrm{SL}(2,\BZ)$. This acts non-trivially on the supercharges and to preserve supersymmetry we need to combine the transformation with a discrete $\U(1)_r$ rotation which acts as a sign flip on the Coulomb branch operator $\tr\Phi^2$. This results, as mentioned before, in a Coulomb branch operator of dimension 4 and, furthermore, the complexified gauge coupling is fixed at the value $\tau=i$.\footnote{Note that $\tau =i$ is the fixed point of the $\mathsf{S}$ action $\tau \rightarrow -\frac{1}{\tau}$. At this value of $\tau$, the supercharges transform by the phase $e^{-i\pi/4}$. Therefore, the discrete $\U(1)_r$ rotation that compensates such an action in order to preserve 4d $\CN=2$ supersymmetry acts on all supercharges as $e^{i\pi/4}$. As a result, these transformations flip the sign of the Coulomb branch operator $\tr \Phi^2$.} This is, however, a different discrete symmetry with respect to the one relevant for us.\footnote{Note, however, that upon compactification on a circle, both $\CT^{r=1}_{\bar{\mathrm{n.min}\, B_3}}$ and the rank-1 theory with Higgs branch $\bar{\mathrm{n.min}\, B_3}$ discussed in \cite{Argyres:2016yzz} give rise to the same 3d $\CN=4$ gauge theory described by the quiver \eref{B3C1O1}. In other words, the 3d theory that comes from compactification is ``blind'' to this subtle difference of the 4d parent theories.} We are therefore led to the conclusion that $\CT^{r=1}_{\bar{\mathrm{n.min}\, B_3}}$ is actually a new rank-1 theory with a Coulomb branch operator of dimension 2 and flavor symmetry $\so(7)$. In the following section, we are going to use this theory as a building block to construct a new rank-2 model.

\subsection{New Rank-2 SCFT with \texorpdfstring{$\SO(4)$}{SO(4)} Global Symmetry}
\label{sec:NewRank2SCFT-2}

Let us now present a new 4d $\CN=2$ rank-2 SCFT with global symmetry algebra $\so(4)$, which we will denote by $\CT^{r=2}_{D_2}$. It can be obtained by a $\BZ_2$ discrete gauging of the theory of class $\mathcal{S}$ of type $A_1$ associated with a genus-1 Riemann surface with two punctures. This class $\CS$ theory admits a Lagrangian description in terms of the $\SU(2) \times \SU(2)$ gauge group with two bifundamental hypermultiplets (see \eg  \cite[Section 5.2]{Hanany:2010qu}). 

The Higgs branch of the $\CT^{r=2}_{D_2}$ theory can be studied by considering a $\BZ_2$ wreathing of the mirror theory of the $A_1$ class $\mathcal{S}$ theory with genus 1 and two punctures, namely
\bes{ \label{wreathedmirrA1g1e2}
\begin{tikzpicture}[baseline,font=\footnotesize,scale=1]
\node (U2) at (0,0) {$\U(2)$};  
\node (U1l) at (-1.2,0) {$\U(1)$}; 
\node (U1r) at (1.2,0) {$\U(1)$}; 
\path (U2) edge [out=45,in=135,looseness=4]  node[midway,above]{$\mathrm{Adj}$} (U2);
\draw (U1l)--(U2)--(U1r);
 \draw [<->,red] (U1l) to [out=-140,in=-40] (U1r);
\end{tikzpicture}
\quad\text{\footnotesize$/\U(1)$}
}
where the loop around the $\U(2)$ node denotes the adjoint hypermultiplet. The Higgs branch Hilbert series of $\CT^{r=2}_{D_2}$ can be computed from the Coulomb branch Hilbert series of this wreathed quiver as follows:
\bes{ \label{HSwreathedmirrA1g1e2}
\mathrm{HS}\left[\text{CB of}\, \eref{wreathedmirrA1g1e2}\right] (z_1, z_2; t)=&\, (1-t^2) \sum_{u<v}\,\, \sum_{m_1 \geq m_2=0}t^{2\Delta} \frac{P_{\U(2)}(m_1,m_2;t)}{(1-t^2)^2} z_1^{m_1+m_2} z_2^{u+v} \\
& +(1-t^2)\sum_{u=v}\,\, \sum_{m_1 \geq m_2=0}t^{2\Delta} \frac{P_{\U(2)}(m_1,m_2;t)}{(1-t^2)(1-t^4)} z_1^{m_1+m_2} z_2^{u+v}~,
}
with
\bes{
2 \Delta = \sum_{i=1}^2 |u-m_i|+ \sum_{i=1}^2 |v-m_i|~,
}
where the prefactor $(1-t^2)$ and the restriction $m_2=0$ are there for modding out an overall $\U(1)$. This also imposes the following constraint on the topological fugacities (see \cite[(3.3)]{Cremonesi:2014vla}):
\bes{ \label{constop1}
z_1z_2=1.
}
The factor $\frac{1}{(1-t^2)^2}$ in the first line is the contributions of the two $\U(1)$ gauge groups, and the factor $\frac{1}{(1-t^2)(1-t^4)}$ in the second line comes from applying the Molien formula \eref{Molien} to $G=\langle (12) \rangle$.  Another way to realize the Coulomb branch of the wreathed quiver is to consider the Coulomb branch of the following quiver \cite{Hanany:2018vph, Hanany:2018cgo, Hanany:2018dvd}:
\bes{ \label{U2adjU2adj}
\begin{tikzpicture}[baseline,font=\footnotesize,scale=1]
\node (U2l) at (0,0) {$\U(2)$};   
\node (U2r) at (2,0) {$\U(2)$}; 
\path (U2l) edge [out=45,in=135,looseness=4] node[midway,above]{$\mathrm{Adj}$} (U2l);
\path (U2r) edge [out=45,in=135,looseness=4] node[midway,above]{$\mathrm{Adj}$} (U2r);
\draw (U2l)--(U2r);
\end{tikzpicture}
\quad\text{\footnotesize$/\U(1)$}
}
The Coulomb branch Hilbert series is also precisely given by \eref{HSwreathedmirrA1g1e2}, since the two terms combine into
\bes{
\scalebox{0.91}{$
\begin{split}
\mathrm{HS}\left[\text{CB of}\, \eref{U2adjU2adj}\right] (z_1, z_2; t)&=(1-t^2) \sum_{v \geq u > -\infty}\limits \,\, \sum_{m_1 \geq m_2=0}\limits t^{2\Delta} P_{\U(2)}(m_1,m_2;t) P_{\U(2)}(v,u;t) z_1^{m_1+m_2} z_2^{u+v} \\ &= \mathrm{HS}\left[\text{CB of}\, \eref{wreathedmirrA1g1e2}\right] (z_1, z_2; t)~.
\end{split}$}
}
The Coulomb branch symmetry of this theory is very interesting and worth discussing in detail.  In the UV, the manifest Coulomb branch symmetry is $\U(1)$. However, the term at order $t^2$ is
\bes{
1+z_1 z_2^2+z_1 z_2+z_2+z_1+\frac{1}{z_2}
\overset{\eref{constop1}}{=} 2(z_1+1+z_1^{-1})~,
}
indicating that the symmetry gets enhanced to $\su(2) \times \su(2) \cong \so(4)$ in the IR.  We see that one element of the subalgebra of $\so(4)$ comes from the trace of the adjoint scalar in the vector multiplet of one of the $\U(2)$ gauge groups, and the other comes from the monopole operator with flux $(m_1, m_2; u, v) = (1,0; 0,1)$.  Note that the latter becomes another Cartan element only after imposing the condition \eref{constop1}. By performing the computation in this way, we can refine only one fugacity (not two fugacities) of $\so(4)$.  Upon setting $z_1=z_2=1$, the Hilbert series can be summed into a closed form as
\bes{ \label{closedformurTD2}
&\mathrm{HS}\left[\text{CB of}\, \eref{wreathedmirrA1g1e2}\right] (t) = \mathrm{HS}\left[\text{CB of}\, \eref{U2adjU2adj}\right] (t) = \frac{1 + 3 t^2 + 11 t^4 + 10 t^6 + 11 t^8 + 
 3 t^{10} + t^{12}}{(1 - t^2)^3 (1 - t^4)^3}~,\\ 
 }
whose expansion reads
\bes{
\mathrm{HS}\left[\text{CB of}\, \eref{wreathedmirrA1g1e2}\right] (t) & = \mathrm{HS}\left[\text{CB of}\, \eref{U2adjU2adj}\right] (t)\\  & = 1 + 6 t^2 + 29 t^4 + 89 t^6 + 236 t^8 + 521 t^{10}+\mathcal{O}(t^{12})  \\
&= \PE \left[ 6 t^2 + 8 t^4 - 15 t^6 - 4 t^8 + 64 t^{10} +\mathcal{O}(t^{12}) \right]~.
}
This is the unrefined Higgs branch Hilbert series of $\CT^{r=2}_{D_2}$.
Observe that the numerator is palindromic and that the pole at $t=1$ is of order $6$, implying that the moduli space is 6 complex (or 3 quaternionic) dimensional.  This is equal to the dimension of the Higgs branch of the $A_1$ class $\mathcal{S}$ theory with genus 1 and two punctures \cite[Section 5.2]{Hanany:2010qu}, as expected, since discrete gauging does not affect the dimension of the Higgs branch.

Let us realize $\CT^{r=2}_{D_2}$ in another way. This $A_1$ class $\mathcal{S}$ theory associated to a genus-1 Riemann surface with two punctures can be realized by gauging the diagonal $\su(2)$ subalgebra of $\su(2)^2$ in the $\su(2)^4 \subset \so(8)$ flavor symmetry of $\SU(2)$ SQCD with $4$ flavors,\footnote{Throughout this paper, we will often use the shortened notation $\su(2)^n$ to indicate $\su(2)_1 \oplus \su(2)_2 \oplus \ldots \oplus \su(2)_n$.} which is the theory of class $\mathcal{S}$ of type $A_1$ associated with a sphere with four punctures. Recall that the $\BZ_2$ discrete gauging of the latter theory gives $\CT^{r=1}_{\bar{\mathrm{n.min}\, B_3}}$. We thus propose that
\bes{ \label{TD2fromgaugedB3}
\CT^{r=2}_{D_2} = \CT^{r=1}_{\bar{\mathrm{n.min}\, B_3}}///{(\su(2) \subset \so(7))}~.
}
Note that the embedding index of $\su(2) \subset \so(7)$ is equal to $2$, which can be computed, for example, from the branching rule
\bes{
[1,0,0]_{\so(7)} \rightarrow [0;0;2]_{\su(2)^3}+[1;1;0]_{\su(2)^3}~,
}
namely 
\bes{
I_{\su(2) \hookrightarrow \so(7)} = \frac{T([2]_{\su(2)})+ 4 T([0]_{\su(2)}) }{T([1,0,0]_{\so(7)})}  =2~.
}
Since the flavor central charge of the $\so(7)$ symmetry is 4 \cite{Argyres:2016yzz}, the beta function of the $\su(2)$ gauge algebra on the right-hand side is zero due to the cancelation between the contribution of the vector multiplet and the embedding index times the flavor central charge \cite{Argyres:2007cn}.  From \eref{TD2fromgaugedB3}, we also see that the scaling dimensions of the Coulomb branch operators of $\CT_{D_2}$ are therefore $2$ and $2$.  Using the description \eref{TD2fromgaugedB3} along with \eref{B3C1O1}, we see that upon compactifying $\CT^{r=2}_{D_2}$ on a circle to 3d, the resulting theory is
\bes{ \label{3dredTD2}
\begin{tikzpicture}[baseline,font=\footnotesize]
\node (SO4) at (3.8,0) {$[\SO(4)]$};
\node (USp2) at (1.9,0) {$\USp(2)$};
\node (O1) at (0,0) {$\O(1)$};
\node (SO3) at (1.9,-1) {$\SO(3)$};
\draw (USp2) -- (SO3);
\draw (SO4) -- (USp2) -- (O1);
\end{tikzpicture}
}
where the $\SO(3)$ gauge symmetry corresponds to the $\su(2)$ gauge algebra that appeared in \eref{TD2fromgaugedB3}.

Let us now discuss the Higgs branch of \eref{TD2fromgaugedB3}, which can be computed via the Higgs branch of \eref{3dredTD2}. The $\SO(3)$ gauge group is not fully Higgsed on the generic point of the Higgs branch due to the fact that there is one flavor of hypermultiplet transforming under its vector representation. As pointed out in \cite[Footnote 7]{Cremonesi:2014uva}, the Higgs branch of the theory $[\USp(2)]-\SO(3)$ is equal to the Higgs branch of the theory $[\USp(2)]-\O(1)$, which is isomorphic to $\BC^2/\BZ_2$. As a result, instead of computing the Higgs branch of \eref{3dredTD2}, we can compute the Higgs branch of the following auxiliary quiver:
\bes{ \label{3dredTD2mod}
\begin{tikzpicture}[baseline,font=\footnotesize]
\node (SO4) at (3.8,0) {$[\SO(4)]$};
\node (USp2) at (1.9,0) {$\USp(2)$};
\node (O1) at (0,0) {$\O(1)$};
\node (O1d) at (1.9,-1) {$\O(1)$};
\draw (USp2) -- (O1d);
\draw (SO4) -- (USp2) -- (O1);
\end{tikzpicture}
}
From this theory, we see that the Higgs branch of $\CT_{D_2}^{r = 2}$ is $4+1+1-3=3$ quaternionic dimensional, as expected.  Let us compute the Higgs branch Hilbert series of $\CT^{r=2}_{D_2}$ from \eref{3dredTD2mod}:
\bes{
&\mathrm{HS}\left[\text{HB of} \, \CT^{r=2}_{D_2}\right] (y_1, y_2; t) = \frac{1}{4} \sum_{\epsilon_1 = \pm 1}\, \sum_{\epsilon_2 = \pm 1} \oint_{|z|=1} \frac{dz}{2\pi i z} (1-z^2)  \\
& \qquad \qquad \qquad \times \frac{\PE \left[\left(y_1+y^{-1}_1\right) \left(y_2+y^{-1}_2\right) \left(z+z^{-1}\right) t+ \epsilon_1 \left(z+z^{-1}\right) t+ \epsilon_2 \left(z+z^{-1}\right) t \right]}{\PE \left[\left(z^2+1+z^{-2}\right) t^2\right]}~,
}
whose expansion in powers of $t$ yields
\bes{
\mathrm{HS}\left[\text{HB of} \, \CT^{r=2}_{D_2}\right] (y_1, y_2; t) =  \PE &\left[ \left([2;0]_{\su(2)^2}+[0;2]_{\su(2)^2}\right) t^2 + \left([2;2]_{\su(2)^2} -1\right) t^4   
  \right.\\ & \left.  \, - \left([2;2]_{\su(2)^2} + [2;0]_{\su(2)^2}+[0;2]_{\su(2)^2} \right) t^6 + \mathcal{O}(t^8) \right]~.
}
Upon setting $y_1=y_2=1$, we recover \eref{closedformurTD2}, as required.

\section{Discrete Gauging and Mixed 't Hooft Anomalies}
\label{sec:disgauging-hooft}

In this section, we investigate via an example how discrete gauging affects the mixed 't Hooft anomaly between a 1-form and a 0-form symmetry. In particular, consider the 4d $\CN=2$ SCFT described by the following quiver:
\begin{align} \label{SU4w4SU2}
    \begin{tikzpicture}[scale=0.86,font=\footnotesize,baseline=0cm]
\node (A1) at (0,0) {$\SU(2n)$};
\node (A2) at (1.2,-1.2) {$\SU(n)$};
\node (A3) at (-1.2,1.2) {$\SU(n)$};
\node (A4) at (1.2,1.2) {$\SU(n)$};
\node (A5) at (-1.2,-1.2) {$\SU(n)$}; 
\draw (A2)--(A1) (A3)--(A1) (A4)--(A1) (A5)--(A1); 
\end{tikzpicture}
\end{align}
This theory has a $\U(1)^4$ 0-form flavor symmetry and a $\BZ_n^{[1]}$ 1-form symmetry (see \cite[(4.28)]{Bhardwaj:2021pfz}).  It also admits a realization in class $\mathcal{S}$ of type $A_{2n-1}$, with four twisted trivial punctures $[2n+1]_t$ and four untwisted minimal punctures $[2n-1,1]$ on a sphere \cite[Section 5.1]{Chacaltana:2012ch}. Quivers of this type have been studied in detail in \cite{DelZotto:2015rca, Closset:2020afy, Kang:2021lic, Carta:2023bqn}. As discussed around \cite[(4.4)]{Carta:2023bqn}, there is a one-dimensional submanifold of the conformal manifold upon which the duality group $\SL(2, \BZ)$ acts, and the $\mathsf{S}$ generator of $\SL(2,\BZ)$ also acts on the global structure of the gauge group. Therefore, upon gauging the $\BZ_n^{[1]}$ 1-form symmetry of this theory, we obtain a non-invertible symmetry \cite{Kaidi:2021xfk}. 

As discussed in \cite[Section 5.1]{Chacaltana:2012ch}, the aforementioned class $\mathcal{S}$ theory can also be constructed as follows. First, we consider a sphere $X$ with punctures $[2n-1,1]$, $[2n-1,1]$, $[2n+1]_t$, $[2n+1]_t$ and $[1^{2n}]$. Then, we weakly gauge two copies of $X$ together via the $[1^{2n}]$ punctures, where the latter corresponds to the gauge algebra $\su(2n)$.  The theory associated with each sphere $X$ corresponds to two copies of 4d $\CN=2$ $\SU(n)$ SQCD with $2n$ flavors, whose flavor symmetry is $\SU(2n)/\BZ_n$, where the center $\BZ_{2n}$ of $\SU(2n)$ is screened by the baryons in the rank-$n$ antisymmetric representation of $\SU(2n)$ to $\BZ_n$. Gauging two copies of $X$ via the gauge group $\SU(2n)$ leads to the $\BZ_n^{[1]}$ 1-form symmetry, as discussed before.

The 3d mirror theory of \eref{SU4w4SU2} is given by
\begin{align}\label{eq:USp2n-4U1}
    \begin{tikzpicture}[scale=0.86,font=\footnotesize,baseline=0cm]
\node (A1) at (0,0) {$\USp(2n)$};
\node (A2) at (1.2,-1.2) {$\U(1)$};
\node (A3) at (-1.2,1.2) {$\U(1)$};
\node (A4) at (1.2,1.2) {$\U(1)$};
\node (A5) at (-1.2,-1.2) {$\U(1)$}; 
\draw (A2)--(A1) (A3)--(A1) (A4)--(A1) (A5)--(A1); 
\path (A1) edge [out=30,in=-30,looseness=6] node[midway,right]{$\mathrm{AS}'$}  (A1);
\end{tikzpicture}
\quad \text{\footnotesize$/\BZ_2$}
\end{align}
where $\mathrm{AS'}$ denotes the hypermultiplet in the rank-2 antisymmetric traceless representation of $\usp(2n)$, i.e. $[0,1,0,\ldots,0]_{\usp(2n)}$. Let us explain how to construct this theory. First, we start from the 3d mirror theory of the 4d theory associated with $X$. Each $[2n-1,1]$ puncture contributes $T_{[2n-1,1]}(\SU(2n)): \U(1)-[2n]$ whose flavor symmetry is $\su(2n)$; the $[1^{2n}]$ punctures give the tail $[2n]-\U(2n-1)-\ldots-\U(2)-\U(1)$ of the quiver; finally, each $[2n+1]_t$ puncture does not contribute to any matter field, but constrains the middle gauge node to be $\usp(2n)$ \cite{Kang:2022zsl}.  As a result, the 3d mirror of the theory associated with $X$ is\footnote{It can be checked that the Higgs and Coulomb branch dimensions of this quiver agree with the Coulomb and Higgs branch dimensions of two copies of 3d $\CN=4$ $\SU(n)$ SQCD with $2n$ flavors.}
\bes{ \label{mirrorX}
    \begin{tikzpicture}[scale=0.86,font=\footnotesize,baseline=0cm]
\node (c) at (0,0) {$\USp(2n)$};
\node (U1a) at (-1.2,1.2) {$\U(1)$};
\node (U1b) at (-1.2,-1.2) {$\U(1)$}; 
\node (t1) at (2.5,0) {$\U(2n-1)$};
\node (t2) at (4,0) {$\cdots$};
\node (t3) at (5.5,0) {$\U(2)$};
\node (t4) at (7,0) {$\U(1)$};
\draw (c)--(U1a); 
\draw (c)--(U1b); 
\draw (c) -- (t1) -- (t2) -- (t3) -- (t4);
\end{tikzpicture}
\quad \text{\footnotesize$/\BZ_2$}
}
A non-trivial test of this result is that, if we ungauge the leftmost two $\U(1)$ gauge nodes, we obtain the quiver depicted in \cite[Figure 60(d)]{Gaiotto:2008ak}, which is indeed the mirror theory of two copies of the $\U(n)$ gauge theory with $2n$ flavors. Now, we gauge the $\SU(2n)$ group of the diagonal subgroup $\SU(2n)/\BZ_n$ of the Coulomb branch symmetry $(\SU(2n)/\BZ_n)^4$ of the two copies of \eref{mirrorX}. Since the end of the tail is the $\USp(2n)$ gauge group, we use the following branching rule from $\su(2n)$ to $\usp(2n)$:
\bes{
[1,0, \ldots 0,1]_{\su(2 n)} \quad \rightarrow \quad [2,0,\ldots,0]_{\usp(2 n)}+[0,1,0,\ldots,0]_{\usp(2 n)}~,
}
where the moment map in the adjoint representation $[1,0,\ldots,0,1]_{\su(2 n)}$ of $\su(2n)$  becomes the moment map in the adjoint representation $[2,0,\ldots,0]_{\usp(2 n)}$ of $\usp(2n)$ along with a hypermultiplet $\mathrm{AS}'$ in the rank-2 antisymmetric traceless representation $[0,1,0,\ldots,0]_{\usp(2 n)}$ of $\usp(2n)$. We note that the mirror theory \eref{eq:USp2n-4U1} has a $\BZ_n^{[1]}$ 1-form symmetry arising from gauging the $\SU(2n)$ group of the Coulomb branch symmetry. 

Let us now explain the notation $/\BZ_2$ in \eref{mirrorX}. The quiver \eref{mirrorX} has an overall $\BZ_2$ that acts trivially on the matter fields. This can be considered as the center of $\USp(2n)$ that is not screened by $\mathrm{AS}'$ whose action on the bifundamental fields can be absorbed by appropriate $\U(1)$ gauge transformations. This gives rise to a $\BZ_2^{[1]}$ 1-form symmetry, which we denote by $(\BZ_2^{[1]})_C$.  For simplicity, we denote by $/\BZ_2$ the gauging of the $(\BZ_2^{[1]})_C$ 1-form symmetry.  The reason why the mirror theory \eref{eq:USp2n-4U1} must involve the gauging of $(\BZ_2^{[1]})_C$ can be understood as follows. In the special case of $n=1$, the original theory \eref{SU4w4SU2} reduces to $\SU(2)$ SQCD with 4 flavors. This theory does not have a 1-form symmetry, and so, in the mirror theory \eref{eq:USp2n-4U1}, the $(\BZ_2^{[1]})_C$ 1-form symmetry must be gauged.  In this case, \eref{eq:USp2n-4U1} is equivalent to the well-known 3d mirror theory of $\SU(2)$ SQCD with 4 flavors \cite{Intriligator:1996ex}, namely the affine $D_4$ quiver
\bes{\label{affineD4}
   \begin{tikzpicture}[scale=0.86,font=\footnotesize,baseline=0cm]
\node (A1) at (0,0) {$\USp(2)$};
\node (A2) at (1.2,-1.2) {$\U(1)$};
\node (A3) at (-1.2,1.2) {$\U(1)$};
\node (A4) at (1.2,1.2) {$\U(1)$};
\node (A5) at (-1.2,-1.2) {$\U(1)$}; 
\draw (A2)--(A1) (A3)--(A1) (A4)--(A1) (A5)--(A1); 
\end{tikzpicture}
\quad \text{\footnotesize$/\BZ_2$}\qquad \quad \text{or} \qquad \quad    \begin{tikzpicture}[scale=0.86,font=\footnotesize,baseline=0cm]
\node (A1) at (0,0) {$\U(2)$};
\node (A2) at (1.2,-1.2) {$\U(1)$};
\node (A3) at (-1.2,1.2) {$\U(1)$};
\node (A4) at (1.2,1.2) {$\U(1)$};
\node (A5) at (-1.2,-1.2) {$\U(1)$}; 
\draw (A2)--(A1) (A3)--(A1) (A4)--(A1) (A5)--(A1); 
\end{tikzpicture} \quad \text{\footnotesize$/\U(1)$}
}

In the following, we will also consider a variant of \eref{eq:USp2n-4U1}, with $n=1$, where $(\BZ_2^{[1]})_C$ is not gauged. We put $/\BZ_2$ in brackets, namely $(/\BZ_2)$, to denote which of the two options we are taking into account, namely whether $(\BZ_2^{[1]})_C$ is gauged or not.  The case of $n=2$ will be discussed in detail in Appendix \ref{sec:Casen2USp4}.

\subsection{Wreathing Quiver \texorpdfstring{\eqref{eq:USp2n-4U1}}{} with \texorpdfstring{$n=1$}{n=1}} \label{sec:disgauging-affD4}

In this case, theory \eref{SU4w4SU2} becomes $\SU(2)$ SQCD with 4 flavors, whose 3d mirror theory is given by \eref{affineD4}.  In the following, we consider also the option where $(\BZ_2^{[1]})_C$ is not gauged, \ie 
\bes{ \label{eq:USp2-4U1}
\begin{tikzpicture}[scale=0.86,font=\footnotesize,baseline=0cm]
\node (A1) at (0,0) {$\USp(2)$};
\node (A2) at (1.2,-1.2) {$\U(1)$};
\node (A3) at (-1.2,1.2) {$\U(1)$};
\node (A4) at (1.2,1.2) {$\U(1)$};
\node (A5) at (-1.2,-1.2) {$\U(1)$}; 
\draw (A2)--(A1) (A3)--(A1) (A4)--(A1) (A5)--(A1); 
\end{tikzpicture}
}
We denote explicitly by $\eref{eq:USp2-4U1}/\BZ_2$ the option where the 1-form symmetry is gauged, \ie  \eref{affineD4}.  

The Coulomb branch Hilbert series of theory \eref{eq:USp2-4U1}$/\BZ_2$ is (see also \cite[Section 2.6.1]{Bourget:2020xdz})
\begin{equation} \label{CBUSp24U1}
    \begin{split}
        & \mathrm{HS}\left[\text{CB of}\, \eref{eq:USp2-4U1}/\BZ_2\right]({\vec z}| \omega; t) \\
        & = \sum_{\epsilon=0}^1 \omega^\epsilon \sum_{u_1 \in \BZ+\frac{\epsilon}{2}} \, \sum_{u_2 \in \BZ+\frac{\epsilon}{2}} \, \sum_{u_3 \in \BZ+\frac{\epsilon}{2}} \, \sum_{u_4 \in \BZ+\frac{\epsilon}{2}} \, \sum_{m \geq \frac{\epsilon}{2}} t^{2\Delta} \left( \prod_{i=1}^4 z_i^{u_i} \right) \frac{P_{\USp(2)}(m;t)}{(1-t^2)^4}~,
    \end{split}
\end{equation}
where $\omega$ (with $\omega^2=1$) is the fugacity for the $\BZ_2^{[0]}$ 0-form symmetry that arises from gauging the 1-form symmetry of \eref{eq:USp2-4U1}, $z_{1,2,3,4}$ are the fugacities for the four $\U(1)$ topological symmetries, and
\bes{
2 \Delta &=  \sum_{j=1}^4 \sum_{s= \pm 1}|s m -u_j| -2|2m|~.
}
In the dressing factor $\frac{P_{\USp(2)}(m;t)}{(1-t^2)^4}$, the denominator is the contribution of each $\U(1)$ gauge group and the numerator, which is the contribution of the $\USp(2)$ gauge group, is given by
\bes{
P_{\USp(2)}(m;t) = \begin{cases}
(1-t^4)^{-1} &\qquad m=0 \\
(1-t^2)^{-1} &\qquad m\neq0
\end{cases}~.
}
The Coulomb branch Hilbert series of theory \eref{eq:USp2-4U1} can then be obtained by gauging the 0-form symmetry associated with $\omega$:
\bes{ \label{gaugeomegaUSp2w4U1}
& \mathrm{HS}\left[\text{CB of}\, \eref{eq:USp2-4U1}\right]({\vec z}; t) = \frac{1}{2} \sum_{\omega=\pm 1} \mathrm{HS}\left[\text{CB of}\, \eref{eq:USp2-4U1}/\BZ_2\right]({\vec z}| \omega; t)~.
}

Let us now discuss the explicit form of the above Hilbert series. Since theory $\eref{eq:USp2-4U1}/\BZ_2$ is a mirror theory of $\SU(2)$ SQCD with 4 flavors, the Coulomb branch Hilbert series of the former can be expressed in terms of $\so(8)$ representations as follows \cite[(3.26)]{Benvenuti:2010pq}:
\bes{
\mathrm{HS}\left[\text{CB of}\, \eref{eq:USp2-4U1}/\BZ_2\right]({\vec z}| \omega; t)  = \sum_{m=0}^\infty [0,m,0,0]_{\so(8)}(\vec y) t^{2m}~,
}
where $\vec y = \vec y (\vec z, \omega)$.\footnote{If we write the character of the vector representation of $\so(8)$ as $\sum_{i=1}^4 (y_i +y_i^{-1})$ and write the character of the fundamental of $\su(2)_i$ as $\mathfrak{z}_i+\mathfrak{z}^{-1}_i$, then a map from $\so(8)$ fugacities to the $\su(2)^4 \times \BZ_2$ can be chosen as 
\bes{
y_1 = \omega \mathfrak{z}_1 \mathfrak{z}_2~, \quad y_2 = \omega \mathfrak{z}_1 \mathfrak{z}_2^{-1}~, \quad y_3 = \mathfrak{z}_3 \mathfrak{z}_4~, \quad y_4 = \mathfrak{z}_3 \mathfrak{z}_4^{-1}~. \nonumber
}
Note that we have $\mathfrak{z}_i = z_i^{\frac{1}{2}}$ in \eref{HSexplUSp2-4U1modZ2}.
\label{foot:mapso8tosu2to4}}
Since only the multiples of the highest weight of the $\so(8)$ adjoint representation appear in the Hilbert series, the flavor symmetry group is $\SO(8)/\BZ_2$ (see also \cite[Section 4]{Cremonesi:2014vla}).  Upon decomposing the $\so(8)$ representations in terms of those of $\su(2)^4$, the above Hilbert series can be rewritten as \cite[(4.7)]{Hanany:2010qu}, where the fugacity $\omega$ can be inserted as a coefficient in front of each term that contains odd highest weights of $\su(2)$.  More compactly, we rewrite this in terms of the HWG as (see \cite[(2.33)]{Bourget:2020xdz})
\bes{ \label{HWGUSp2-4U1modZ2}
\scalebox{0.95}{$
\begin{split}
&\mathrm{HWG}\left[\text{CB of}\, \eref{eq:USp2-4U1}/\BZ_2\right](m_1, \ldots, m_4| \omega; t)\\
&= \PE\left[ \left(   m_1^2+m_2^2+m_3^2+m_4^2  + \omega \, m_1 m_2 m_3 m_4 \right) t^2+ \left(1+ \omega \, m_1 m_2 m_3 m_4 \right) t^4 - m_1^2 m_2^2 m_3^2m_4^2 t^8 \right]~,
\end{split}$}
}
where $m_{1,2,3,4}$ are the fugacities for the highest weights of the representations of $\su(2)^4$.  Expanding in terms of $\su(2)^4$ fugacities, we have 
\bes{ \label{HSexplUSp2-4U1modZ2}
\scalebox{0.93}{$
\begin{split}
&\mathrm{HS}\left[\text{CB of}\, \eref{eq:USp2-4U1}/\BZ_2\right]({\vec z}| \omega; t) \\
&= \PE \left[ \left\{\sum_{i=1}^4 \left(z_i+1+z^{-1}_i\right) +\omega  \sum_{s_1, \ldots, s_4 = \pm 1} z_1^{\frac{1}{2} s_1} z_2^{\frac{1}{2} s_2} z_3^{\frac{1}{2} s_3} z_4^{\frac{1}{2} s_4} \right\} t^2 \right.\\
&\left.\quad \quad \, \, - \left\{ \sum_{1\leq i < j \leq 4} \left(z_i+1+z^{-1}_i\right) \left(z_j+1+z^{-1}_j\right) + 4 +3 \omega \sum_{s_1, \ldots, s_4 = \pm 1} z_1^{\frac{1}{2} s_1} z_2^{\frac{1}{2} s_2} z_3^{\frac{1}{2} s_3} z_4^{\frac{1}{2} s_4} \right\} t^4 +\ldots \right]~.
\end{split}$}
}
In terms of the $\su(2)^4$ symmetry algebra, we conclude that the faithful flavor symmetry group of $\eref{eq:USp2-4U1}/\BZ_2$ is (see \cite[(4.29)]{Bhardwaj:2021ojs})
\bes{
\CF_{\eref{eq:USp2-4U1}/\BZ_2} = \frac{\SU(2)_1 \times \SU(2)_2 \times \SU(2)_3\times \SU(2)_4}{(\BZ_2)_{12} \times (\BZ_2)_{23} \times (\BZ_2)_{34}}~,
}
where each factor $(\BZ_2)_{ij}$ in the denominator acts by flipping the signs of the fugacities $z^{\frac{1}{2}}_i \rightarrow - z^{\frac{1}{2}}_i$ and $z^{\frac{1}{2}}_j \rightarrow - z^{\frac{1}{2}}_j$ simultaneously. This action simultaneously flips the sign of the fundamental representations of $\su(2)_i$ and $\su(2)_j$, and therefore leaves the Hilbert series and the HWG invariant.

Gauging the $\BZ_2^{[0]}$ 0-form symmetry associated with $\omega$ by summing over $\omega=\pm 1$ in \eref{HWGUSp2-4U1modZ2}, we obtain the highest weight generating function associated with \eref{gaugeomegaUSp2w4U1} as (see \cite[(2.34)]{Bourget:2020xdz})
\begin{equation} 
\label{HWGUSp2-4U1}
\scalebox{0.96}{$\displaystyle\begin{aligned}
&\mathrm{HWG}\left[\text{CB of}\, \eref{eq:USp2-4U1}\right](m_1,\ldots, m_4; t)\\ &= \PE\left[\left(m_1^2+m_2^2+m_3^2+m_4^2\right) t^2 + \left(1+m_1^2 m_2^2 m_3^2 m_4^2\right) t^4 + m_1^2 m_2^2 m_3^2 m_4^2 t^6 - m_1^4 m_2^4 m_3^4 m_4^4 t^{12} \right]~.
\end{aligned}$}
\end{equation}
We see that the flavor symmetry algebra of theory \eref{eq:USp2-4U1} is $\su(2)^4$, in contrast to that of $\eref{eq:USp2-4U1}/\BZ_2$ which is $\so(8)$.  Since only even powers of $m_i$ appear in the HWG,  the faithful flavor symmetry group of \eref{eq:USp2-4U1} is actually
\bes{
\CF_{\eref{eq:USp2-4U1}} = \SO(3)^4 \cong \left(\CF_{\eref{eq:USp2-4U1}/\BZ_2}\right)/(\BZ_2)_{i}~,\quad i \in \{1, 2, 3, 4\}~,
} 
where $(\BZ_2)_{i}$ denotes the $\BZ_2$ action that flips sign of the fundamental representation of $\SU(2)_i$, \ie  the center of $\SU(2)_i$. Setting $z_{1,2,3,4}=1$, we obtain the closed form of the unrefined Hilbert series as
\bes{
&\mathrm{HS}\left[\text{CB of}\, \eref{eq:USp2-4U1}\right](t)=\\
&\frac{1 + 7 t^2 + 101 t^4 + 244 t^6 + 666 t^8 + 650 t^{10} + 666 t^{12} + 
 244 t^{14} + 101 t^{16} + 
 7 t^{18} + t^{20}}{(1 - t)^{10} (1 + t)^{10} (1 + t^2)^5}~.
}

We remark that these Hilbert series can be obtained by studying the Higgs branch of the following theory:
\bes{ \label{SU2w4flvsplit}
\begin{tikzpicture}[baseline,font=\footnotesize]
\node (SO4) at (0,0) {$[\SO(4)]$};
\node (SU2) at (1.8,0) {$\SU(2)$};
\node (SO4r) at (4.6,0) {$[\SO(4)]$};
\draw (SO4) -- (SU2);
\draw[blue] (SU2) to node[midway,above] {\textcolor{blue} {$(\BZ_2^{[0]})_\omega$}} (SO4r);
\end{tikzpicture}
}
where we start from $\SU(2)$ SQCD with 4 flavors of hypermultiplets in the fundamental representation, split the hypermultiplets into two equal sets, and then consider the $\BZ_2$ action (associated with $\omega$) that acts on one of these sets. The Higgs branch Hilbert series of this theory can be obtained as follows:
\bes{
\scalebox{0.99}{$
\begin{split}
&\mathrm{HS}\left[\text{HB of}\, \eref{SU2w4flvsplit}\right](\fz_1, \ldots, \fz_4| \omega; t)\\
& =\oint_{|z|=1} \frac{dz}{2\pi i z} (1-z^2) \frac{\PE\left[\left(z+z^{-1}\right) \left\{\omega \left( \fz_1+\fz_1^{-1} \right)\left( \fz_2+\fz_2^{-1} \right) + \left( \fz_3+\fz_3^{-1} \right)\left( \fz_4+\fz_4^{-1} \right) \right\} t \right]}{\PE \left[\left(z^2+1+z^{-2}\right) t^2\right]}~.
\end{split}$}
}
Indeed, we find that
\bes{
\mathrm{HS}\left[\text{HB of}\, \eref{SU2w4flvsplit}\right](z_1^{\frac{1}{2}}, \ldots, z_4^{\frac{1}{2}}| \omega; t) &= \mathrm{HS}\left[\text{CB of}\, \eref{eq:USp2-4U1}/\BZ_2\right](z_1, \ldots, z_4| \omega; t)~, \\
\frac{1}{2} \sum_{\omega = \pm 1} \mathrm{HS}\left[\text{HB of}\, \eref{SU2w4flvsplit}\right](z_1^{\frac{1}{2}}, \ldots, z_4^{\frac{1}{2}}| \omega; t) &= \mathrm{HS}\left[\text{CB of}\, \eref{eq:USp2-4U1}\right](z_1, \ldots, z_4; t)~.
}
On the other hand, the Coulomb branch of $\eref{SU2w4flvsplit}/(\BZ_2^{[0]})_\omega$ is equal to the Higgs branch of \eref{eq:USp2-4U1}.  We thus propose that \eref{SU2w4flvsplit} is a mirror theory of $\eref{eq:USp2-4U1}/\BZ_2$, and $\eref{SU2w4flvsplit}/(\BZ_2^{[0]})_\omega$ is a mirror theory of \eref{eq:USp2-4U1}. We will examine theory \eref{SU2w4flvsplit} in detail, using the superconformal index as the main tool, in Section \ref{app:mirrordualsD4}.

Comparing \eref{HWGUSp2-4U1modZ2} and \eref{HWGUSp2-4U1}, we see that upon gauging the $(\BZ_2^{[1]})_C$ 1-form symmetry of theory \eref{eq:USp2-4U1}, the $\su(2)_i$ representations with odd highest weight appear, so the flavor symmetry group can no longer be $\SO(3)^4$.  This means that there is a mixed anomaly between $(\BZ_2^{[1]})_C$ and the flavor symmetry $\CF_{\eref{eq:USp2-4U1}}$, characterized by the following anomaly theory:
\bes{ \label{mixedanomn1}
i \pi \int_{M_4} B^{(2)} \cup w_2(\CF_{\eref{eq:USp2-4U1}})~,
}
where $B^{(2)}$ is the 2-form background gauge field associated with $(\BZ_2^{[1]})_C$ and $w_2(\CF_{\eref{eq:USp2-4U1}})$ is the second Stiefel--Whitney class that obstructs the lift of the $\CF_{\eref{eq:USp2-4U1}}$ bundle to the $\CF_{\eref{eq:USp2-4U1}/\BZ_2}$ bundle. Here, $M_4$ is the 4d bulk whose boundary is the 3d spacetime in which the theory in question lives.

Let us now study discrete gauging of theories \eref{eq:USp2-4U1} and $\eref{eq:USp2-4U1}/\BZ_2$ by performing the $\ZZ_2$, $\ZZ_3$ and $S_4$ wreathing with respect to the four $\U(1)$ gauge nodes. We explicitly indicate how the range of the summations of the gauge fluxes is restricted and how the dressing factors in \eref{CBUSp24U1} get modified. Note that such discrete gauging of theory $\eref{eq:USp2-4U1}/\BZ_2$ has already been studied in \cite{Hanany:2020jzl, Bourget:2020bxh}.
We will also study how the mixed anomaly \eref{mixedanomn1} changes upon discrete gauging.  The readers can find details on wreathing by other groups, including $S_3$ and $\BZ_4$, in Appendix \ref{sec:Casen2USp4}, where we discuss the case of quiver \eqref{eq:USp2n-4U1} with $n=2$. 

\subsubsection*{\texorpdfstring{$\ZZ_2$}{Z2} Wreathing} 
\label{sec:wreathZ2USp2}

Let us perform the $\BZ_2$ wreathing on the $\U(1)$ gauge nodes corresponding to the magnetic fluxes $u_1$ and $u_2$. In \eref{CBUSp24U1}, the gauge fluxes get restricted and the dressing factor is modified as follows.
\begin{equation}
\renewcommand*{\arraystretch}{1.4}
    \begin{array}{c|c}
        \text{Restriction} & \text{Dressing factor} \\
        \hline
        u_1<u_2 & (1-t^2)^{-4}P_{\USp(2)}(m;t)\\
        u_1=u_2 & (1-t^2)^{-3}(1-t^4)^{-1}P_{\USp(2)}(m;t)
    \end{array}
\end{equation}
We also set $z_1=z_2 \equiv x$. Note that the nodes associated with the magnetic fluxes $u_3$ and $u_4$ are left untouched.  After wreathing, the Coulomb branch Hilbert series is the same as that of the following theory:
\bes{ \label{wreathnminB3}
\begin{tikzpicture}[baseline,font=\footnotesize,scale=1]
\node (U2l) at (0,0) {$\USp(2)$}; 
\node (U2r) at (2,0) {$\U(2)$}; 
\node (U1a) at (-1,1) {$\U(1)$}; 
\node (U1b) at (1,1) {$\U(1)$}; 
\path (U2r) edge [out=45,in=-45,looseness=6] node[midway,right]{$\mathrm{Adj}$} (U2r);
\draw (U2l)--(U2r);
\draw (U2l) -- (U1a);
\draw (U2l) -- (U1b);
\end{tikzpicture}
\quad\text{\footnotesize$(/\BZ_2)$}
} 
where the fugacity associated to the topological symmetry of the $\U(2)$ node with an adjoint hypermultiplet is denoted by $x$, and those of the two $\U(1)$ gauge nodes are denoted by $z_3$ and $z_4$.  The notation $(/\BZ_2)$ denotes the options whether $(\BZ^{[1]}_2)_C$ is gauged or not.

This quiver has an enhanced $\so(7)$ global symmetry algebra. In fact, the Coulomb branch in question is isomorphic to the closure of the next-to-minimal orbit of $B_3$, $\bar{\mathrm{n.min}\,B_3}$, whose HWG is given by
\bes{
\mathrm{HWG} \left[\,\bar{\mathrm{n.min}\,B_3}\,\right] = \PE\left[\mu_2 t^2 + \mu_1^2 t^4 \right]~,
}
where $\mu_1$ and $\mu_2$ are the fugacities for the highest weights of the vector and adjoint representations of $\so(7)$.  We can rewrite this in terms of the highest weights of representations of $\su(2)^3 \times \BZ_2$ as
\bes{
\scalebox{0.97}{$
\begin{split}
& \mathrm{HWG}\left[\text{CB of}\, \left[\eref{eq:USp2-4U1}{/\BZ_2}\right]_{\text{wr} \, \BZ_2}\right] (m_x, m_3, m_4| \omega; t) = \mathrm{HWG}\left[\text{CB of}\, \eref{wreathnminB3}{/\BZ_2}\right] (m_x, m_3, m_4| \omega; t) \\
& =\PE \left[\left(m_x^2+m_3^2+m_4^2+ \omega m_3 m_4   m_x^2\right) t^2 + \left\{ \omega  \left(m_3 m_4 m_x^2-m_3 m_4\right) \right.\right. \\
&\left.\left.\quad\quad \, \, \, -m_3^2 m_x^4-m_4^2 m_x^4-m_3^2 m_x^2-m_3^2 m_4^2 m_x^2-m_4^2 m_x^2 -2 m_x^2-2 m_3^2-2 m_4^2-2 \right\} t^4 +\ldots \right]~,
\end{split}$}
}
where $m_x$, $m_3$, $m_4$ are the fugacities of the highest weight of $\su(2)_x \oplus \su(2)_3 \oplus \su(2)_4$.  The first $\su(2)$ factor is that associated with fugacity $x$ and the latter two are those left untouched upon wreathing. The closed form of the unrefined Coulomb branch Hilbert series reads \cite[Figure 9]{Bourget:2020bxh}
\bes{
\mathrm{HS}\left[\text{CB of}\, \left[\eref{eq:USp2-4U1}{/\BZ_2}\right]_{\text{wr} \, \BZ_2}\right](t)&=\mathrm{HS}\left[\text{CB of}\, \eref{wreathnminB3}{/\BZ_2}\right](t)\\
&=\frac{1 + 11 t^2 + 30 t^4 + 30 t^6 + 11 t^8 + t^{10}}{(1 - t)^{10} (1 + t)^{10}}~.
}
If we sum over $\omega = \pm 1$, \ie  gauging the $\BZ_2^{[0]}$ 0-form symmetry associated with $\omega$, we arrive at theory \eqref{eq:USp2-4U1} wreathed by $\BZ_2$, or equivalently at \eref{wreathnminB3} (without $/\BZ_2$), which has the $(\BZ_2^{[1]})_C$ 1-form symmetry.  The latter has a global symmetry $\SO(3)^3$, since only even powers of $m_{x}$, $m_3$ and $m_4$ appear in the HWG. In this case, the unrefined Coulomb branch Hilbert series can be expressed in closed form as
\bes{ \label{closedformwrZ2}
&\mathrm{HS}\left[\text{CB of}\, \eref{eq:USp2-4U1}_{\text{wr} \, \BZ_2}\right](t)=\mathrm{HS}\left[\text{CB of}\, \eref{wreathnminB3}\right](t)\\
&= \frac{1 + 4 t^2 + 63 t^4 + 112 t^6 + 352 t^8 + 280 t^{10} + 352 t^{12} + 112 t^{14} + 63 t^{16} + 4 t^{18} + t^{20}}{(1 - t)^{10} (1 + t)^{10} (1 + t^2)^5} ~.
}

Let us comment on the mixed anomaly. Observe that gauging the $(\BZ_2^{[1]})_C$ 1-form symmetry leads to the terms with odd powers in both $m_3$ and $m_4$ in the HWG. In terms of the symmetry algebra $\su(2)^3$, the faithful global symmetry groups are
\bes{ \label{FUSp2-4U1wrZ2}
 \CF_{{\left[{\eqref{eq:USp2-4U1}}/\BZ_2\right]}_{\text{wr} \, \BZ_2}} &= \SO(3)_x \times \frac{\SU(2)_3 \times \SU(2)_4}{(\BZ_2)_{34}}~, \\
 \CF_{{\eqref{eq:USp2-4U1}}_{\text{wr} \, \BZ_2}} &= \SO(3)_x \times \SO(3)_3 \times \SO(3)_4 \cong \left(\CF_{{\left[{\eqref{eq:USp2-4U1}}/\BZ_2\right]}_{\text{wr} \, \BZ_2}}\right)/(\BZ_2)_{i}~,\quad i \in \{3, 4\}~.
}
This means that there is a mixed anomaly between $(\BZ_2^{[1]})_C$ and the flavor symmetry $\CF_{{\eqref{eq:USp2-4U1}}_{\text{wr} \, \BZ_2}}$, characterized by the following anomaly theory:
\bes{ \label{mixedanomn2}
i \pi \int_{M_4} B^{(2)} \cup w_2(\CF_{{\eqref{eq:USp2-4U1}}_{\text{wr} \, \BZ_2}})~,
}
where $B^{(2)}$ is the 2-form background gauge field associated with $(\BZ_2^{[1]})_C$ and $w_2(\CF_{{\eqref{eq:USp2-4U1}}_{\text{wr} \, \BZ_2}})$ is the second Stiefel--Whitney class that obstructs the lift of the $\CF_{{\eqref{eq:USp2-4U1}}_{\text{wr} \, \BZ_2}}$ bundle to the $\CF_{{\left[{\eqref{eq:USp2-4U1}}/\BZ_2\right]}_{\text{wr} \, \BZ_2}}$ bundle.

\subsubsection*{\texorpdfstring{$\BZ_3$}{Z3} Wreathing}

 In order to analyze the Coulomb branch of the $\BZ_3$ wreathing of theory \eref{eq:USp2-4U1}$/\BZ_2$ from equation \eref{CBUSp24U1}, the gauge fluxes get restricted and the dressing factor is modified as follows.
\begin{equation}
\renewcommand*{\arraystretch}{1.4}
    \begin{array}{c|c}
        \text{Restriction} & \text{Dressing factor} \\
        \hline
        u_1<u_2\geq u_3 & (1-t^2)^{-4}P_{\USp(2)}(m;t)\\
        u_1=u_2=u_3 & (1-t^2+t^4)\left(1-t^2\right)^{-3} \left(1-t^6\right)^{-1}P_{\USp(2)}(m;t)
    \end{array}
\end{equation}
We also set $z_1=z_2=z_3 \equiv x$. Note that the dressing factor in the last line, apart from the factor $P_{\USp(2)}(m;t)$, can be obtained using the discrete Molien formula \eref{Molien}, with $G = \langle (123) \rangle$.

As a result of wreathing, we obtain the closure of the double cover of the subregular orbit of $G_2$ \cite{Achar2012GeometricSS} (see also \cite{Hanany:2020jzl, Bourget:2020bxh}), whose HWG is
\bes{ \label{HWGDCsubregG2}
\PE [\nu _1 t^2 +\left(\nu _1+\nu _2^2\right) t^4 + 2 \nu _2^3 t^6-\nu _2^6 t^{12} ]~,
}
where $\nu_1$ and $\nu_2$ are fugacities for the adjoint and fundamental representation of $G_2$. The unrefined Hilbert series is given by (see \cite[Figure 9]{Bourget:2020bxh})
\bes{
\mathrm{HS}\left[\text{CB of}\, \left[\eref{eq:USp2-4U1}/\BZ_2\right]_{\text{wr} \, \BZ_3}\right](t) 
&=\frac{1 + 4 t^2 + 23 t^4 + 23 t^6 + 4 t^8 + t^{10}}{(1 - t)^{10} (1+t)^{10}}\\   
&= 1 + 14 t^2 + 118 t^4 + 693 t^6 + 3094 t^8 +\ldots~.
}
Such Hilbert series can be computed from the Coulomb branch Hilbert series of the non-simply laced quiver:
\bes{
\U(3) \Rrightarrow \U(2)-\U(1)
}
where the overall $\U(1)$ can be modded out from one of the gauge nodes on the right of the arrow, namely either the rightmost $\U(1)$ gauge node as in \cite[Table 6]{Hanany:2020jzl}, or from the $\U(2)$ gauge node.  For the latter, we obtain
\bes{ \label{quivG2SU2modZ2}
\U(3) \Rrightarrow (\SU(2)/\BZ_2)-\U(1)
} 
where $(\SU(2)/\BZ_2)$ means that the gauge group is taken to be $\U(2)/\U(1) \cong \SU(2)/\BZ_2$.
The HWG \eref{HWGDCsubregG2} can be rewritten in terms of the highest weights of representations of $\su(2)_x \times \su(2)_4$ as 
\bes{
\scalebox{0.9}{$
\begin{split}
&\mathrm{HWG}\left[\text{CB of}\, \left[\eref{eq:USp2-4U1}/\BZ_2\right]_{\text{wr} \, \BZ_3}\right] (m_x,m_4| \omega; t) = \mathrm{HWG}\left[\text{CB of}\, \eref{quivG2SU2modZ2}\right] (m_x,m_4| \omega; t)\\ &= 
\PE \left[ \left(m_x^2+m_4^2+ \omega m_4 m_x^3\right) t^2 + \left\{\omega  \left(3 m_4 m_x^3+m_4 m_x\right)-m_x^6-m_4^2 m_x^4-m_x^2-m_4^2-1 \right\} t^4+\ldots\right]~,
\end{split}$}
}
where we remark that the fugacity $\omega$ can be regarded as coming from $/\BZ_2$ in quiver \eref{quivG2SU2modZ2}, and that there is no overall $\U(1)$ to be modded out in the latter.\footnote{Explicitly, the Coulomb branch Hilbert series of \eref{quivG2SU2modZ2} is given by
\bes{ \nonumber
\scalebox{0.93}{$
\begin{split}
&\mathrm{HS}\left[\text{CB of}\, \eref{quivG2SU2modZ2}\right] (z_x,z_4| \omega; t) = \sum_{\epsilon =0}^1 \omega^{\epsilon} \sum_{\overset{m_1 \geq m_2 \geq m_3}{m_{1,2,3} \in \BZ+\frac{\epsilon}{2}}}\,\, \sum_{n \in \BZ_{\geq 0}+\frac{\epsilon}{2}}  \,\, \sum_{u \in \BZ+\frac{\epsilon}{2}} t^{2\Delta}\,\, \frac{ P_{\U(3)}(m_1,m_2,m_3;t) P_{\USp(2)}(n;t)}{(1-t^2)} \,\,z_4^{m_1+m_2+m_3} z_x^{u} ~,
\end{split}$}
}
with 
\bes{\nonumber
2\Delta= \sum_{i=1}^3 (|3m_i-n|+|3m_i+n|) + |n-u| +|-n-u| - 2\sum_{1\leq j<k \leq 3} |m_j-m_k| - 2|2n|~.
}\label{foot:G2modZ2}
}

If we sum over $\omega = \pm 1$, \ie  we gauge the $\BZ_2^{[0]}$ 0-form symmetry associated with $\omega$, we arrive at the theory that has a $(\BZ_2^{[1]})_C$ 1-form symmetry.  The latter has a global symmetry $\SO(3)^2$, since only even powers of $m_{x}$ and $m_4$ appear in the HWG.  The Hilbert series after gauging can also be obtained from the Coulomb branch of the following quiver:
\bes{ \label{quivG2SU2}
\U(3) \Rrightarrow \SU(2)-\U(1)~,
}
where the $(\BZ_2^{[1]})_C$ symmetry can be regarded as coming from the $\SU(2)$ gauge group in this quiver.\footnote{Explicitly, the Coulomb branch Hilbert series of \eref{quivG2SU2} can be obtained from the expression in Footnote \ref{foot:G2modZ2} by summing over $\omega = \pm 1$ and dividing the result by $2$. Equivalently, we can restrict the sums in Footnote \ref{foot:G2modZ2} to be only over integral magnetic fluxes.}
In this case, the unrefined Hilbert series is given by
\bes{
&\mathrm{HS}\left[\text{CB of}\, \eref{eq:USp2-4U1}_{\text{wr} \, \BZ_3}\right](t) = \mathrm{HS}\left[\text{CB of}\, \eref{quivG2SU2}\right](t)\\
&=\frac{1 + t^2 + 37 t^4 + 76 t^6 + 218 t^8 + 230 t^{10} + 218 t^{12} +76 t^{14} + 37 t^{16} + t^{18}+ t^{20}}{(1 - t)^{10} (1 + t)^{10} (1 + t^2)^5} \\
&=1 + 6 t^2 + 62 t^4 + 341 t^6 + 1558 t^8 + \ldots ~.
}

In fact, in terms of the symmetry algebra $\su(2)^2$, the faithful global symmetry groups are
\bes{
 \CF_{{\left[\eqref{eq:USp2-4U1}/\BZ_2\right]}_{\text{wr} \, \BZ_3}} &=  \frac{\SU(2)_x \times \SU(2)_4}{(\BZ_2)_{x4}}~, \\
 \CF_{{\eqref{eq:USp2-4U1}}_{\text{wr} \, \BZ_3}} &= \SO(3)_x \times \SO(3)_4 \cong \left(\CF_{\left[{\eqref{eq:USp2-4U1}/\BZ_2}\right]_{\text{wr} \, \BZ_3}}  \right)/(\BZ_2)_4~.
}
The change of the flavor symmetry groups caused by gauging $(\BZ_2^{[1]})_C$ implies that there is a mixed anomaly between $(\BZ_2^{[1]})_C$ and the flavor symmetry $\CF_{{\eqref{eq:USp2-4U1}}_{\text{wr} \, \BZ_3}}$, characterized by the following anomaly theory:
\bes{ \label{mixedanomn3}
i \pi \int_{M_4} B^{(2)} \cup w_2(\CF_{{\eqref{eq:USp2-4U1}}_{\text{wr} \, \BZ_3}})~,
}
where $B^{(2)}$ is the 2-form background gauge field associated with $(\BZ_2^{[1]})_C$ and $w_2(\CF_{{\eqref{eq:USp2-4U1}}_{\text{wr} \, \BZ_3}})$ is the second Stiefel--Whitney class that obstructs the lift of the $\CF_{{\eqref{eq:USp2-4U1}}_{{\text{wr} \, \BZ_3}}}$ bundle to the $\CF_{{\left[{\eqref{eq:USp2-4U1}}/\BZ_2\right]}_{\text{wr} \, \BZ_3}}$ bundle.

\subsubsection*{\texorpdfstring{$S_4$}{S4} Wreathing}

In \eref{CBUSp24U1}, the gauge fluxes get restricted, and the dressing factor is modified as follows.
\begin{equation}
\renewcommand*{\arraystretch}{1.4}
    \begin{array}{c|c}
       \text{Restriction} & \text{Dressing factor} \\
        \hline
        u_1<u_2<u_3<u_4 & (1-t^2)^{-4}P_{\USp(2)}(m;t)\\
        u_1=u_2<u_3<u_4 & (1-t^2)^{-3}(1-t^4)^{-1}P_{\USp(2)}(m;t) \\
        u_1<u_2=u_3<u_4 & (1-t^2)^{-3}(1-t^4)^{-1}P_{\USp(2)}(m;t) \\
        u_1<u_2<u_3=u_4 & (1-t^2)^{-3}(1-t^4)^{-1}P_{\USp(2)}(m;t) \\
        u_1=u_2=u_3<u_4 & (1-t^2)^{-2}(1-t^4)^{-1}(1-t^6)^{-1}P_{\USp(2)}(m;t) \\
        u_1<u_2=u_3=u_4 & (1-t^2)^{-2}(1-t^4)^{-1}(1-t^6)^{-1}P_{\USp(2)}(m;t) \\
        u_1=u_2<u_3=u_4 & (1-t^2)^{-2}(1-t^4)^{-2}P_{\USp(2)}(m;t) \\
        u_1=u_2=u_3=u_4 & (1-t^2)^{-1}(1-t^4)^{-1}(1-t^6)^{-1}(1-t^8)^{-1}P_{\USp(2)}(m;t) \\
    \end{array}
\end{equation}
We also set $z_1=z_2=z_3=z_4=x$. Note that the dressing factor in the last line, apart from the factor $P_{\USp(2)}(m;t)$, can be obtained using the discrete Molien formula \eref{Molien}, with $G = \langle (12), (23), (34) \rangle$, which amounts to setting $u_1=u_2$, $u_2=u_3$ and $u_3=u_4$. The Hilbert series computed by wreathing is equal to the Coulomb branch Hilbert series of the following theory:
\begin{align} \label{quivUSp2U4adj}
\begin{tikzpicture}[baseline,font=\footnotesize,scale=1]
\node (U2l) at (0,0) {$\USp(2)$}; 
\node (U2r) at (2,0) {$\U(4)$}; 
\path (U2r) edge [out=45,in=-45,looseness=6] node[midway,right]{$\mathrm{Adj}$} (U2r);
\draw (U2l)--(U2r);
\end{tikzpicture}
\quad\text{\footnotesize$(/\BZ_2)$}
\end{align}
Upon gauging the $(\BZ_2^{[1]})_C$ 1-form symmetry, the theory $\eref{quivUSp2U4adj}/\BZ_2$ has an enhanced global symmetry $\su(3)$, and the HWG of the Coulomb branch Hilbert series can be written as
\bes{
&\mathrm{HWG}\left[\text{CB of}\, \eref{quivUSp2U4adj}/\BZ_2\right] (m_1,m_2; t)\\
&=\PE\left[ m_1 m_2 t^2+ \left(m_1^2+m_1 m_2+m_2^2+1\right) t^4 + \left(2 m_1 m_2^2+2 m_1^2 m_2+3\right) t^6 \right.\\
&\left. \qquad \qquad \, \, \, \, \, \, \, \, \, \, \, \, \, \, \, + \left(m_1^4+m_2^4+2 m_1^2 m_2^2+m_1+m_2\right) t^8+\ldots\right]~,
}
where $m_1^p m_2^q$ denotes the fugacity for the highest weight of the representation $[p,q]_{\su(3)}$ of $\su(3)$.  It can also be rewritten in terms of the highest weights of the representation of $\su(2)_x \times \BZ_2$ as
\bes{
&\mathrm{HWG}\left[\text{CB of}\, \left[\eref{eq:USp2-4U1}/\BZ_2\right]_{\text{wr} \, S_4}\right] (m_x| \omega; t) = \mathrm{HWG}\left[\text{CB of}\, \eref{quivUSp2U4adj}/\BZ_2\right] (m_x| \omega; t)\\
&=\PE\left[ \left(m^2+  m^4 \omega \right) t^2 +\left\{2 m^4+3 +\omega \left(2 m^4+m^2+1\right) \right\} t^4  \right.\\
&\left. \qquad \, + \left\{m^6+3 m^4+2 m^2+1 +\omega \left(2 m^6+2 m^4+2 m^2+3\right) \right\} t^6 +\ldots\right]~,
}
where $m$ is the fugacity for the highest weight of the fundamental representation of $\su(2)_x$. 

If we sum over $\omega=\pm 1$, we arrive at the theory \eref{quivUSp2U4adj} without $/\BZ_2$.  Since only even powers of $m$ appear in the resulting HWG, the theory has an $\SO(3)_x$ global symmetry.

In terms of the symmetry algebra $\su(2)_x$, the faithful global symmetry of both $\eref{quivUSp2U4adj}/\BZ_2$ and $\eref{quivUSp2U4adj}$ is $\SO(3)_x$. Since this is unchanged upon gauging $(\BZ_2^{[1]})_C$, we conclude that there is no mixed anomaly between $(\BZ_2^{[1]})_C$ and $\SO(3)_x$.

\section{Discrete Gauging of \texorpdfstring{$\SU(2)$}{SU(2)} with \texorpdfstring{$4$}{4} Flavors via Superconformal Index}
\label{app:mirrordualsD4}
In this section, we consider $\SU(2)$ SQCD with 4 flavors and discrete gauging thereof from the perspective of the superconformal index.  We will see that the index naturally allows us to consider discrete gauging by all $\BZ_2$ and $\BZ_2 \times \BZ_2$ subgroups of the permutation group $S_4$ of four objects.  Interestingly, the Coulomb and Higgs branch limits of the index reproduce the Hilbert series that come from the corresponding wreathing studied in \cite[Figure 9 and Figure 11]{Bourget:2020bxh}. A benefit of the index is that it provides a clear physical origin of each discrete symmetry involved in the gauging. In particular, we will see that these discrete symmetries involve the $\BZ_2$ symmetry that acts on a subset of the hypermultiplets and various charge conjugation symmetries associated with the flavor symmetries. To compare the result with the wreathing of \eref{affineD4}, we will treat the SQCD theory in question as a 3d $\CN=4$ gauge theory.

Let us start from $\SU(2)$ SQCD with 4 flavors, which we will represent as
\bes{ \label{USp2wSO8pflv}
\begin{tikzpicture}[baseline,font=\footnotesize,scale=1]
\node (USp2) at (0,0) {$\USp(2)$};
\node (SO8) at (2,0) {$[\SO(8)_\chi]$};
\draw (USp2) -- (SO8);
\end{tikzpicture}
}
where the subscript $\chi$ denotes the $\BZ_2^{[0]}$ $0$-form charge conjugation symmetry associated with the $\so(8)$ flavor symmetry.  Its index can be derived as in \cite{Aharony:2013dha, Aharony:2013kma, Beratto:2021xmn, Mekareeya:2022spm}. For $\chi = 1$, it is given by
\bes{ \label{indUSp2wSO8pflv}
\CI_{\eqref{USp2wSO8pflv}}(\vec{y}, \vec{m}| a, n_a| \chi = 1; x) 
 = &\,\frac{1}{2} \sum_{l \in \BZ} \oint \frac{d z}{2 \pi i z} \CZ^{\USp(2)}_{\text{vec}}\left(z; l; x\right) \prod_{s = {0, \pm 1}} \CZ^{1}_{\chi} \left(z^{2 s} a^{-2}; 2 s l -2 n_a; x\right) \\ 
&  \times \prod_{i = 1}^{4} \prod_{s_1, s_2 = \pm 1} \CZ^{1/2}_{\chi} \left(z^{s_1} y_i^{s_2} a; s_1 l + s_2 m_i + n_a; x\right) ~,
}
where the $\USp(2)$ vector multiplet contribution is
\bes{
&\CZ^{\USp(2)}_{\text{vec}}(z; l; x)
={x^{-{\left|2 l\right|}}} \prod_{{s}={\pm{1}}}{\left({1}-{\left(-{1}\right)^{2 {s}{l}}}{z^{2 s}}{x^{2 \left|{s}{l}\right|}}\right)}
}
and the contribution of the chiral multiplet of $R$-charge $R$ is
\bes{
\CZ^R_{\chi}(z; l;x) = \left( x^{1-R} z^{-1} \right)^{|l|/2} \prod_{j=0}^\infty \frac{1-(-1)^l z^{-1} x^{|l|+2-R+2j}}{1-(-1)^l z\,  x^{|l|+R+2j}}~,
}
where the subscript $\chi$ here stands for ``chiral'' and should not be confused with the fugacity $\chi$ above. Moreover, we denote by $(\vec{y}, \vec{m})$ the (fugacities, background magnetic fluxes) for the $\so(8)$ flavor symmetry algebra and by $(a, n_a)$ those for the axial symmetry. The index for $\chi = -1$ can be derived from \eqref{indUSp2wSO8pflv} by setting $y_4 = 1$, $y_4^{-1} = -1$ and $m_4 = 0$. From now on, we will set $m_i = n_a = 0$ and drop their dependence from the index.

The mirror theory of $\SU(2)$ SQCD with 4 flavors is given by the affine $D_4$ quiver \eref{affineD4}. If we denote by $\mathfrak{z}_{1, 2, 3, 4}$ the topological fugacities associated with \eqref{affineD4}, mirror symmetry implies the following equality between indices:
\bes{
\CI_{\eqref{USp2wSO8pflv}}(y_1, y_2, y_3, y_4| a| \chi = 1; x) = \CI_{\eqref{affineD4}}(\mathfrak{z}_1, \mathfrak{z}_2, \mathfrak{z}_3, \mathfrak{z}_4| a^{-1}|\omega = 1; x)~,
}
where $y_{1, 2, 3, 4}$ and $\mathfrak{z}_{1, 2, 3, 4}$ are related according to the fugacity map reported in Footnote \ref{foot:mapso8tosu2to4}, with $\omega = 1$.

For reference, we provide the expansion of the index up to order $x^2$, which can be expressed in terms of the characters of representations of $\so(8)$ as
\bes{ \label{indexUSp2wSO8pflv}
\CI_{\eqref{USp2wSO8pflv}}(\vec{y}| a| \chi = 1; x) = &\, 1+{{a^2 {[0,1,0,0]}_{\so(8)}}}{x} \\ & +\left\{{{a^4 {[0,2,0,0]}_{\so(8)}}+ 2 a^{-4}-\left({{[0,1,0,0]}_{\so(8)}}+1\right)}\right\}{x^2}+\ldots~.
}
We can take the Higgs and Coulomb branch limits of the index \eref{indUSp2wSO8pflv} in order to study the Higgs and Coulomb branches of \eqref{USp2wSO8pflv}.  We define \cite{Razamat:2014pta}
\bes{ \label{CBHBlimits}
&h = x^{1/2} a~, ~ c = x^{1/2} a^{-1}~, \\ 
&\text{or equivalently} \quad  x = h c ~,~ a = (h/c)^{1/2}~,
}
and substitute them in the index \eref{indUSp2wSO8pflv}.  In the Higgs branch limit we send $c \rightarrow 0$ and keep $h$ fixed, whereas in the Coulomb branch limit we send $h \rightarrow 0$ and keep $c$ fixed. In agreement with mirror symmetry, for $\chi = 1$ these reproduce
\bes{ \label{HBCBofSU2w4flv}
\text{HB of \eqref{USp2wSO8pflv}} &= \text{CB of \eqref{affineD4}} = \bar{\mathrm{min}\,D_4}~, \\
\text{CB of \eqref{USp2wSO8pflv}} &= \text{HB of \eqref{affineD4}} = \BC^2/\hat{D}_4~,
}
where the Hilbert series of $\BC^2/\hat{D}_4$ is given by
\bes{ \label{HSC2modD4}
\mathrm{HS} \left[\BC^2/\hat{D}_4 \right](t) = \PE\left[2 t^4 + t^6 - t^{12}\right]~.
}

\subsubsection*{Rewriting \eref{USp2wSO8pflv} in Two Ways}

For our purposes, we rewrite \eqref{USp2wSO8pflv} in two ways. In the first way, we split the $\so(8)$ flavor symmetry of theory \eqref{USp2wSO8pflv} into $\so(7)$ and $\so(1)$ as
\bes{ \label{USp2SO7SO1}
\begin{tikzpicture}[baseline,font=\footnotesize]
\node (SO1) at (0,0) {$[\SO(1)_{\chi_1}]$};
\node (USp2) at (1.8,0) {$\USp(2)$};
\node (SO7) at (3.6,0) {$[\SO(7)_{\chi_2}]$};
\draw (SO1) -- (USp2);
\draw (USp2) -- (SO7);
\end{tikzpicture}
}
where the advantage of such description is that it possesses two different charge conjugation symmetries $(\BZ_2^{[0]})_{\chi_1}$ and $(\BZ_2^{[0]})_{\chi_2}$, instead of a single one. The fugacities $\chi_1$ and $\chi_2$ corresponding to these two appear in a manifest way in the index of theory, which reads
\bes{ \label{indUSp2SO7SO1}
\CI_{\eqref{USp2SO7SO1}}(\vec{y}| a| \chi_1, \chi_2; x) 
 = &\,\frac{1}{2} \sum_{l \in \BZ} \oint \frac{d z}{2 \pi i z} \CZ^{\USp(2)}_{\text{vec}}\left(z; l; x\right) \prod_{s = {0, \pm 1}} \CZ^{1}_{\chi} \left(z^{2 s} a^{-2}; 2 s l; x\right) \\
&  \times\prod_{s = \pm 1} \CZ^{1/2}_{\chi} \left(z^s \chi_1 a; s l; x\right) \CZ^{1/2}_{\chi} \left(z^s \chi_2 a; s l; x\right) \\
&\times\prod_{i = 1}^{3} \prod_{s_1, s_2 = \pm 1} \CZ^{1/2}_{\chi} \left(z^{s_1} y_i^{s_2} a; s_1 l; x\right) ~.
}
For reference, we report the index \eqref{indUSp2SO7SO1} with $y_i=1$ as follows:
\bes{ \label{indunrefSO1SO7}
\scalebox{1}{$
\begin{split}
\CI_{\eqref{USp2SO7SO1}}(a| \chi_1, \chi_2; x) = &\, 1+ a^2 \left(15+ 6 \chi_1 +6 \chi_2 + \chi_1 \chi_2\right) x \\ &  + \left[ a^4 \left(125  + 70 \chi_1 + 70 \chi_2 +35 \chi_1 \chi_2\right) + a^{-4} \left(1+\chi_1 \chi_2\right) \right.\\ & \quad \,\,\left.-\left(16 + 6 \chi_1 + 6 \chi_2 +\chi_1 \chi_2\right) \right]x^2+\ldots~.
\end{split}$}
}
Let us comment on the Coulomb branch of theory \eqref{USp2wSO8pflv}. As discussed in \eref{HBCBofSU2w4flv}, this is isomorphic to $\BC^2/\hat{D}_4$.  The two generators that contribute to the order $t^4$ of \eqref{HSC2modD4} come from the coefficient of $a^{-4}$ at order $x^2$ in the index. One is the Casimir operator of the $\USp(2)$ gauge group, denoted by $\tr(\varphi^2)$, with $\varphi$ the scalar in the vector multiplet. The other is the monopole operator, denoted by $\mathfrak{M}$.  From the coefficients of  $a^{-4} x^2$ in \eref{indunrefSO1SO7}, we see that $\mathfrak{M}$ carries the fugacity $\chi_1 \chi_2$, \ie  it is charged under the combination of the charge conjugation symmetries associated with $\SO(1)$ and $\SO(7)$, whereas $\tr(\varphi^2)$ is neutral under these symmetries.  The generator corresponding to the term at order $t^6$ in \eqref{HSC2modD4} is $G_6 \equiv \mathfrak{M} \varphi$.  These generators satisfy the relation at order $t^{12}$, namely
\bes{
G_6^2+ \mathfrak{M}^2 \tr(\varphi^2) = [\tr(\varphi^2)]^3~,
}
which is the defining equation of $\BC^2/\hat{D}_4$.

The second way is to rewrite $\SU(2)$ SQCD with 4 flavors as follows:
\bes{ \label{USp2wSO8pflvsplit}
\begin{tikzpicture}[baseline,font=\footnotesize]
\node (SO4) at (0,0) {$[\SO(4)_{\chi_L}]$};
\node (USp2) at (1.8,0) {$\USp(2)$};
\node (SO4r) at (4.6,0) {$[\SO(4)_{\chi_R}]$};
\draw (SO4) -- (USp2);
\draw[blue] (USp2) to node[midway,above] {\textcolor{blue}{$(\BZ_2^{[0]})_\omega$}} (SO4r);
\end{tikzpicture}
}
which can be obtained from \eqref{USp2wSO8pflv} by splitting the $\so(8)$ flavor symmetry into two equal sets and then acting with a discrete $(\BZ_2^{[0]})_{\omega}$ $0$-form symmetry, whose fugacity is denoted by $\omega$, only on one of these sets. We will denote by $y_{1,2}$ the $\so(4)$ fugacities of the flavor node on the right, namely the one which is acted by $(\BZ_2^{[0]})_\omega$, and by $y_{3,4}$ the $\so(4)$ fugacities of the flavor node on the left. For $\chi_L=\chi_R = 1$, the index reads
\bes{ \label{indUSp2wSO8pflvb}
\scalebox{0.90}{$\displaystyle
\begin{aligned}
    \CI_{\eqref{USp2wSO8pflvsplit}}(y_1, y_2, y_3, y_4| a| \chi_R=1, \chi_L = 1| \omega; x) 
 = &\, \frac{1}{2} \sum_{l \in \BZ} \oint \frac{d z}{2 \pi i z} \CZ^{\USp(2)}_{\text{vec}}\left(z; l; x\right) \prod_{s = {0, \pm 1}} \CZ^{1}_{\chi} \left(z^{2 s} a^{-2}; 2 s l; x\right) \\ 
&  \times \prod_{i = 1}^{2} \prod_{s_1, s_2 = \pm 1} \CZ^{1/2}_{\chi} \left(\omega z^{s_1} y_i^{s_2} a; s_1 l ; x\right) \\
& \times \prod_{i = 3}^{4} \prod_{s_1, s_2 = \pm 1} \CZ^{1/2}_{\chi} \left(z^{s_1} y_i^{s_2} a; s_1 l ; x\right) ~,
\end{aligned}$
}
}
and the index for $\chi_R = -1$ (resp. $\chi_L = -1$) can be derived from \eqref{indUSp2wSO8pflvb} by setting $y_2 = 1$, $y_2^{-1} = -1$ (resp. $y_4 = 1$, $y_4^{-1} = -1$). 
The advantage of considering theory \eqref{USp2wSO8pflvsplit} is that it can be refined with three fugacities associated to discrete $0$-form symmetries. In particular, the extra fugacity $\omega$ associated to the $(\BZ_2^{[0]})_{\omega}$ $0$-form symmetry can be gauged in order to obtain a theory with a dual $1$-form symmetry $(\BZ_2^{[1]})_{C}$ corresponding to the center of the $\USp(2)$ gauge group in the left quiver depicted in \eref{USp2wSO8pflvsplitmodZ20}. Indeed, the equality
\bes{ \label{indAffineD4mirrorUSp2wSO8pflvsplit}
 \CI_{\eqref{USp2wSO8pflvsplit}}(y_1, y_2, y_3, y_4| a| \chi_R = 1, \chi_L = 1| \omega; x)=\CI_{\eqref{affineD4}}(\mathfrak{z}_1, \mathfrak{z}_2, \mathfrak{z}_3, \mathfrak{z}_4| a^{-1}| \omega; x)
 }
implies the following mirror duality:
\bes{ \label{USp2wSO8pflvsplitmodZ20}
\begin{tikzpicture}[baseline,font=\footnotesize]
\node (A1) at (-3,0) {$\USp(2)$};
\node (A1label) at (-3,-2) {\eref{eq:USp2-4U1}};
\node (A2) at (-1.8,-1.2) {$\U(1)$};
\node (A3) at (-4.2,1.2) {$\U(1)$};
\node (A4) at (-1.8,1.2) {$\U(1)$};
\node (A5) at (-4.2,-1.2) {$\U(1)$}; 
\draw (A2)--(A1) (A3)--(A1) (A4)--(A1) (A5)--(A1);
\draw [<->] (-0.5,0) to (1.5,0);
\node (SO4) at (3,0) {$[\SO(4)_+]$};
\node (USp2) at (4.8,0) {$\USp(2)$};
\node (SO4r) at (7.6,0) {$[\SO(4)_+]$};
\node (A1label) at (4.8,-1.95) {\eqref{USp2wSO8pflvsplit}$_{\chi_R = \chi_L = 1}/(\BZ_2^{[0]})_{\omega}$};
\draw (SO4) -- (USp2);
\draw[blue] (USp2) to node[midway,above] {\textcolor{blue}{$(\BZ_2^{[0]})_\omega$}} (SO4r);
\node at (9.5,0) {$/(\BZ_2^{[0]})_\omega$};
\end{tikzpicture}
}
where we use the notation $\SO(n)_\pm$ to specify that the fugacity associated to the $\BZ_2$ charge conjugation symmetry of the theory is equal to $\pm 1$ respectively.

\subsubsection*{Discrete Gauging}

In the rest of this section, starting from theories \eqref{USp2SO7SO1} and \eqref{USp2wSO8pflvsplit}$/(\BZ_2^{[0]})_{\omega}$, we will show that, upon gauging different combinations of the $\BZ_2^{[0]}$ $0$-form symmetries of the aforementioned theories, we are able to obtain the mirror duals of \eqref{affineD4} and \eqref{eq:USp2-4U1} wreathed by all $\BZ_2$ and $\BZ_2 \times \BZ_2$ subgroups of $S_4$, namely 
\bi
\item $\BZ_2 = \langle (12) \rangle$, 
\item double transposition $= \langle (12)(34) \rangle$, which is also isomorphic to $\BZ_2$, 
\item non-normal Klein $= \langle (12),(34) \rangle$, which is isomorphic to $\BZ_2 \times \BZ_2$, and
\item normal Klein $= \langle (12)(34), (13)(24) \rangle$, which is also isomorphic to $\BZ_2 \times \BZ_2$.
\ei
We will discuss the computation of the indices in the following, where for and {\it only for} the double transposition and normal Klein subgroups, it is necessary to treat
\bes{
\eta \equiv \chi_1 \chi_2 ~, \quad \xi \equiv \chi_R \chi_L
}
as {\it independent} fugacities from $\chi_{1,2}$ and $\chi_{R,L}$ in \eqref{USp2SO7SO1} and \eqref{USp2wSO8pflvsplit}$/(\BZ_2^{[0]})_{\omega}$, respectively. Note that $\eta$ and $\xi$ are not introduced as independent fugacities for the cases of $\BZ_2$ and the non-normal Klein subgroups. Let us summarize the important results in Table \eqref{eq:wrTable}:
\bes{ 
\scalebox{0.77}{
\begin{tabular}{c||c|c||c|c||c}
Wreathing & Mirror description & Global & Mirror description & Global & CB \\  
 by $\Gamma$ & of \eqref{affineD4}$_{\text{wr} \, \Gamma}$  & symmetry & of \eqref{eq:USp2-4U1}$_{\text{wr} \, \Gamma}$ & symmetry & $\BC^2/\Omega$\\
\hline
\noalign{\vskip 2mm} 
Trivial & $\eqref{USp2wSO8pflv}_{\chi=1}$ $\equiv$ $\eqref{USp2SO7SO1}_{\chi_1=\chi_2=1}$ & $\so(8)$ & $\eqref{USp2wSO8pflvsplit}_{\chi_R=\chi_L=1}$$/(\BZ_2^{[0]})_{\omega}$ & $\su(2)^4$ & $\hat{D}_4$\\[2mm]
$\BZ_2$ & $\eqref{USp2SO7SO1}_{\chi_2=1}$$/(\BZ_2^{[0]})_{\chi_1}$ & $\so(7)$ & $\eqref{USp2wSO8pflvsplit}_{\chi_L=1}$$/(\BZ_2^{[0]})_{\omega, \chi_R}$ & $\su(2)^3$ & $\hat{D}_6$ \\[2mm]
Double transp. & \eqref{USp2SO7SO1}$_{\eta = 1}/(\BZ_2^{[0]})_{\chi_1, \chi_2}$ & $\su(4) \oplus \u(1)$ & \eqref{USp2wSO8pflvsplit}$_{\xi = 1}/(\BZ_2^{[0]})_{\omega, \chi_R, \chi_L}$ & $\su(2)^2$ & $\hat{D}_4$ \\[2mm]
Non-normal Klein & \eqref{USp2SO7SO1}$/(\BZ_2^{[0]})_{\chi_1, \chi_2}$ & $\su(4)$ & \eqref{USp2wSO8pflvsplit}$/(\BZ_2^{[0]})_{\omega, \chi_R, \chi_L}$ & $\su(2)^2$ & $\hat{D}_6$ \\[2mm]
Normal Klein & \eqref{USp2SO7SO1}$_{\eta = 1, \chi_2 = -1}/(\BZ_2^{[0]})_{\chi_1}$ & $\usp(4)$ & \eqref{USp2wSO8pflvsplit}$_{\xi = 1, \chi_L = -1}/(\BZ_2^{[0]})_{\omega, \chi_R}$ & $\su(2)$ & $\hat{D}_4$
\end{tabular}
\label{eq:wrTable}}
}
where the notation $/(\BZ^{[0]}_2)_{a,b,c}$ means that we gauge the $0$-form symmetries associated with the fugacities $a$, $b$ and $c$.  In the last column, we list the Kleinian singularities, which are isomorphic to the Coulomb branch of the mirror theories of \eqref{affineD4} and \eqref{eq:USp2-4U1} wreathed by $\Gamma$; they are the same for both theories. We also remark that, by global symmetry, we mean the one associated with the 3d $\CN=4$ flavor current multiplet which can be read off from the order $x$ of the index.\footnote{However, in the 3d $\CN=2$ language, one has to include also the $\U(1)$ axial symmetry associated with the fugacity $a$.}  

Interestingly, there is a hierarchy in the global symmetries: reading the table from bottom to top, the symmetry gets larger at each step. On the other hand, reading the table from left to right (i.e. moving from theory \eqref{USp2SO7SO1} to \eqref{USp2wSO8pflvsplit}$/(\BZ_2^{[0]})_{\omega}$ and related wreathings), the symmetry is decomposed into a number of $\su(2)$ factors.   

\subsection{Mirrors of the Affine \texorpdfstring{$D_4$}{D4} Quiver \eref{affineD4} Wreathed by \texorpdfstring{$\Gamma$}{Gamma}}

Let us start deriving the results anticipated in Table \eqref{eq:wrTable} by focusing on theory \eqref{USp2SO7SO1}, which allows us to reproduce the mirror description arising from wreathing the affine $D_4$ quiver \eqref{affineD4} by a subgroup $\Gamma$ of $S_4$, where $\Gamma$ can be taken to be one of the $\BZ_2$ or $\BZ_2 \times \BZ_2$ subgroups of $S_4$ as mentioned above.

\subsubsection*{The Case \texorpdfstring{$\Gamma = \BZ_2$}{Z2}}

As claimed in \cite{Bourget:2020bxh}, the mirror dual of the $\BZ_2 = \langle (12) \rangle$ wreathing of the affine $D_4$ quiver, namely \eqref{affineD4}$_{\text{wr} \, \BZ_2}$, is described by
\bes{ \label{USp2SO7O1}
\begin{tikzpicture}[baseline,font=\footnotesize]
\node (O1) at (0,0) {$\O(1)$};
\node (USp2) at (1.8,0) {$\USp(2)$};
\node (SO7) at (3.6,0) {$[\SO(7)_+]$};
\draw (O1) -- (USp2);
\draw (USp2) -- (SO7);
\end{tikzpicture}
}

We now consider this wreathing from the perspective of discrete gauging of \eref{USp2SO7SO1}. Indeed, as pointed out in \cite[Section 3.2]{KobakSwann}, this corresponds to gauging $\O(1)$ inside $\so(8)$ in theory \eqref{USp2wSO8pflv}. This can be realized starting from the theory \eqref{USp2SO7SO1} by gauging the $(\BZ_2^{[0]})_{\chi_1}$ symmetry and setting $\chi_2=1$. The index of theory \eqref{USp2SO7O1} can be obtained as
\bes{ \label{indUSp2SO7O1}
\CI_{\eqref{USp2SO7O1}}(\vec{y}| a; x) = \frac{1}{2} \sum_{\chi_1 = \pm 1} \CI_{\eqref{USp2SO7SO1}}(\vec{y}| a| \chi_1, \chi_2 = 1; x)~.
}
In agreement with mirror symmetry, the flavor symmetry algebra of \eqref{USp2SO7O1} is $\so(7)$, as manifest from the expansion of the index
\bes{ \label{indexUSp2SO7O1}
\CI_{\eqref{USp2SO7O1}}(\vec{y}| a; x)  = &\, {1}+{{a^2 {[0,1,0]}_{\so(7)}}} x \\ &+\left\{{a^4 \left({{[2,0,0]}_{\so(7)}}+ {{[0,2,0]}_{\so(7)}}\right)+ a^{-4}-\left({{[0,1,0]}_{\so(7)}}+1\right)}\right\}{x^2}+\ldots~,
}
and, taking the Higgs and Coulomb branch limits of the index as in \eqref{CBHBlimits}, we find
\bes{
\text{HB of \eqref{USp2SO7O1}} &= \text{CB of } \eqref{affineD4}_{\text{wr} \, \BZ_2} = \bar{\mathrm{n. min}\,B_3}~, \\
\text{CB of \eqref{USp2SO7O1}} &= \text{HB of } \eqref{affineD4}_{\text{wr} \, \BZ_2} = \BC^2/\hat{D}_6~,
}
where the Hilbert series of the latter is
\bes{ \label{HSC2modD6}
\mathrm{HS}\left[\BC^2/\hat{D}_6\right](t) = \PE\left[t^4 + t^8 + t^{10} - t^{20}\right]~.
}

Let us compare the Coulomb branches of \eqref{affineD4} and \eqref{affineD4}$_{\text{wr} \, \BZ_2}$ by contrasting this result with that discussed below \eref{indunrefSO1SO7}. Since the charge conjugation symmetry associated with $\chi_1$ is gauged, we see that $\mathfrak{M}$ and $\mathfrak{M} \varphi$ are no longer gauge invariant, but $\mathfrak{M}^2$ and $\mathfrak{M}^2 \varphi$ are.  In summary, the generators associated with the order $t^4$, $t^8$ and $t^{10}$ of \eref{HSC2modD6} are respectively $\tr(\varphi^2)$, $\mathfrak{M}^2$ and $G_{10} \equiv \mathfrak{M}^2 \varphi$. They satisfy the relation at order $t^{20}$, namely
\bes{
G_{10}^2 + \mathfrak{M}^4 \tr(\varphi^2) = [\tr(\varphi^2)]^5 ~,
}
which is indeed the defining relation of $\BC^2/\hat{D}_6$.

\subsubsection*{The Case \texorpdfstring{$\Gamma = $}{Gamma} Double Transposition}
We can proceed by studying the mirror description of wreathing \eqref{affineD4} by the double transposition subgroup, corresponding to $\Gamma = \langle (12)(34) \rangle$. In terms of theory \eqref{USp2SO7SO1}, this can be interpreted as defining a new independent fugacity $\eta \equiv \chi_1 \chi_2$ before gauging $(\BZ_2^{[0]})_{\chi_1}$ and $(\BZ_2^{[0]})_{\chi_2}$. We remark that, when $(\BZ_2^{[0]})_{\chi_1}$ and $(\BZ_2^{[0]})_{\chi_2}$ are gauged, $\eta$ remains untouched. The index of the theory is defined as
\bes{ \label{indwrDTmirr}
\CI_{\eqref{USp2SO7SO1}_\eta/(\BZ_2^{[0]})_{\chi_1, \chi_2}}(\vec{y}| a; x) = \frac{1}{4} \sum_{\chi_1, \chi_2 = \pm 1} \CI_{\eqref{USp2SO7SO1}}(\vec{y}| a| \chi_1, \chi_2; x)|_{\eta \equiv \chi_1 \chi_2}~.
}
Up to order $x^2$, for $\eta = 1$, it reads
\bes{ \label{indexwrDTmirr}
\CI_{\eqref{USp2SO7SO1}_{\eta = 1}/(\BZ_2^{[0]})_{\chi_1, \chi_2}}(\vec{u}| a; x)  = &\, {1}+{a^2 \left({{[1,0,1]}_{\su(4)}} + 1 \right)}{x}\\ & +\left\{a^4 \left({{[2,0,2]}_{\su(4)}}+3 {{[0,2,0]}_{\su(4)}}+ {{[1,0,1]}_{\su(4)}}+1\right) \right. \\ &\quad \left.+ 2 a^{-4} -\left({{[1,0,1]}_{\su(4)}} + 2\right)\right\} x^2 + \ldots~,
}
with $\vec{u} = \vec{u}(\vec{y})$.\footnote{If we set the $\u(1)$ fugacity to unity and we write the character of the fundamental representation of $\su(4)$ as $u_1 + u_2 u_1^{-1} + u_3 u_2^{-1} + u_3^{-1}$, then we can choose the following fugacity map from $\so(8)$ to $\su(4) \oplus \u(1)$:
\bes{
y_1 = u_2~, \quad y_2 = u_1 u_3^{-1}~, \quad y_3 = u_2 u_1^{-1} u_3^{-1}~, \quad y_4 = 1~. \nonumber
}
\label{foot:mapso8tosu4u1}} As can be seen from the coefficient of $x$, the global symmetry is $\su(4) \oplus \u(1)$. Note that even though the rank of the global symmetry is $4$, the description that we use allows us to see only three Cartan elements whose fugacities are denoted by $u_{1,2,3}$.

Alternatively, the index \eref{indexwrDTmirr} can also be obtained from \eref{indexUSp2wSO8pflv} by exploiting the following branching rules from $\so(8)$ to $\su(4) \oplus \u(1)$:
\bes{
\scalebox{0.98}{$\displaystyle
\renewcommand*{\arraystretch}{1.1}
\begin{array}{rcclclclcl}
[0,1,0,0]_{\so(8)} & \to &   & { [0,0,0]_{\su(4)}(0)} & + & {\red [0,1,0]_{\su(4)}(2)} & + & {\red [0,1,0]_{\su(4)}(-2)} & + & [1,0,1]_{\su(4)}(0)\,,\\\relax
[0,2,0,0]_{\so(8)} & \to &  & [0,0,0]_{\su(4)}(0)    & + & {\red [0,1,0]_{\su(4)}(2)} & + & {\red [0,1,0]_{\su(4)}(-2)} & + & { [1,0,1]_{\su(4)}(0)}\\
                   &     & + & [0,2,0]_{\su(4)}(4)    & + & { [0,2,0]_{\su(4)}(0)}     & + & [0,2,0]_{\su(4)}(-4)        & + &{\red [1,1,1]_{\su(4)}(2)}\\
                   &     & + & {\red [1,1,1]_{\su(4)}(-2)} & + & [2,0,2]_{\su(4)}(0)\,,
\end{array}$}
}
where the terms in {\red red} contain odd highest weights of $\su(4)$ and are thus projected out by gauging $(\BZ_2^{[0]})_{\chi_1, \chi_2}$.  With this approach, we can refine the index with respect to the fugacity of $\u(1)$ in $\su(4) \oplus \u(1)$.  For example, if we denote this fugacity by $b$, the term $3 {{[0,2,0]}_{\su(4)}}$ in \eref{indexwrDTmirr} can be refined as
\bes{
(b^4+1+b^{-4}) {{[0,2,0]}_{\su(4)}}~.
}

Taking the Higgs and Coulomb branch limits as described in \eref{CBHBlimits}, we obtain the Higgs branch Hilbert series
\bes{
&\mathrm{HS}\left[\text{HB of \eqref{USp2SO7SO1}$_{\eta = 1}/(\BZ_2^{[0]})_{\chi_1, \chi_2}$}\right](\vec{u}; t) = \mathrm{HS}\left[\text{CB of } \eqref{affineD4}_{\text{wr $\langle (12)(34) \rangle$}}\right](\vec{u}; t) \\ &= \PE\left[\left({{[1,0,1]}_{\su(4)}} + 1\right) t^2 + \left\{2 {{[0,2,0]_{\su(4)}}}-\left({{[1,0,1]_{\su(4)}}}-1\right)\right\} t^4 + \ldots\right]
}
and the Coulomb branch Hilbert series
\bes{
&\mathrm{HS}\left[\text{CB of \eqref{USp2SO7SO1}$_{\eta = 1}/(\BZ_2^{[0]})_{\chi_1, \chi_2}$}\right](t) = \mathrm{HS}\left[\text{HB of } \eqref{affineD4}_{\text{wr $\langle (12)(34) \rangle$}}\right](t) \\ &= \mathrm{HS}\left[\BC^2/\hat{D}_4\right](t) = \PE\left[2 t^4 + t^6 - t^{12}\right]~.
}
The former can be expressed in closed form, upon setting $u_i = 1$, as
\bes{
&\mathrm{HS}\left[\text{HB of {\eqref{USp2SO7SO1}$_{\eta = 1}/(\BZ_2^{[0]})_{\chi_1, \chi_2}$}}\right] (t) = \mathrm{HS}\left[\text{CB of } \eqref{affineD4}_{\text{wr $\langle (12) (34) \rangle$}}\right] (t) \\ &= \frac{1 + 11 t^2 + 85 t^4 + 280 t^6 + 602 t^8 + 730 t^{10} + 602 t^{12} + 280 t^{14} + 85 t^{16} + 11 t^{18} + t^{20}}{(1 - t)^{10} (1 + t)^{10} (1 + t^2)^5} ~.
}
Both the Higgs and the Coulomb branch Hilbert series agree with the results reported in \cite[Figure 9 and Figure 11]{Bourget:2020bxh}.

\subsubsection*{The Case \texorpdfstring{$\Gamma = $}{Gamma} Non-normal Klein}

The non-normal Klein four-subgroup of $S_4$ is generated by $\langle(12),(34)\rangle$, which, in the field theoretic language we have developed, can be translated into gauging both $(\BZ_2^{[0]})_{\chi_1}$ and $(\BZ_2^{[0]})_{\chi_2}$ independently in \eqref{USp2SO7SO1}. This is reflected in the fact that we do not introduce the variable $\eta$ in this case, which means that the index of the theory is given by
\bes{ \label{indwrNNKmirr}
\CI_{\eqref{USp2SO7SO1}/(\BZ_2^{[0]})_{\chi_1, \chi_2}}(\vec{y}| a; x) = \frac{1}{4} \sum_{\chi_1, \chi_2 = \pm 1} \CI_{\eqref{USp2SO7SO1}}(\vec{y}| a| \chi_1, \chi_2; x)~.
}
Up to order $x^2$, it reads
\bes{ \label{indexwrNNKmirr}
\CI_{\eqref{USp2SO7SO1}/(\BZ_2^{[0]})_{\chi_1, \chi_2}}(\vec{u}| a; x)  
= &\,{1}+{{a^2 {[1,0,1]}_{\su(4)}}} x  \\ &+ \left\{a^4 \left({{[2,0,2]}_{\su(4)}}+2 {{[0,2,0]}_{\su(4)}}+1\right)\right. \\ &\left.\quad+a^{-4}-\left({{[1,0,1]}_{\su(4)}}+1\right)\right\} x^2 + \ldots~,
}
with $\vec{u} = \vec{u}(\vec{y})$, as explained in Footnote \ref{foot:mapso8tosu4u1}. The coefficient of order $x$ confirms that the flavor symmetry algebra is indeed $\su(4)$.

This index can also be obtained from \eref{indexUSp2SO7O1} by using the following branching rule from $\so(7)$ to $\su(4)$ and projecting out all of the terms which are killed by gauging $(\BZ_2^{[0]})_{\chi_2}$:  
\bes{
[0,1,0]_{\so(7)}&\to{\red [0,1,0]_{\su(4)}}+[1,0,1]_{\su(4)}~,\\
[2,0,0]_{\so(7)}&\to[0,0,0]_{\su(4)}+{\red [0,1,0]_{\su(4)}}+[0,2,0]_{\su(4)}~,\\
[0,2,0]_{\so(7)}&\to[0,2,0]_{\su(4)}+{\red [1,1,1]_{\su(4)}}+[2,0,2]_{\su(4)}~,
}
where we highlighted in {\red red} the representations of $\su(4)$ which are projected out after gauging $(\BZ_2^{[0]})_{\chi_2}$, specifically those containing odd highest weights of $\su(4)$.

Using \eqref{CBHBlimits}, the Higgs and Coulomb branch limits reproduce the Higgs branch Hilbert series
\bes{
&\mathrm{HS}\left[\text{HB of {\eqref{USp2SO7SO1}$/(\BZ_2^{[0]})_{\chi_1, \chi_2}$}}\right] (\vec{u}; t) = \mathrm{HS}\left[\text{CB of } \eqref{affineD4}_{\text{wr $\langle (12), (34) \rangle$}}\right] (\vec{u}; t) \\ &= \PE\left[[1,0,1]_{\su(4)} t^2 + \left({{[0,2,0]}_{\su(4)}} - {{[1,0,1]}_{\su(4)}}\right) t^4 + \ldots\right]
}
and the Coulomb branch Hilbert series
\bes{
&\mathrm{HS}\left[\text{CB of {\eqref{USp2SO7SO1}$/(\BZ_2^{[0]})_{\chi_1, \chi_2}$}}\right] (t) = \mathrm{HS}\left[\text{HB of } \eqref{affineD4}_{\text{wr $\langle (12), (34) \rangle$}}\right] (t) \\ &= \mathrm{HS}\left[\BC^2/\hat{D}_6\right](t) = \PE\left[t^4 + t^8 + t^{10} - t^{20}\right]~.
}
The Coulomb branch is in agreement with \cite[Figure 9 and Figure 11]{Bourget:2020bxh}, as well as the unrefined Higgs branch Hilbert series, which admits the following closed formula:
\bes{
&\mathrm{HS}\left[\text{HB of {\eqref{USp2SO7SO1}$/(\BZ_2^{[0]})_{\chi_1, \chi_2}$}}\right] (t) = \mathrm{HS}\left[\text{CB of } \eqref{affineD4}_{\text{wr $\langle (12), (34) \rangle$}}\right] (t) \\ &= \frac{1 + 10 t^2 + 55 t^4 + 150 t^6 + 288 t^8 + 336 t^{10} + 288 t^{12} + 150 t^{14} + 55 t^{16} + 10 t^{18} + t^{20}}{(1 - t)^{10} (1 + t)^{10} (1 + t^2)^5} ~.
}

\subsubsection*{The Case \texorpdfstring{$\Gamma = $}{Gamma} Normal Klein}

Finally, let us turn our attention to $\Gamma = \langle (12)(34) , (13)(24) \rangle$. In terms of theory \eqref{USp2SO7SO1}, this corresponds to the situation in which we first define $\eta \equiv \chi_1 \chi_2$ as an independent fugacity and then set $\chi_2 = -1$ and gauge $(\BZ_2^{[0]})_{\chi_1}$. Again, we remark that $\eta$ is not affected by the process of gauging $(\BZ_2^{[0]})_{\chi_1}$.\\
The index of the theory reads
\bes{ \label{indwrNKmirr}
\CI_{\eqref{USp2SO7SO1}_{\eta, \, \chi_2 = -1}/(\BZ_2^{[0]})_{\chi_1}}(\vec{y}| a; x) &= \frac{1}{2} \sum_{\chi_1 = \pm 1} \CI_{\eqref{USp2SO7SO1}}(\vec{y}| a| \chi_1, \chi_2; x)|_{\eta \equiv \chi_1 \chi_2, \, \chi_2 = -1}~,
}
which, for $\eta = 1$, can be expanded in terms of the characters of representations of $\usp(4)$ as
\bes{ \label{indexwrNKmirr}
\scalebox{0.98}{$
\begin{split}
\CI_{\eqref{USp2SO7SO1}_{\eta = 1, \chi_2 = -1}/(\BZ_2^{[0]})_{\chi_1}}(\vec{v}| a; x) = &\, {1}+{{a^2 {[2,0]}_{\usp(4)}}}{x}\\ &  +\left\{a^4 \left({[4,0]}_{\usp(4)}+{[2,1]}_{\usp(4)}+{[0,2]}_{\usp(4)}+{[0,1]}_{\usp(4)}+1\right) \right. \\ &\left. \quad +2 a^{-4}-\left({[2,0]}_{\usp(4)}+1\right)\right\} x^2 + \ldots~,
\end{split}$}
}
with $\vec{v} = \vec {v}(\vec{y})$.\footnote{If we write the character of the fundamental representation of $\usp(4)$ as $\sum_{i=1}^2 (v_i +v_i^{-1})$, then, in order to express \eqref{indwrNKmirr} in the form \eqref{indexwrNKmirr}, we can choose the following fugacity map:
\bes{
y_1 = v_1 v_2~, \quad y_2 = v_1 v_2^{-1}~, \quad y_3 = 1~. \nonumber
}}
The coefficient of $x$ indicates that the flavor symmetry algebra is indeed $\mathfrak{usp}(4)$. Upon using \eqref{CBHBlimits}, we find that the Higgs branch limit of \eqref{indwrNKmirr} is given by
\bes{
&\mathrm{HS}\left[\text{HB of {\eqref{USp2SO7SO1}$_{\eta = 1, \chi_2 = -1}/(\BZ_2^{[0]})_{\chi_1}$}}\right] (\vec{v}; t) = \mathrm{HS}\left[\text{CB of } \eqref{affineD4}_{\text{wr $\langle (12) (34), (13) (24) \rangle$}}\right] (\vec{v}; t) \\ &= \PE\left[{[2,0]}_{\usp(4)} t^2 + \left(2 {[0,2]}_{\usp(4)}+{[0,1]}_{\usp(4)}+ 2\right) t^4+\ldots\right]~,
}
whereas the Coulomb branch limit yields
\bes{
&\mathrm{HS}\left[\text{CB of {\eqref{USp2SO7SO1}$_{\eta = 1, \chi_2 = -1}/(\BZ_2^{[0]})_{\chi_1}$}}\right] (t) = \mathrm{HS}\left[\text{HB of } \eqref{affineD4}_{\text{wr $\langle (12) (34), (13) (24) \rangle$}}\right] (t) \\ &= \mathrm{HS}\left[\BC^2/\hat{D}_4\right](t) = \PE\left[2 t^4 + t^6 - t^{12}\right]~.
}
Moreover, the unrefined Higgs branch Hilbert series can be expressed in closed form as
\bes{
&\mathrm{HS}[\text{HB of {\eqref{USp2SO7SO1}$_{\eta = 1, \chi_2 = -1}/(\BZ_2^{[0]})_{\chi_1}$}}] (t) = \mathrm{HS}[\text{CB of } \eqref{affineD4}_{\text{wr $\langle (12) (34), (13) (24) \rangle$}}] (t) \\ &= \frac{1 + 5 t^2 + 45 t^4 + 130 t^6 + 314 t^8 + 354 t^{10} + 314 t^{12} + 130 t^{14} + 45 t^{16} + 5 t^{18} + t^{20}}{(1 - t)^{10} (1 + t)^{10} (1 + t^2)^5} ~.
}
Once again, these results are in agreement with \cite[Figure 9 and Figure 11]{Bourget:2020bxh}.
\subsection{Mirrors of Quiver \eref{eq:USp2-4U1} Wreathed by \texorpdfstring{$\Gamma$}{Gamma}}
Let us now analyze theory \eqref{USp2wSO8pflvsplit}$/(\BZ_2^{[0]})_\omega$, which allows us to reproduce the details reported in Table \eqref{eq:wrTable} and, therefore, obtain new information about the mirror description \eqref{eq:USp2-4U1} and its wreathing by a subgroup $\Gamma$ of $S_4$. As before, $\Gamma$ is taken to be one among the $\BZ_2$, double transposition, normal Klein and non-normal Klein subgroups.

In the case where $\Gamma$ is trivial, we have already pointed out that the mirror dual of \eqref{eq:USp2-4U1} is theory \eqref{USp2wSO8pflvsplit}$/(\BZ_2^{[0]})_\omega$ with $\chi_R = \chi_L = 1$, see \eqref{USp2wSO8pflvsplitmodZ20}. The index can be computed as described in \eref{indUSp2wSO8pflvb} with $\omega$ being summed over $\pm 1$. The result up to order $x^2$ is given by
\bes{
\scalebox{0.99}{$
\begin{split}
&\CI_{\eqref{USp2wSO8pflvsplit}/(\BZ_2^{[0]})_{\omega}}(\vec{\mathfrak{z}}| a| \chi_R = 1, \chi_L = 1; x) \\&=   1+a^2 \left(\left[2; 0; 0; 0\right]_{\su(2)^4} + \left[0; 2; 0; 0\right]_{\su(2)^4} + \left[0; 0; 2; 0\right]_{\su(2)^4} + \left[0; 0; 0; 2\right]_{\su(2)^4} \right){x} \\ & \quad \, \, \, \, \,+ \left\{a^4 \left(\left[4; 0; 0; 0\right]_{\su(2)^4} + \left[0; 4; 0; 0\right]_{\su(2)^4} + \left[0; 0; 4; 0\right]_{\su(2)^4} + \left[0; 0; 0; 4\right]_{\su(2)^4} \right.\right.\\& \quad \,\,\,\, \,  +\left.\left.\left[2; 2; 2; 2\right]_{\su(2)^4} + \left[2; 2; 0; 0\right]_{\su(2)^4} + \left[0; 0; 2; 2\right]_{\su(2)^4} + \left[2; 0; 2; 0\right]_{\su(2)^4} \right.\right.\\&\quad \,\,\,\, \, +\left.\left. \left[2; 0; 0; 2\right]_{\su(2)^4} + \left[0; 2; 2; 0\right]_{\su(2)^4} + \left[0; 2; 0; 2\right]_{\su(2)^4} + 1\right) + 2 a^{-4} \right.\\& \quad \, \, \, \, \,- \left.\left(\left[2; 0; 0; 0\right]_{\su(2)^4} + \left[0; 2; 0; 0\right]_{\su(2)^4} + \left[0; 0; 2; 0\right]_{\su(2)^4} + \left[0; 0; 0; 2\right]_{\su(2)^4} + 1\right)\right\} x^2 + \ldots~,
\end{split}$}
}
where $\vec{\mathfrak{z}} = \vec{\mathfrak{z}} (\vec{y})$, see \eqref{indAffineD4mirrorUSp2wSO8pflvsplit}. Note that the coefficient of $x$ indicates that the flavor symmetry is $\su(2)^4$.  This index can also be obtained from \eref{indexUSp2wSO8pflv} by exploiting the branching rule from $\so(8)$ to $\su(2)^4$:
\bes{
\scalebox{0.85}{$\displaystyle
\renewcommand*{\arraystretch}{1.1}
\begin{array}{rcclclclclcl}
[0,1,0,0]_{\so(8)} & \to &   & [2;0;0;0]_{\su(2)^4}        & + & [0;2;0;0]_{\su(2)^4}        & + & [0;0;2;0]_{\su(2)^4}        & + & [0;0;0;2]_{\su(2)^4}        & + & {\blue [1;1;1;1]_{\su(2)^4}}\,,\\\relax
[0,2,0,0]_{\so(8)} & \to &   & [0;0;0;0]_{\su(2)^4}        & + & {\blue[1;1;1;1]_{\su(2)^4}} & + & [2;2;0;0]_{\su(2)^4}        & + & [2;0;2;0]_{\su(2)^4}        & + & [2;0;0;2]_{\su(2)^4}           \\
                   &     & + & [0;2;2;0]_{\su(2)^4}        & + & [0;2;0;2]_{\su(2)^4}        & + & [0;0;2;2]_{\su(2)^4}        & + & [4;0;0;0]_{\su(2)^4}        & + & [0;4;0;0]_{\su(2)^4}           \\
                   &     & + & [0;0;4;0]_{\su(2)^4}        & + & [0;0;0;4]_{\su(2)^4}        & + & {\blue[3;1;1;1]_{\su(2)^4}} & + & {\blue[1;3;1;1]_{\su(2)^4}} & + & {\blue[1;1;3;1]_{\su(2)^4}}    \\
                   &     & + & {\blue[1;1;1;3]_{\su(2)^4}} & + & [2;2;2;2]_{\su(2)^4}~,
\end{array}$}
}
where the terms in {\blue blue} contain odd highest weights of $\su(2)_{1, 2}$ and they are projected out upon gauging $(\BZ_2^{[0]})_\omega$.  Using \eref{CBHBlimits}, we obtain the Higgs and Coulomb branch limits for $\chi_R = \chi_L = 1$ as follows:
\bes{
\text{HB of \eqref{USp2wSO8pflvsplit}}/(\BZ_2^{[0]})_{\omega} &= \text{CB of \eqref{eq:USp2-4U1}}~, \\
\text{CB of \eqref{USp2wSO8pflvsplit}}/(\BZ_2^{[0]})_{\omega} &= \text{HB of \eqref{eq:USp2-4U1}} = \BC^2/\hat{D}_4~.
}

Now, let us proceed by considering $\Gamma$ non-trivial. To systematically derive all the wreathings we are interested in, it is useful to consider the expansion of the index of theory \eqref{USp2wSO8pflvsplit}$/(\BZ_2^{[0]})_\omega$ for generic $\chi_R$ and $\chi_L$ by pointing out the explicit dependence on the two charge conjugation fugacities. For instance, the coefficient at order $x$ reads
\bes{\label{indexUSp2wSO8flvsplimodZ2ordx}
a^2 \left\{\frac{\left(1 + \chi_R\right)}{2} \left[\left(y_1^2 + 1 + y_1^{-2}\right) + \left(y_1 \leftrightarrow y_2\right)\right] + \frac{\left(1 + \chi_L\right)}{2} \left[\left(y_3^2 + 1 + y_3^{-2}\right) + \left(y_3 \leftrightarrow y_4\right)\right] \right\}~.
}
Note that, if we gauge the symmetry associated with $\chi_R$ (resp. $\chi_L$) by summing the fugacity over $\pm 1$, this leads to the coefficient $\frac{1}{2}$ in front of the characters in \eref{indexUSp2wSO8flvsplimodZ2ordx}. This means that, in such a case, the symmetry is broken to the diagonal subalgebra of $\su(2)_1 \oplus \su(2)_2$ (resp. $\su(2)_3 \oplus \su(2)_4$). In the index, this amounts to setting $\mathfrak{z}_1 = \mathfrak{z}_2 \equiv f_1$ (resp. $\mathfrak{z}_3 = \mathfrak{z}_4 \equiv f_2$). After doing so, the index becomes well-defined again. Explicitly, if we set $\mathfrak{z}_1 = \mathfrak{z}_2 \equiv f_1$ and $\mathfrak{z}_3 = \mathfrak{z}_4 \equiv f_2$, the index for generic $\chi_L$ and $\chi_R$ can be written as\footnote{If we denote the diagonal combination of $\su(2)_i$ and $\su(2)_j$ as $\su(2)_{i j}$, then $\left[a; b\right]_{\su(2)^2}$ stands for the character of the representation of $\su(2)_{1 2} \oplus \su(2)_{3 4}$, with highest weights $a$ and $b$ under $\su(2)_{1 2}$ and $\su(2)_{3 4}$ respectively.}
\bes{ \label{indexUSp2wSO8flvsplimodZ2}
\scalebox{0.9}{$\displaystyle
\begin{aligned}
\CI_{\eqref{USp2wSO8pflvsplit}/(\BZ_2^{[0]})_{\omega}}(\vec{f}| a| \chi_R, \chi_L; x) = &\, {1}+a^2 \left\{\left(1 + \chi_R\right) \left[2; 0\right]_{\su(2)^2} + \left(1 + \chi_L\right) \left[0; 2\right]_{\su(2)^2} \right\}{x} \\ 
&  +\left\{a^4 \left[\left[4; 4\right]_{\su(2)^2} + \chi_R \left[2; 4\right]_{\su(2)^2} + \chi_L \left[4; 2\right]_{\su(2)^2} + \left(3 + \chi_R\right) \left[4; 0\right]_{\su(2)^2} \right.\right.\\
&  +\left.\left.\left(3 + \chi_L\right) \left[0; 4\right]_{\su(2)^2} + \left(1 + \chi_R + \chi_L + 2 \chi_R \chi_L\right) \left[2; 2\right]_{\su(2)^2} \right.\right.\\
&  +\left.\left.2 \chi_R \left[2; 0\right]_{\su(2)^2} + 2 \chi_L \left[0; 2\right]_{\su(2)^2} + 4 \right] + \left(1 + \chi_R \chi_L\right) a^{-4} \right.\\
& -\left.\left[\left(1 + \chi_R\right) \left[2; 0\right]_{\su(2)^2} + \left(1 + \chi_L\right) \left[0; 2\right]_{\su(2)^2} + 1\right]\right\} x^2 + \ldots~.
\end{aligned}$}
}
This will be useful for obtaining the indices for the theories listed in \eqref{eq:wrTable}.

\subsubsection*{The Case \texorpdfstring{$\Gamma = \BZ_2$}{Z2}}

Analogously to what we did when dealing with theory \eqref{USp2SO7SO1}, to implement wreathing by $\Gamma = \BZ_2 = \langle (12) \rangle$, we gauge the $(\BZ_2^{[0]})_{\chi_R}$ symmetry of theory \eqref{USp2wSO8pflvsplit}$/(\BZ_2^{[0]})_{\omega}$. At the level of the index, once we set $\chi_L = 1$, this amounts to\footnote{Given that theory \eqref{USp2wSO8pflvsplit}$/(\BZ_2^{[0]})_{\omega}$ is symmetric under swapping $\left(\chi_R; y_1, y_2 \right) \leftrightarrow \left(\chi_L; y_3, y_4 \right)$, the mirror description corresponding to the $\BZ_2$ wreathing of theory \eqref{eq:USp2-4U1} can be realized equivalently by gauging $(\BZ_2^{[0]})_{\chi_L}$ and setting $\chi_R = 1$.}
\bes{ \label{indUSp2wSO9flvsplitZ2}
\CI_{\eqref{USp2wSO8pflvsplit}_{\chi_L = 1}/(\BZ_2^{[0]})_{\omega, \chi_R}}(\vec{y}| a; x) = \frac{1}{2} \sum_{\chi_R = \pm 1} \CI_{\eqref{USp2wSO8pflvsplit}/(\BZ_2^{[0]})_{\omega}}(\vec{y}| a| \chi_R, \chi_L = 1; x)~.
}
In order to compute \eqref{indUSp2wSO9flvsplitZ2}, it is necessary to take the diagonal combination of $\su(2)_1 \oplus \su(2)_2$ since the latter is broken into the diagonal subalgebra.  As discussed above, the corresponding fugacity is denoted by $f_1$.  However, $\su(2)_3$ and $\su(2)_4$ remain unbroken, and the corresponding fugacities are $\fz_{3,4}$. Using \eref{indUSp2wSO8pflvb} and the discussion below, we obtain the following index, which can be written in terms of the characters of representations of $\su(2)^3$ as follows:\footnote{We denote by $\left[a; b; c\right]$ the character of the representation of $\su(2)_{1 2} \oplus \su(2)_3 \oplus \su(2)_4$ with highest weights $a, b$ and $c$ under $\su(2)_{1 2}, \su(2)_3$ and $\su(2)_4$ respectively.}
\bes{ \label{indexUSp2wSO8flvspliZ2}
\scalebox{0.89}{$\displaystyle
\begin{aligned}
\CI_{\eqref{USp2wSO8pflvsplit}_{\chi_L = 1}/(\BZ_2^{[0]})_{\omega, \chi_R}}(f_1, \mathfrak{z}_3, \mathfrak{z}_4| a; x) = &\,{1}+a^2 \left( \left[2; 0; 0\right]_{\su(2)^3} + \left[0; 2; 0\right]_{\su(2)^3} + \left[0; 0; 2\right]_{\su(2)^3}\right){x} \\ & +\left\{a^4 \left(\left[4; 2; 2\right]_{\su(2)^3} + 2 \left[4; 0; 0\right]_{\su(2)^3} + \left[0; 4; 0\right]_{\su(2)^3} +\left[0; 0; 4\right]_{\su(2)^3} \right.\right.\\& +\left.\left. \left[2; 2; 0\right]_{\su(2)^3}+\left[2; 0; 2\right]_{\su(2)^3} +2 \left[0; 2; 2\right]_{\su(2)^3} + 2 \right) + a^{-4} \right. \\& -\left. \left(\left[2; 0; 0\right]_{\su(2)^3}  +\left[0; 2; 0\right]_{\su(2)^3} + \left[0; 0; 2\right]_{\su(2)^3} + 1\right)\right\} x^2 + \ldots~,
\end{aligned}$}
}
where one can check that, after gauging $\chi_R$ and setting $\chi_L = 1$, \eqref{indexUSp2wSO8flvspliZ2} is equivalent to \eqref{indexUSp2wSO8flvsplimodZ2} upon taking the diagonal combination of $\su(2)_3$ and $\su(2)_4$ and taking tensor products of $\su(2)_{3 4}$:
\bes{
\left[4; 2; 2\right]_{\su(2)^3} \to & \left[4; 4\right]_{\su(2)^2} + \left[4; 2\right]_{\su(2)^2} + \left[4; 0\right]_{\su(2)^2}~, \\
\left[0; 2; 2\right]_{\su(2)^3} \to & \left[0; 4\right]_{\su(2)^2} + \left[0; 2\right]_{\su(2)^2} + \left[0; 0\right]_{\su(2)^2}~.
}
Now, we can take the Higgs and Coulomb branch limits of the index as described in \eqref{CBHBlimits}: the Higgs branch Hilbert series reads
\bes{
&\mathrm{HS}\left[\text{HB of \eqref{USp2wSO8pflvsplit}}_{\chi_L = 1}/(\BZ_2^{[0]})_{\omega, \chi_R}\right] (f_1, \mathfrak{z}_3, \mathfrak{z}_4; t) = \mathrm{HS}\left[\text{CB of } \eqref{eq:USp2-4U1}_{\text{wr} \, \BZ_2}\right] (f_1, \mathfrak{z}_3, \mathfrak{z}_4; t) \\ &= \PE\left[\left({[2;0;0]}_{\su(2)^3} + {[0;2;0]}_{\su(2)^3} + {[0;0;2]}_{\su(2)^3}\right) t^2 \right.\\&\left. \quad  \, \, \, \, \, \, \, +\left(\left[4; 2; 2\right]_{\su(2)^3} + \left[4; 0; 0\right]_{\su(2)^3} + \left[0; 2; 2\right]_{\su(2)^3} - 1\right) t^4 + \ldots\right]~,
}
whereas the Coulomb branch Hilbert series is given by
\bes{
&\mathrm{HS}\left[\text{CB of \eqref{USp2wSO8pflvsplit}}_{\chi_L = 1}/(\BZ_2^{[0]})_{\omega, \chi_R}\right] (t) = \mathrm{HS}\left[\text{HB of } \eqref{eq:USp2-4U1}_{\text{wr} \, \BZ_2}\right] (t) \\ &=\mathrm{HS}\left[\BC^2/\hat{D}_6\right](t)= \PE\left[t^4 + t^8 + t^{10} - t^{20}\right]~.
}
Moreover, once unrefined, the Higgs branch Hilbert series can be expressed in closed form as
\bes{
\mathrm{HS}\left[\text{HB of \eqref{USp2wSO8pflvsplit}}_{\chi_L = 1}/(\BZ_2^{[0]})_{\omega, \chi_R}\right] (t) = \mathrm{HS}[\text{CB of } \eqref{eq:USp2-4U1}_{\text{wr} \, \BZ_2}] (t) = \eqref{closedformwrZ2} ~.
}
\subsubsection*{The Case \texorpdfstring{$\Gamma = $}{Gamma} Double Transposition}
The mirror description of wreathing \eqref{eq:USp2-4U1} by the double transposition subgroup of $S_4$, namely $\Gamma = \langle (12)(34) \rangle$, can be derived by defining a new independent fugacity $\xi \equiv \chi_R \chi_L$ before gauging $(\BZ_2^{[0]})_{\chi_R}$ and $(\BZ_2^{[0]})_{\chi_L}$. In this theory, the fugacity $\xi$ plays the same role as $\eta$ in theory \eqref{USp2SO7SO1}.

The index, which is given by
\bes{ \label{indwrDTmirrmodZ2}
\CI_{\eqref{USp2wSO8pflvsplit}_{\xi}/(\BZ_2^{[0]})_{\omega, \chi_R, \chi_L}}(\vec{y}| a; x) = \frac{1}{4} \sum_{\chi_R, \chi_L = \pm 1} \CI_{\eqref{USp2wSO8pflvsplit}/(\BZ_2^{[0]})_\omega}(\vec{y}| a| \chi_R, \chi_L; x)|_{\xi \equiv \chi_R \chi_L}~,
}
can be expanded in powers of $x$ using \eqref{indexUSp2wSO8flvsplimodZ2} with $\chi_R \chi_L \equiv \xi$. Up to order $x^2$, for $\xi = 1$, it reads
\bes{ 
\scalebox{0.94}{$\displaystyle
\begin{aligned}
    \CI_{\eqref{USp2wSO8pflvsplit}_{\xi = 1}/(\BZ_2^{[0]})_{\omega, \chi_R, \chi_L}}(\vec{f}| a; x) = &\,{1}+a^2 \left( \left[2; 0\right]_{\su(2)^2} +  \left[0; 2\right]_{\su(2)^2} \right){x} \\ & +\left\{a^4 \left(\left[4; 4\right]_{\su(2)^2} + 3 \left[4; 0\right]_{\su(2)^2} +3 \left[0; 4\right]_{\su(2)^2} +3 \left[2; 2\right]_{\su(2)^2} + 4 \right)  \right.\\&  +\left. 2 a^{-4}-\left(\left[2; 0\right]_{\su(2)^2} +  \left[0; 2\right]_{\su(2)^2} + 1\right)\right\} x^2 + \ldots~,
\end{aligned}
$}
}
from which, using \eqref{CBHBlimits}, we recover the Higgs branch Hilbert series
\bes{
&\mathrm{HS}\left[\text{HB of \eqref{USp2wSO8pflvsplit}}_{\xi = 1}/(\BZ_2^{[0]})_{\omega, \chi_R, \chi_L}\right] (\vec{f}; t) = \mathrm{HS}\left[\text{CB of } \eqref{eq:USp2-4U1}_{\text{wr} \, \langle (12)(34) \rangle}\right] (\vec{f}; t) \\ &= \PE\left[\left({[2;0]}_{\su(2)^2} + {[0;2]}_{\su(2)^2}\right) t^2\right.\\&\left. \quad  \, \, \, \, \, \, \, +\left({[4;4]}_{\su(2)^2} + 2 {[4;0]}_{\su(2)^2}+ 2 {[0;4]}_{\su(2)^2}  +2 {[2;2]}_{\su(2)^2} + 2\right) t^4 + \ldots\right]
}
and the Coulomb branch Hilbert series
\bes{
&\mathrm{HS}\left[\text{CB of \eqref{USp2wSO8pflvsplit}}_{\xi = 1}/(\BZ_2^{[0]})_{\omega, \chi_R, \chi_L}\right] (t) = \mathrm{HS}\left[\text{HB of } \eqref{eq:USp2-4U1}_{\text{wr} \, \langle (12)(34) \rangle}\right] (t) \\
& =\mathrm{HS}\left[\BC^2/\hat{D}_4\right](t)= \PE\left[2 t^4 + t^6 - t^{12}\right]~.
}
Furthermore, the unrefined Higgs branch Hilbert series admits the following closed expression:
\bes{
&\mathrm{HS}[\text{HB of \eqref{USp2wSO8pflvsplit}}_{\xi = 1}/(\BZ_2^{[0]})_{\omega, \chi_R, \chi_L}] (t) = \mathrm{HS}[\text{CB of } \eqref{eq:USp2-4U1}_{\text{wr} \, \langle (12)(34) \rangle}] (t) \\ &= \frac{1 + t^2 + 61 t^4 + 94 t^6 + 378 t^8 + 274 t^{10} + 378 t^{12} + 94 t^{14} + 61 t^{16} + t^{18} + t^{20}}{(1 - t)^{10} (1 + t)^{10} (1 + t^2)^5} ~.
}
\subsubsection*{The Case \texorpdfstring{$\Gamma = $}{Gamma} Non-normal Klein}
Wreathing quiver \eqref{eq:USp2-4U1} by $\Gamma = \langle (12),(34) \rangle$ is equivalent to gauging both $(\BZ_2^{[0]})_{\chi_R}$ and $(\BZ_2^{[0]})_{\chi_L}$ independently, without introducing the fugacity $\xi$, in theory \eqref{USp2wSO8pflvsplit}$/(\BZ_2^{[0]})_\omega$. This situation can be analyzed using the index
\bes{ \label{indwrNNKmirrmodZ2}
\CI_{\eqref{USp2wSO8pflvsplit}/(\BZ_2^{[0]})_{\omega, \chi_R, \chi_L}}(\vec{y}| a; x) = \frac{1}{4} \sum_{\chi_R, \chi_L = \pm 1} \CI_{\eqref{USp2wSO8pflvsplit}/(\BZ_2^{[0]})_\omega}(\vec{y}| a| \chi_R, \chi_L; x)~,
}
whose expansion, from \eqref{indexUSp2wSO8flvsplimodZ2}, reads
\bes{ 
\scalebox{0.99}{$\displaystyle
\begin{aligned}
\CI_{\eqref{USp2wSO8pflvsplit}/(\BZ_2^{[0]})_{\omega, \chi_R, \chi_L}}(\vec{f}| a; x) = &\, {1}+a^2 \left\{\left[2; 0\right]_{\su(2)^2} + \left[0; 2\right]_{\su(2)^2} \right\}{x} \\ & +\left\{a^4 \left(\left[4; 4\right]_{\su(2)^2} + 3 \left[4; 0\right]_{\su(2)^2} + 3 \left[0; 4\right]_{\su(2)^2} + \left[2; 2\right]_{\su(2)^2} + 4 \right) \right.\\& +\left.a^{-4} -\left(\left[2; 0\right]_{\su(2)^2} + \left[0; 2\right]_{\su(2)^2} + 1\right)\right\} x^2 + \ldots~.
\end{aligned}$}
}
Using \eqref{CBHBlimits}, we obtain the Higgs branch limit of \eqref{indwrNNKmirrmodZ2} as
\bes{
\scalebox{0.98}{$
\begin{split}
&\mathrm{HS}\left[\text{HB of \eqref{USp2wSO8pflvsplit}}/(\BZ_2^{[0]})_{\omega, \chi_R, \chi_L}\right] (\vec{f}; t) = \mathrm{HS}\left[\text{CB of } \eqref{eq:USp2-4U1}_{\text{wr} \, \langle (12), (34) \rangle}\right] (\vec{f}; t) \\ &= \PE\left[\left({[2;0]}_{\su(2)^2} + {[0;2]}_{\su(2)^2}\right) t^2 +\left({[4;4]}_{\su(2)^2} + 2 {[4;0]}_{\su(2)^2}+ 2 {[0;4]}_{\su(2)^2}+ 2\right) t^4 + \ldots\right]~,
\end{split}$}
}
whereas the Coulomb branch limit of \eqref{indwrNNKmirrmodZ2} is given by
\bes{
&\mathrm{HS}\left[\text{CB of \eqref{USp2wSO8pflvsplit}}/(\BZ_2^{[0]})_{\omega, \chi_R, \chi_L}\right] (t) = \mathrm{HS}\left[\text{HB of } \eqref{eq:USp2-4U1}_{\text{wr} \, \langle (12), (34) \rangle}\right] (t) \\
& =\mathrm{HS}\left[\BC^2/\hat{D}_6\right](t) = \PE\left[t^4 + t^8 + t^{10} - t^{20}\right]~.
}
The unrefinement of the former can be written in closed form as
\bes{
&\mathrm{HS}\left[\text{HB of \eqref{USp2wSO8pflvsplit}}/(\BZ_2^{[0]})_{\omega, \chi_R, \chi_L}\right] (t) = \mathrm{HS}\left[\text{CB of } \eqref{eq:USp2-4U1}_{\text{wr} \, \langle (12), (34) \rangle}\right] (t) \\ &= \frac{1 + t^2 + 43 t^4 + 37 t^6 + 208 t^8 + 92 t^{10} + 208 t^{12} + 37 t^{14} + 43 t^{16} + t^{18} + t^{20}}{(1 - t)^{10} (1 + t)^{10} (1 + t^2)^5} ~.
}
\subsubsection*{The Case \texorpdfstring{$\Gamma = $}{Gamma} Normal Klein}
Finally, we can gain new insights into the $\Gamma = \langle (12)(34) , (13)(24) \rangle$ wreathing of quiver \eqref{eq:USp2-4U1} by exploiting the mirror description \eqref{USp2wSO8pflvsplit}$/(\BZ_2^{[0]})_{\omega}$ with the independent fugacity $\xi$ defined as before, $\chi_L = -1$ and $(\BZ_2^{[0]})_{\chi_R}$ gauged. In such a case, the index is given by
\bes{ \label{indwrNKmirrmodZ2}
\CI_{\eqref{USp2wSO8pflvsplit}_{\xi, \, \chi_L = -1}/(\BZ_2^{[0]})_{\omega, \chi_R}}(\vec{y}| a; x) = \frac{1}{2} \sum_{\chi_R = \pm 1} \CI_{\eqref{USp2wSO8pflvsplit}/(\BZ_2^{[0]})_\omega}(\vec{y}| a| \chi_R, \chi_L; x)|_{\xi \equiv \chi_L \chi_R, \, \chi_L = -1}
}
and its expansion for $\xi = 1$ can be evaluated thanks to \eqref{indexUSp2wSO8flvsplimodZ2}:
\bes{ 
\scalebox{0.9}{$\displaystyle
\begin{aligned}
\CI_{\eqref{USp2wSO8pflvsplit}_{\xi = 1, \chi_L = -1}/(\BZ_2^{[0]})_{\omega, \chi_R}}(f| a; x) = &\,{1}+a^2 \left[2\right]_{\su(2)} {x}  \\ &+\left\{a^4 \left(\left[8\right]_{\su(2)} + 7 \left[4\right]_{\su(2)} + 7\right) + 2 a^{-4} - \left(\left[2\right]_{\su(2)} + 1\right)\right\} x^2 + \ldots~,
\end{aligned}$}
}
where we observe that the flavor symmetry is just $\su(2)$, with $\su(2)_{1 2} \oplus \su(2)_{3 4}$ broken to the diagonal combination, with $f_1 = f_2 \equiv f$. Using \eqref{CBHBlimits}, we find that the Higgs and Coulomb branch limits of \eqref{indwrNKmirrmodZ2} reproduce the Higgs branch Hilbert series, which is given by
\bes{
&\mathrm{HS}\left[\text{HB of \eqref{USp2wSO8pflvsplit}}_{\xi = 1, \chi_L = -1}/(\BZ_2^{[0]})_{\omega, \chi_R}\right] (f; t) = \mathrm{HS}\left[\text{CB of } \eqref{eq:USp2-4U1}_{\text{wr} \, \langle (12)(34) , (13)(24) \rangle}\right] (f; t) \\ &= \PE\left[{[2]}_{\su(2)} t^2 + \left({[8]}_{\su(2)} + 6 {[4]}_{\su(2)} + 6\right) t^4 + \ldots\right]~,
}
and the following Coulomb branch Hilbert series:
\bes{
&\mathrm{HS}\left[\text{CB of \eqref{USp2wSO8pflvsplit}}_{\xi = 1, \chi_L = -1}/(\BZ_2^{[0]})_{\omega, \chi_R}\right] (t) = \mathrm{HS}\left[\text{HB of } \eqref{eq:USp2-4U1}_{\text{wr} \, \langle (12)(34) , (13)(24) \rangle}\right] (t) \\&=\mathrm{HS}\left[\BC^2/\hat{D}_4\right](t)= \PE\left[2 t^4 + t^6 - t^{12}\right]~.
}
We also provide a closed formula for the unrefined Higgs branch Hilbert series, which reads
\bes{
&\mathrm{HS}[\text{HB of \eqref{USp2wSO8pflvsplit}}_{\xi = 1, \chi_L = -1}/(\BZ_2^{[0]})_{\omega, \chi_R}] (t) = \mathrm{HS}[\text{CB of } \eqref{eq:USp2-4U1}_{\text{wr} \, \langle (12)(34), (13)(24) \rangle}] (t) \\ &= \frac{1 - 2 t^2 + 41 t^4 + 19 t^6 + 234 t^8 + 86 t^{10} + 234 t^{12} + 19 t^{14} + 41 t^{16} -2 t^{18} + t^{20}}{(1 - t)^{10} (1 + t)^{10} (1 + t^2)^5} ~.
}

\section{Twisted Compactification of \texorpdfstring{$\SU(n)$}{SU(n)} SQCD with \texorpdfstring{$2n$}{2n} Flavors}
\label{sec:twistedcompactificationsSUN}
In this section, we consider the compactification on a circle with a $\BZ_2$ outer-automorphism twist of 4d $\CN=2$ $\SU(n)$ SQCD with $2n$ flavors of hypermultiplets in the fundamental representation.  The method we adopt is very similar to that discussed in \cite[(3.1)]{Bourget:2020asf} and \cite{Bourget:2020mez}, where 4d $\CN=2$ SCFTs arising from 5d $\CN = 1$ SCFTs compactified with a $\BZ_k$ twist were studied. In which case, it was pointed out that if one starts with the magnetic quiver $\mathsf{Q}'$ of the 5d theory containing $k$ identical simply-laced legs, then the magnetic quiver $\mathsf{Q}$ of the 4d $\CN = 2$ theory is obtained by folding the $k$ legs of $\mathsf{Q}'$. The quiver $\mathsf{Q}$ contains a non-simply laced edge with multiplicity $k$.  The Higgs branch of the parent theory can be computed from the Coulomb branch of $\mathsf{Q}$ using the prescription given in \cite{Cremonesi:2014xha}.

Our main proposal is that the Coulomb branch of the following non-simply-laced quiver describes the Higgs branch of the 3d theory arising from the $\BZ_2$ twisted compactification of 4d $\CN=2$ $\SU(n)$ SQCD with $2n$ flavors:
\bes{ \label{TCSUNw2N}
\begin{tikzpicture}[baseline,font=\footnotesize,scale=1]
\node (N) at (0.5,0) {$\U(n)$};
\node (1u) at (1,1.1) {$[(\U(1))]$};
\node (1d) at (1,-1.1) {$\U(1)$};
\node (Nm1) at (-1.4,0) {$\U(n-1)$};
\node (dots) at (-3,0) {$\cdots$};
\node (2) at (-4.2,0) {$\U(2)$};
\node (1) at (-5.6,0) {$\U(1)$};
\draw (1) to (2) to (dots) to (Nm1);
\draw (N) to (1u);
\draw (N) to (1d);
\draw[implies-,double equal sign distance] (Nm1) -- (N);
\end{tikzpicture}
}
where an overall $\U(1)$ should be modded out from one of the $\U(1)$ gauge nodes on the right of $\Leftarrow$, denoted by $[(\U(1))]$.  This quiver can be obtained by taking the 3d mirror theory of $\SU(n)$ SQCD with $2n$ flavors depicted in \eref{mirrSUnw2nx} and folding it with respect to the vertical axis of symmetry. The latter introduces the non-simply laced edge. The direction of the arrow can be fixed by considering the special case of $n=2$ as follows. The Higgs branch of the 4d theory, namely $\SU(2)$ SQCD with 4 flavors, is isomorphic to the closure of the minimal nilpotent orbit of $\so(8)$, \ie  $\bar{\mathrm{min}\,D_4}$.  As discussed in \cite{Dey:2016qqp} and \cite[Section 4 and Figure 19]{Bourget:2020bxh}, this is folded to the closure of the minimal nilpotent orbit of $\so(7)$, \ie  $\bar{\mathrm{min}\,B_3}$. In fact, the latter is described by the Coulomb branch of \eref{TCSUNw2N} with $n=2$, which is indeed the affine Dynkin diagram of $\so(7)$ \cite{Cremonesi:2014xha}.

We can also obtain the quiver \eref{TCSUNw2N} in the case of $n=3$ by exploiting the Argyres--Seiberg duality \eref{ArgyresSeiberg}, with the assumption that the $\BZ_2$ twist acts only on the strongly-coupled sector.  Recall that the Higgs branch of the rank-1 $E_6$ theory is isomorphic to $\bar{\mathrm{min}\,E_6}$. As discussed in \cite{Bourget:2020bxh} and \cite[(3.3)]{Bourget:2020asf}, the $\BZ_2$ folding of the latter leads to $\bar{\mathrm{min}\,F_4}$, which is described by the Coulomb branch of the affine Dynkin diagram of $F_4$ \cite{Cremonesi:2014xha}:
\bes{
\begin{tikzpicture}[baseline,font=\footnotesize]
\node (3) at (0.4,0) {$\U(3)$};
\node (2) at (1.8,0) {$\U(2)$};
\node (1) at (3.2,0) {$[(\U(1))]$};
\node (2l) at (-1,0) {$\U(2)$};
\node (1l) at (-2.4,0) {$\U(1)$};
\draw (1l) -- (2l);
\draw[implies-,double equal sign distance] (2l) -- (3);
\draw (3) -- (2) -- (1);
\end{tikzpicture}
}
Let us now proceed further in a similar way as discussed around \eref{mirrR0na}.  In particular, the first step is to couple {\it two} flavors of hypermultiplets to $\su(2) \subset \mathfrak{f}_4$ in $\bar{\mathrm{min}\,F_4}$ and gauge this $\su(2)$ symmetry algebra.  In the second step, we then give a mass to one of such flavors. Thus, it is expected that the $\BZ_2$ twisted compactification of $\SU(3)$ SQCD with 6 flavors is obtained after integrating out the massive hypermultiplet.  In terms of the magnetic quiver, the first step corresponds to considering the following theory:
\bes{
\begin{tikzpicture}[baseline,font=\footnotesize]
\node (3) at (0.4,0) {$\U(3)$};
\node (2) at (1.8,0) {$\U(2)$};
\node (1) at (3.2,0) {$[(\U(1))]$};
\node (2l) at (-1,0) {$\U(2)$};
\node (1l) at (-2.4,0) {$\U(1)$};
\node (1d) at (1.8,-1) {$\U(1)$};
\draw (1l) -- (2l);
\draw (2) -- (1d);
\draw[implies-,double equal sign distance] (2l) -- (3);
\draw (3) -- (2) -- (1);
\end{tikzpicture}
}
The second step corresponds to turning on the FI parameter to one of the $\U(1)$ gauge nodes on the right. Flowing to the IR, we then obtain the quiver \eref{TCSUNw2N} with $n=3$, as required (see \cite{Bourget:2020mez, vanBeest:2021xyt, Bourget:2023uhe}).

The above argument can be generalized to higher $n$ via \eref{genASduality} in a straightforward fashion. We propose that the magnetic quiver for the $\BZ_2$ twisted compactification of the $R_{0,n}$ theory is
\bes{ \label{TCR0N}
\begin{tikzpicture}[baseline,font=\footnotesize]
\node (N) at (0.5,0) {$\U(n)$};
\node (1u) at (1.9,0) {$\U(2)$};
\node (1d) at (3.3,0) {$[(\U(1))]$};
\node (Nm1) at (-1.4,0) {$\U(n-1)$};
\node (dots) at (-3,0) {$\cdots$};
\node (2) at (-4.2,0) {$\U(2)$};
\node (1) at (-5.6,0) {$\U(1)$};
\draw (1) to (2) to (dots) to (Nm1);
\draw (N) to (1u) to (1d);
\draw[implies-,double equal sign distance] (Nm1) -- (N);
\end{tikzpicture}
}
The rest of the procedure then follows as for the case of $n=3$.

The Coulomb branch symmetry of theory \eref{TCSUNw2N} is $\so(7)$ for $n=2$ and $\usp(2n) \oplus \u(1)_b$ for $n\geq 3$. The HWG of the Coulomb branch Hilbert series is given by 
\bes{ \label{HWGTCSUNSQCD}
\mathrm{HWG}\left[\text{CB of \eref{TCSUNw2N}}\right] =\PE \left[t^2+\left(b^{n}\nu_{n}+b^{-n}\nu_{n} \right) t^n +\sum_{k=1}^{n-1}\nu_{k}^2t^{2k} \right]~,
}
where $\nu_k^p$ denotes the highest weight fugacity of the representation $[0^{k-1},p,0^{n-k}]_{\usp(2n)}$ of $\usp(2n)$, and $b$ is the fugacity for the $\u(1)_b$ symmetry.  Here, we normalize the charge under $\u(1)_b$ in such a way that the operators that come from the baryons in 4d $\SU(n)$ SQCD have charge $n$, \ie  equal to that of the baryonic charge.  It is instructive to compare this HWG to that of the Higgs branch of $\SU(n)$ SQCD with $2n$ flavors \cite[(5.52)]{Bourget:2019rtl}:
\begin{equation} \label{HWGSQCD}
\PE \left[t^2+\left(b^{n}\mu_{n}+b^{-n}\mu_{n} \right) t^n +\sum_{k=1}^{n-1}\mu_{k} \mu_{2 n - k} t^{2k} \right] ~,
\end{equation}
where $b$ is the fugacity for the baryonic symmetry and $\mu_{k_1}^{p_1} \mu_{k_2}^{p_2}$ denotes the highest weight fugacity of the representation $[0^{k_1 - 1},p_1,0^{k_2 - k_1 -1},p_2,0^{2 n - k_2 -1}]_{\su(2n)}$ of $\su(2n)$. 
In the following, we write down explicitly the representations under which generators and relations of the Higgs branch transform for the cases of $n=2$ and $n=3$.  We see that, for general $n$, the Higgs branch generators of $\SU(n)$ SQCD that are {\it not} projected out by the twisted compactification transform in the following representations of $\usp(2n) \oplus \u(1)_b$.
\begin{itemize}
    \item \textbf{Mesons} at order $t^2$:  $[2,0,\ldots,0]_{\usp(2n)}(0)+ [0,0,\ldots,0]_{\usp(2n)}(0)$.
    \item \textbf{Baryons} at order $t^{n}$: $[0,\ldots,0,1]_{\usp(2n)}(\pm n)$, where $(\pm n)$ are the baryonic $\u(1)_b$ charges.
\end{itemize}

\subsection*{The Case of $n=2$}

Let us expand the HWG \eref{HWGSQCD} in power series in $t$ and decompose the $\su(2n)$ representations into those of $\usp(2n)$.  Taking the plethystic logarithm, we can extract the information of the generators and the relations of the moduli space.  We can perform a similar computation starting from \eref{HWGTCSUNSQCD} and compare the two results.

The generators of the Higgs branch of $\SU(2)$ SQCD with 4 flavors correspond to the coefficient at order $t^2$ and transform in the following representations of $\usp(4) \oplus \u(1)_b$:
\bes{
 \quad {[0,1]_{\usp(4)}}( \pm 2) +{\red {[0,1]_{\usp(4)}}(0)} +{[2,0]_{\usp(4)}}(0)+ {\red [0,0]_{\usp(4)}(\pm 2)}+[0,0]_{\usp(4)}(0)~,
}
whereas the relations (at order $t^4$) transform under the following representations:
\bes{
\renewcommand*{\arraystretch}{1.1}
\begin{array}{clclclcl}
 & {\red [0,0]_{\usp(4)}(\pm 2, 0)} & + & 2\times [0,0]_{\usp(4)}(0) & + &[0,0]_{\usp(4)}(\pm 4) & + &{\red [2,0]_{\usp(4)}(\pm 2, 0)} \\
+ & [2,0]_{\usp(4)}(\pm 2,0) & +
& {\red 2 \times [0,1]_{\usp(4)}(0)} & + & [0,1]_{\usp(4)}(\pm 2,0) & + &[0,2]_{\usp(4)}(0)~,
\end{array}
}
where we denoted in {\red red} the representations that are projected out by the twist, \ie  they are {\it absent} in theory \eref{TCSUNw2N}. Note that $\mathbf{r}(\pm q ,0)$ is the shorthand notation for $\mathbf{r}(q)+\mathbf{r}(-q)+ \mathbf{r}(0)$.

\subsection*{The Case of $n=3$}

The Higgs branch generators of $\SU(3)$ with 6 flavors transform in the following representations of $\usp(6) \oplus \u(1)_b$.
\begin{itemize}
    \item \textbf{Mesons} at order $t^2$: ${\red [0,1,0]_{\usp(6)}(0)}+[2,0,0]_{\usp(6)}(0)+[0,0,0]_{\usp(6)}(0)$~.
    \item \textbf{Baryons} at order $t^3$: ${\red [1,0,0]_{\usp(6)}(\pm 3)} +[0,0,1]_{\usp(6)}(\pm 3)$~.
\end{itemize}
Again, we denoted in {\red red} the representations that are removed by the twisted compactification. The relations that are satisfied by the generators transform in the following representations.
\begin{itemize}
    \item At order $t^4$, the relation is ${\red [2,0,0]_{\usp(6)}(0)}+[0,1,0]_{\usp(6)}(0)+[0,0,0]_{\usp(6)}(0)$. 
    \item At order $t^5$, the relation is 
    \[
    \begin{split}
        &{\red 3 \times [1,0,0]_{\usp(6)}(\pm 3)}+[0,0,1]_{\usp(6)}(\pm 3)+{\red [1,1,0]_{\usp(6)}(\pm 3)}+[1,1,0]_{\usp(6)}( \pm 3)\fstop
    \end{split}\] 
    \item At order $t^6$, the relation is 
    \[
\begin{array}{clcllcllcl}
 &{\red [0,1,0]_{\usp(6)}(\pm 6)} & + & \red{2 \times} & {\red [0,1,0]_{\usp(6)}(0)} & + & \red {2 \times} & {\red [1,0,1]_{\usp(6)}(0)} & + & {\red [0,0,0]_{\usp(6)}(0)} \\
       + & [2,0,0]_{\usp(6)}(\pm 6, 0) & + & & [0,0,2]_{\usp(6)}(0) & + & &[0,2,0]_{\usp(6)}(0) & + &[0,0,0]_{\usp(6)}(0)\fstop
\end{array}
\] 
\end{itemize}
We remark that, in addition to the aforementioned relations, there is an extra term contributing $+[0,1,0]_{\usp(6)}(0)$ at order $t^6$ to the plethystic logarithm of the Coulomb branch Hilbert series of theory \eref{TCSUNw2N} for $n = 3$.

\subsubsection*{Difference Between Discrete Gauging and Twisted Compactification}

Let us comment on the difference between discrete gauging discussed in the preceding sections and twisted compactification discussed in this section. In the latter, we start from a 4d SCFT and end up with a 3d theory; for example, 4d $\CN=2$ $\SU(2)$ SQCD with 4 flavors becomes a 3d $\CN=4$ theory whose Coulomb branch is isomorphic to $\bar{\mathrm{min}\,B_3}$.  On the contrary, the discrete gauging of a 4d SCFT gives rise to another 4d SCFT, which of course can be dimesionally reduced on a circle to obtain a 3d $\CN=4$ theory, which {\it may differ} from the aforementioned one. For example, as discussed around \eref{B3C1O1}, a $\BZ_2$ discrete gauging of 4d $\CN=2$ $\SU(2)$ SQCD with 4 flavors leads to a rank-1 4d $\CN=2$ theory whose Higgs branch is isomorphic to $\bar{n.\mathrm{min}\,B_3}$, previously denoted by $\CT^{r=1}_{\bar{n.\mathrm{min}\,B_3}}$. Upon reduction on a circle, this leads to a 3d $\CN=4$ theory whose Higgs branch is still isomorphic to $\bar{n.\mathrm{min}\,B_3}$ and Coulomb branch is isomorphic to $\BC^2/\hat{D}_6$.

\section{Comments on Twisted Compactifications of \texorpdfstring{$\mathcal{N}=4$}{N=4} SYM Theories}
\label{sec:twistedcompactification-SYM}

Let us consider the 4d $\mathcal{N}=4$ super Yang--Mills (SYM) theory with gauge group $\U(N)$ or $\SU(N)$. In the following, we view the SYM theory as a 4d $\CN=2$ gauge theory, where the $\CN=4$ vector multiplet decomposes into one $\CN=2$ hypermultiplet and an $\CN=2$ vector multiplet. Let us denote by $\Phi_1$ and $\Phi_2$ the chiral multiplets that reside in the hypermultiplet.  The vacuum equation implies that $\Phi_1$ and $\Phi_2$ commute with each other. The Higgs branch of the theory with the $\U(N)$ gauge group is therefore isomorphic to the $N$-fold symmetric product of $\BC^2$, $\Sym^N(\BC^2)$. For the $\SU(N)$ gauge group, the free hypermultiplet that arises from the traces of $\Phi_1$ and $\Phi_2$ is absent.

The $\BZ_k$ twisted compactification of $\SU(N)$ SYM by a non-invertible defect was studied in \cite[Section 6]{Kaidi:2022uux}.  In particular, the authors of that reference studied the $\BZ_k$ action which arises from an appropriate combination of the R-symmetry action $\rho$ and the $\BZ_k \subset \SL(2,\BZ)$ transformation $\sigma$ in such a way that the combined action of $(\rho, \sigma)$ preserves $\CN=6$ supersymmetry in the 3d theory.  Explicitly, if we denote the four complex chiral supercharges by $Q^i_\alpha$ with $i=1, \ldots, 4$ the index of the fundamental representation of the $\SU(4)$ R-symmetry and $\alpha=1,2$, then $\sigma$ acts on $Q^i_\alpha$ as 
\bes{
\sigma: \quad Q^i_\alpha \rightarrow e^{-i \pi/k} Q^i_\alpha = \omega_k^{-1/2} Q^i_\alpha~,
}
with $\omega_k \equiv e^{2\pi i/k}$. On the other hand, up to a conjugation, $\rho$ can be chosen to be in the maximal torus of $\Spin(6) \cong \SU(4)$ R-symmetry and can be represented by
\bes{ \label{actionrho}
\rho =(\omega_k^{r_1}, \omega_k^{r_2}, \omega_k^{r_3})~, \qquad \omega_k \equiv e^{2\pi i/k}~, ~r_{1,2,3} \in \BZ~,
}
which can be considered as a simultaneous rotation in the three orthogonal planes in $\BC^3 \cong \BR^6$; see \cite[(3.9)]{Argyres:2016yzz}.  In this presentation, the action of $\rho$ on the four chiral supercharges is
\bes{ \label{actionrhoQ}
\Big(\omega_k^{(r_1+r_2+r_3)/2}, \omega_k^{(r_1-r_2-r_3)/2}, \omega_k^{(-r_1+r_2-r_3)/2}, \omega_k^{(-r_1-r_2+r_3)/2} \Big)~.
}
The choice of \cite[Section 6]{Kaidi:2022uux} corresponds to $(r_1, r_2, r_3) = (-1,-1,-1)$ and so \eref{actionrhoQ} becomes $(\omega_k^{-3/2},\omega_k^{1/2},\omega_k^{1/2},\omega_k^{1/2})$; upon combining it with the action of $\sigma$ stated above, this indeed preserves three complex supercharges, as required.

At the level of the moduli space, it was pointed out that $\BZ_k$ acts on the collection of three complex scalar fields ${\vec \phi}_j$ (with $j=1, \ldots, N$) in the vector multiplet of 4d $\CN=4$ SYM as
\bes{\label{actionZk}
{\vec \phi}_j \,\, \rightarrow \,\, \omega_k {\vec \phi}_j~.
}
This can be appropriately combined with the Weyl group action of $\SU(N)$ and it gives rise to a certain complex reflection group (see Appendix \ref{app:complexreflectiongroup} for more details). By examining the group action on the moduli space, such an $\CN=6$ theory is identified with the ABJ(M) theory \cite{Aharony:2008ug, Aharony:2008gk} $\U(N+x)_k \times \U(N)_{-k}$ such that\footnote{Strictly speaking, for the non-invertible defects, $k$ is restricted to be $3$, $4$ or $6$. However, in the following, we consider general values of $k$.}
\bes{ \label{condtwist}
N = n k+x,\quad \text{with $x<k$}~.
}
The total moduli space of this theory is $\BC^{4n}/G(k,1,n)$, where $G(k,1,n)$ is the complex reflection group in question.
If we view this theory as a 3d $\CN=4$ theory, then the Higgs and Coulomb branches are both isomorphic to 
\bes{ \label{eq:Symn2Zk}
\Sym^n(\BC^2/\BZ_k) \cong \BC^{2 n}/G(k,1,n)~.
}
This can be checked using the Higgs and Coulomb branch limits \cite{Razamat:2014pta} of the index (see the discussion around \eqref{CBHBlimits}) of the corresponding ABJ(M) theory (see \eg \cite{Beratto:2021xmn, Mekareeya:2022spm, Comi:2023lfm}).

In the following, we instead consider the choice $(r_1, r_2, r_3) = (1, 1, -1)$ in \eref{actionrho}, in which case \eref{actionrhoQ} becomes $(\omega_k^{1/2},\omega_k^{1/2},\omega_k^{1/2},\omega_k^{-3/2})$; upon combining it with the action of $\sigma$ stated above, this still preserves three complex supercharges, as it should be.  This choice, on the other hand, is more convenient for the study of the Higgs branch of the 4d $\CN=4$ SYM theory. In particular, the scalar component of the two chiral multiplets $\Phi_1$ and $\Phi_2$ in the hypermultiplet parametrizes a complex plane $\BC^2$ in the orthogonal space $\BC^3 \cong \BR^6$.  The $\BZ_k$ action on $\Phi_{1,2}$ can be chosen as
\bes{ \label{actionphi12}
\Phi_1 \,\, \rightarrow \,\, \omega_k \Phi_1~, \qquad \Phi_2 \,\, \rightarrow \,\, \omega_k^{-1} \Phi_2~.
}

The main point of this section is that, if we consider the action \eref{actionphi12} on the commuting adjoint chiral fields $\Phi_1$ and $\Phi_2$ in the hypermultiplet of 4d $\CN=4$ SYM and examine the chiral ring that is invariant under this $\BZ_k$ action, the resulting Higgs branch moduli space may contain {\it not only} $\Sym^n(\BC^2/\BZ_k)$ as discussed in \eqref{eq:Symn2Zk}, but also a {\it radical ideal}\footnote{See \eg \cite{eisenbud1995} and \cite[Appendix C]{Bourget:2019rtl}.} due to the presence of nilpotent operators. Note that, even though $\Sym^n(\BC^2/\BZ_k)$ is still a component with the largest dimension of the Higgs branch, the radical should not be neglected.  The significance of the latter in supersymmetric gauge theories is discussed, for example, in \cite{Cremonesi:2015lsa, Bourget:2019rtl}.  Subsequently, we give explicit examples of such radical ideals for $N=3$ and $N=4$.\footnote{For the case of $N=2$, the radical ideal is not present, and so we omit the discussion of this case.}  

In the following discussion, we will use extensively the Hilbert series of $\Sym^n(\BC^2/\BZ_k)$, which can be computed using the method given in Appendix \ref{app:complexreflectiongroup}, or simply using the following plethystic exponential \cite{Benvenuti:2006qr, Feng:2007ur}:
\bes{ \label{eq:genSymNC2}
\PE\left[\frac{1-t^{2k}}{(1-t^k z^k)(1-t^2)(1-t^k z^{-k})}\nu\right]~.
}
Upon performing the series expansion in $\nu$, the coefficient of $\nu^n$ gives the Hilbert series of $\Sym^n(\BC^2/\BZ_k)$, as required.

\subsection{\texorpdfstring{$\SU(3)$}{SU(3)} SYM}

Let us consider the $\U(3)$ SYM theory. The Higgs branch moduli space is generated by three sets of generators:
\begin{equation} \label{genN3}
    P_a =\tr(\Phi_a) \coma Q_{ab} =\tr(\tilde{\Phi}_a \tilde{\Phi}_b) \coma S_{abc} = \tr(\tilde{\Phi}_a \tilde{\Phi}_b \tilde{\Phi}_c)\coma \quad a, b, c=1,2~,
\end{equation}
with
\bes{
\tilde{\Phi}_a = \Phi_a - \frac{1}{3}\tr(\Phi_a) \ID_{3\times 3}~.
}
They satisfy the following set of relations:\footnote{These generators and relations have, in fact, been worked out by mathematicians in \cite{NTa, 2023arXiv230706098E}, where, in their notation, $\Phi_1=X$, $\Phi_2 =Y$, $\tilde{\Phi}_1=A$, $\tilde{\Phi}_2 =B$.  Note that the problem of identifying the generators of $\Sym^N(\BC^2)$ and computing the explicit relations among them is closely related to the study of the $N$-th Calogero-Moser space \cite{PROCESI1976306, R1974}. For example, the case of $N=2$ was studied in \cite{EG2002, KMN2021}. \label{foot:notation}}
\begin{equation} \label{relN3}
    \begin{split}
        R^{(1)}_a := &\,\epsilon^{i_1j_1}\epsilon^{i_2j_2}Q_{i_1i_2}S_{j_1j_2a} = 0~,\\
        R^{(2)}_{ab} := &\,\epsilon^{i_1j_1}\epsilon^{i_2j_2}\left(Q_{i_1i_2}Q_{j_1a}Q_{j_2b}+3S_{i_1i_2a}S_{j_1j_2b}\right) = 0~,
    \end{split}
\end{equation}
where the indices $i$'s and $j$'s run from $1$ to $2$.\footnote{These relations can be checked easily by using the fact that $\Phi_1$ and $\Phi_2$ commute, and hence they are simultaneously diagonalizable.} Note that $P_1$ and $P_2$ drop out of the relations $R^{(1)}=0$ and $R^{(2)}=0$. Indeed, for the $\SU(3)$ SYM theory, we can simply set $P_{1}=P_2=0$ in the above discussion.

Let $\CR$ be the polynomial ring in the variables \eref{genN3}, and $\CI$ be the ideal that contains the relations \eref{relN3}. We can compute the Hilbert series of the quotient ring $\CR/\CI$ using \texttt{Macaulay2} \cite{M2}, and the result is
\begin{equation}
    \begin{split}
        \mathrm{HS}(z; t) &= \frac{1+t^2+\left(z+\frac{1}{z}\right) t^3 +t^4+t^6}{\left(1-t^3 z^3\right) \left(1-t^2 z^2\right) (1-t z) \left(1-\frac{t}{z}\right) \left(1-\frac{t^2}{z^2}\right) \left(1-\frac{t^3}{z^3}\right)} \\
        &= \PE \left[ [1]_{\su(2)} t + [2]_{\su(2)} t^2 + [3]_{\su(2)} t^4 - [1]_{\su(2)} t^5 - [2]_{\su(2)} t^6+\ldots \right]~.
    \end{split}
    \label{eq:Sym3C2}
\end{equation}
The first three positive terms in the $\PE$ correspond to the generators $P$, $Q$ and $S$, while the last two negative terms in the $\PE$ correspond to the relations $R^{(1)}=0$ and $R^{(2)}=0$. This is indeed the Hilbert series of $\Sym^3(\BC^2)$, which can be obtained from the $\nu^3$ coefficient of \eqref{eq:genSymNC2} with $k=1$, as expected.

Let us now study the Higgs branch of the 3d theory that arises from the $\BZ_k$ twisted compactification of the 4d $\CN=4$ $\SU(3)$ SYM theory. For $N=3$ and $k >1$, \eref{condtwist} admits two solutions, namely $(k, n, x) = \left\{(3,1,0), \, (2,1,1)\right\}$.

\subsubsection{\texorpdfstring{$\ZZ_3$}{Z3} Twisted Compactification}

Let us first consider a $\ZZ_3$ twisted compactification, corresponding to $(k, n, x) = (3,1,0)$.  From \eref{eq:Symn2Zk}, the expected Higgs branch moduli space is $\mathbb{C}^2/\ZZ_3$.

We consider the action \eref{actionphi12} and set to zero all the generators that are not invariant under a $\ZZ_3$ transformation, namely $P_1, P_2, Q_{11}, Q_{22}, S_{112}, S_{221}$, while leave the other generators untouched. Computationally, we take $\CR$ to be the ring of polynomials in the variables \eref{genN3}, as before, but with the new ideal $\CI$ to be as follows 
\begin{equation}
    \mathcal{I} = \left( P_1,P_2,Q_{11},Q_{22},S_{112},S_{221},Q_{12}^3-3S_{111}S_{222}\right)\coma
\end{equation}
where $R^{(1)}=0$ is just a trivial identity and the last equation in $\mathcal{I}$ comes from $R^{(2)}=0$. Upon computing the Hilbert series of $\CR/\CI$, the result is
\begin{equation}
    \mathrm{HS}(z;t) = \frac{1+t^2+t^4}{\left(1-t^3 z^3\right) \left(1-\frac{t^3}{z^3}\right)}=\frac{1-t^6}{\left(1-t^3 z^3\right) \left(1-t^2\right) \left(1-\frac{t^3}{z^3}\right)}\coma
\end{equation}
which is the moduli space of $\mathbb{C}^2/\ZZ_3$, as expected. In this case, we do not have any radical ideal.

\subsubsection{\texorpdfstring{$\ZZ_2$}{Z2} Twisted Compactification}

Another possible twist is a $\ZZ_2$ twist, corresponding to $(k, n, x) = (2,1,1)$ From \eref{eq:Symn2Zk}, the expected Higgs branch moduli space is $\mathbb{C}^2/\ZZ_2$. 

One could proceed as before, \ie by setting to zero the generators that are not invariant under the $\ZZ_2$ action \eref{actionphi12}. These are precisely the elements of the set $\CZ=\{P_1, P_2, S_{111}, S_{112}, S_{221}, S_{222} \}$. Let $\CR$ be the polynomial ring in variables \eref{genN3}, as before, and let the ideal $\CI$ be as follows: $\CI = \CZ \cup \{ R^{(1)}|_{\CZ=0} \} \cup \{ R^{(2)}|_{\CZ=0} \}$, where $R^{(i)}|_{\CZ=0}$ means that we set all the elements in $\CZ$ to zero in $R^{(i)}$. Computing the Hilbert series of $\CR/\CI$ using \texttt{Macaulay2}, we obtain
\bes{
\scalebox{0.99}{$
\begin{split}
     \mathrm{HS}(z;t) &= \frac{1+t^2+t^4-\left(z^2+\frac{1}{z^2}\right) t^6 + t^8}{\left(1-t^2 z^2\right) \left(1-\frac{t^2}{z^2}\right)}= \frac{1-\left(z^2+\frac{1}{z^2}+1\right) t^6 + \left(z^2+\frac{1}{z^2}+1\right) t^8 -t^{10}}{\left(1-t^2 z^2\right) \left(1-t^2\right) \left(1-\frac{t^2}{z^2}\right)}~,
\end{split}$}
}
which is clearly not the Hilbert series of $\mathbb{C}^2/\ZZ_2$, and its numerator is not even palindromic. This suggests that the quotient ring in question is not irreducible (see \eg \cite{Forcella:2008bb}). Indeed, performing the primary decomposition (\eg using the command \texttt{primaryDecomposition} in \texttt{Macaulay2}), we see that there are two irreducible components.
\begin{itemize}
    \item The first is
\begin{equation}
    \mathcal{I}_1 = \left(P_1,P_2,S_{111},S_{112},S_{221},S_{222},Q_{12}^2-Q_{11}Q_{22}\right)\coma
\end{equation}
with Hilbert series:
\begin{equation}
    \mathrm{HS}_1(z;t) = \frac{1+t^2}{\left(1-t^2 z^2\right) \left(1-\frac{t^2}{z^2}\right)} = \frac{1-t^4}{\left(1-t^2 z^2\right) (1-t^2)\left(1-\frac{t^2}{z^2}\right)}\coma
\end{equation}
corresponding to the $\mathbb{C}^2/\ZZ_2$ moduli space generated by $Q_{12}$, $Q_{11}$ and $Q_{22}$. 
 This branch is as expected from \eref{eq:Symn2Zk}. 
\item The second one is
\begin{equation}
    \mathcal{I}_2 = \left(P_1,P_2,Q_{11},Q_{22},S_{111},S_{112},S_{221},S_{222}, Q_{12}^3\right)\coma
\end{equation}
with Hilbert series
\begin{equation}
    \mathrm{HS}_2(t) = 1+t^2+t^4\coma
\end{equation}
corresponding to the identity operator, $Q_{12}$ and $Q_{12}^2$. This is a radical ideal and $Q_{12}$ is the nilpotent operator satisfying $Q_{12}^3=0$.
\end{itemize}
The two irreducible components intersect at the origin, which is a fat point:
\begin{equation}
    \mathrm{HS}(z;t) = \mathrm{HS}_1(z;t) + \mathrm{HS}_2(t) -(1+t^2)\fstop
\end{equation}
The presence of the $t^2$ in the last term signals that the intersection is not only at the identity, but also at order $2$.

\subsection{\texorpdfstring{$\SU(4)$}{SU(4)} SYM}

Let us now consider $\U(4)$ SYM, whose Higgs branch is $\Sym^4(\BC^2)$. The coordinate ring is discussed in \cite[Theorem 4.1]{2023arXiv230706098E}.  There are 14 generators that can be written in an $\su(2)$ covariant way as
\begin{equation}\label{eq:SU4gen}
    P_a =\tr(\Phi_a) \coma Q_{ab} =\tr(\tilde{\Phi}_a \tilde{\Phi}_b) \coma S_{abc} = \tr(\tilde{\Phi}_a \tilde{\Phi}_b \tilde{\Phi}_c)\coma T_{abcd}=\tr(\tilde{\Phi}_a \tilde{\Phi}_b \tilde{\Phi}_c \tilde{\Phi}_d)~,
\end{equation}
with $a, b, c, d=1,2$, and 
\bes{
\tilde{\Phi}_a = \Phi_a - \frac{1}{4} \tr(\Phi_a) \ID_{4 \times 4}~.
}
They are subject to the 15 relations listed in \cite[Appendix A.3]{2023arXiv230706098E}.\footnote{The notation is as described in Footnote \ref{foot:notation}.}  Let $\CR$ be the polynomial ring in the variables \eref{eq:SU4gen}, and $\CI$ be the ideal that contains the aforementioned relations. We can compute the Hilbert series of the quotient ring $\CR/\CI$ using \texttt{Macaulay2}.  The result is
\bes{
\scalebox{0.99}{$
\begin{split}
        \mathrm{HS}(z;t) &= \frac{1}{\left(1-t^4 z^4\right) \left(1-t^3 z^3\right) \left(1-t^2 z^2\right) \left(1-t z\right) \left(1-\frac{t}{z}\right) \left(1-\frac{t^2}{z^2}\right) \left(1-\frac{t^3}{z^3}\right) \left(1-\frac{t^4}{z^4}\right)} \\
        & \quad \times\left [ 1+t^2+\left(z+\frac{1}{z}\right) t^3 + \left(z^2+\frac{1}{z^2}+2\right) t^4 + \left(z+\frac{1}{z}\right) t^5 + \left(z^2+\frac{1}{z^2}+2\right) t^6\right. \\
        &\left. \qquad \quad \, + \left(z+\frac{1}{z}\right) t^7 + \left(z^2+\frac{1}{z^2}+2\right) t^8+ \left(z+\frac{1}{z}\right) t^9+t^{10}+t^{12} \right] \\
&= \PE \left[ \sum_{s=1}^4 [s]_{\su(2)}t^s - [2]_{\su(2)}t^6-\left([3]_{\su(2)}+[1]_{\su(2)}\right)t^7 - \left([4]_{\su(2)}+1\right)t^8 + \ldots\right]~.
\end{split}   $}
    \label{eq:Sym4C2}
}
The positive terms correspond to the generators \eref{eq:SU4gen}, and the negative terms tell us how the relations transform under $\su(2)$. This Hilbert series is that of $\text{Sym}^4(\BC^2)$, which can be obtained from the $\nu^4$ coefficient of \eqref{eq:genSymNC2} with $k=1$ as expected.  For the $SU(4)$ SYM theory, we simply set $P_1$ and $P_2$ to zero in the above discussion.

Let us now study the Higgs branch of the 3d theory that arises from the $\BZ_k$ twisted compactification of the 4d $\CN=4$ $\SU(4)$ SYM theory. For $N=4$ and $k >1$, \eref{condtwist} admits three solutions, namely $(k, n, x) = \left\{(4,1,0), \, (3,1,1), \, (2,2,0)\right\}$.

\subsubsection{\texorpdfstring{$\ZZ_4$}{Z4} Twisted Compactification}

Let us first consider a $\ZZ_4$ twisted compactification, corresponding to $(k, n, x) = (4,1,0)$. From \eref{eq:Symn2Zk}, the expected Higgs branch moduli space is $\mathbb{C}^2/\ZZ_4$. 

We consider the action \eref{actionphi12} with $k=4$ and set all generators that are not invariant under such a transformation, namely those in the set $\CZ = \{ P_1, P_2, Q_{11}, Q_{22}, S_{111}, S_{112}, S_{221}, S_{222}, T_{1112}, T_{1222} \}$, to zero while we leave the other generators untouched. Adding the elements in $\CZ$ to the ideal $\CI$ containing the 15 relations in \cite[Appendix A.3]{2023arXiv230706098E}, we obtain a new ideal $\CI' = \CZ \cup \CI$. Computing the Hilbert series of $\CR/\CI'$, we obtain
\begin{equation}
\begin{split}
     \mathrm{HS}(z;t) &= \frac{1+t^2+2 t^4+t^6- \left(z^4+\frac{1}{z^4}\right) t^8 + t^{12}}{\left(1-t^4 z^4\right) \left(1-\frac{t^4}{z^4}\right)} \\
     & = \frac{1+t^4-t^6- \left(z^4+\frac{1}{z^4}+1\right)t^8+ \left(z^4+\frac{1}{z^4}\right)t^{10}+t^{12}-t^{14}}{\left(1-t^4 z^4\right) (1-t^2)\left(1-\frac{t^4}{z^4}\right)}\coma
\end{split}
\end{equation}
which is not the Hilbert series of $\mathbb{C}^2/\ZZ_4$. The numerator is also not palindromic. Indeed, upon performing the primary decompositions, we see that there are two irreducible components.
\begin{itemize}
    \item The first is
    \begin{equation}
        \begin{split}
        \mathcal{I}_1 = &\,\left(P_1,P_2,Q_{11},Q_{22},S_{111},S_{112},S_{221},S_{222},T_{1112},T_{1222},\right.\\ &\, \left.Q_{12}^2-4T_{1212},T_{1212}^2-T_{1111}T_{2222}\right)~,    
        \end{split}
        \end{equation}
    with Hilbert series:
    \begin{equation}
        \mathrm{HS}_1(z;t) = \frac{1+t^2+t^4+t^6}{\left(1-t^4 z^4\right) \left(1-\frac{t^4}{z^4}\right)}  = \frac{1-t^8}{\left(1-t^4 z^4\right) (1-t^2)\left(1-\frac{t^4}{z^4}\right)}~,
    \end{equation}
    corresponding to $\mathbb{C}^2/\ZZ_4$ generated by $Q_{12}$, $T_{1111}$ and $T_{2222}$.
    \item The second is
    \begin{equation}
        \begin{split}
        \mathcal{I}_2 = &\,\left(P_1,P_2,Q_{11},Q_{22},S_{111},S_{112},S_{221},S_{222},T_{1111},T_{1112},T_{2221},T_{2222},\right.\\ &\, \left.Q_{12}^3-4Q_{12} T_{1212},T_{1212}^2,Q_{12}^2T_{1212}\right)~,    
        \end{split}
        \end{equation}
    with Hilbert series:
    \begin{equation}
        \mathrm{HS}_2(t) = 1+t^2+2t^4+t^6\coma
    \end{equation}
    corresponding to a radical ideal of order $6$. 
\end{itemize}
The two irreducible components intersect at the origin, which is a fat point of order $t^6$:
\begin{equation}
    \mathrm{HS}(z;t) = \mathrm{HS}_1(z;t)+\mathrm{HS}_2(t)-(1+t^2+t^4+t^6)\fstop
\end{equation}

\subsubsection{\texorpdfstring{$\ZZ_3$}{Z3} Twisted Compactification}

We now consider a $\ZZ_3$ twisted compactification, which corresponds to $(k, n, x) =(3,1,1)$. From \eref{eq:Symn2Zk}, the expected Higgs branch moduli space is $\mathbb{C}^2/\ZZ_3$. We set to zero the generators that are not invariant under the $\ZZ_3$ action \eref{actionphi12}, that is, $P_1, P_2, Q_{11}, Q_{22}, S_{112}, S_{221},T_{1111},T_{2222}, T_{1112}, T_{2221}$. Adding these to the original ideal $\CI$ containing the 15 relations in \cite[Appendix A.3]{2023arXiv230706098E} and recomputing the quotient ring, we obtain
\begin{equation}
    \mathrm{HS}(z;t) = \frac{1+t^2+2 t^4+t^6- \left(z^3+\frac{1}{z^3}\right) t^7 - \left(z^3+\frac{1}{z^3}\right) t^9+t^{10}+t^{12}}{\left(1-t^3 z^3\right) \left(1-\frac{t^3}{z^3}\right)}\coma
\end{equation}
which is not the Hilbert series of $\mathbb{C}^2/\ZZ_3$. In this case, the primary decomposition is even more interesting, because the fat point is also charged under the $\su(2)$  symmetry. We obtain two decompositions as follows.
\begin{itemize}
    \item The first is
     \begin{equation}
        \begin{split}
        \mathcal{I}_1 = &\,\left(P_1,P_2,Q_{11},Q_{22},S_{112},S_{221},T_{1111},T_{1112},T_{2221},T_{2222},\right.\\ &\, \left.Q_{12}^2-3T_{1212},S_{111}S_{222}-Q_{12}T_{1212}\right)~,    
        \end{split}
        \end{equation}
    with Hilbert series:
    \begin{equation}
        \mathrm{HS}_1(z;t) = \frac{1+t^2+t^4}{\left(1-t^3 z^3\right) \left(1-\frac{t^3}{z^3}\right)}=\frac{1-t^6}{\left(1-t^3 z^3\right) \left(1-t^2\right) \left(1-\frac{t^3}{z^3}\right)}\coma
    \end{equation}
    corresponding to the $\mathbb{C}^2/\ZZ_3$ generated by $Q_{12}$, $S_{111}$ and $S_{222}$.
    \item The second is
    \begin{equation}
        \begin{split}
        \mathcal{I}_2 = &\left(P_1,P_2,Q_{11},Q_{22},S_{112},S_{221},T_{1111},T_{1112},T_{1212},T_{2221},T_{2222},\right. \\ & \, \left.S_{111}^2,S_{222}^ 2,Q_{12}^3+S_{111}S_{222},Q_{12}^2S_{222},Q_{12}^2S_{111},Q_{12}S_{111}S_{222}\right)~,
        \end{split}
    \end{equation}
    with Hilbert series:
    \begin{equation}
        \mathrm{HS}_2(z;t) = 1+t^2+ \left(z^3+\frac{1}{z^3}\right) t^3 +t^4+ \left(z^3+\frac{1}{z^3}\right) t^5 +t^6\coma
    \end{equation}
    corresponding to a radical ideal of order $6$. For example, the terms at order $t^3$ are the contributions of the nilpotent operators $S_{111}$ and $S_{222}$.
\end{itemize}
The two irreducible components intersect at the origin, which is a fat point of order $t^5$:
\begin{equation}
    \mathrm{HS}(z;t) = \mathrm{HS}_1(z;t)+\mathrm{HS}_2(z;t)-\left[1+t^2+ \left(z^3+\frac{1}{z^3}\right) t^3 + \left(z^3+\frac{1}{z^3}\right) t^5\right]\fstop
\end{equation}

\subsubsection{\texorpdfstring{$\ZZ_2$}{Z2} Twisted Compactification}

Finally, we can consider a $\ZZ_2$ twisted compactification, corresponding to $(k,n,x)=(2,2,0)$. From \eref{eq:Symn2Zk}, the expected Higgs branch is $\text{Sym}^2(\mathbb{C}^2/\ZZ_2)$. In this case, the generators that are not invariant under \eref{actionphi12} are $P_1,P_2,S_{111},S_{112},S_{221},S_{222}$. As before, we set them to zero by adding them to the original ideal $\CI$ that contains the 15 relations in \cite[Appendix A.3]{2023arXiv230706098E}. The resulting Hilbert series of the new quotient ring is
\begin{equation}
    \mathrm{HS}(z;t) = \frac{1+t^2+ \left(z^2+\frac{1}{z^2}+2\right) t^4+t^6+t^8}{\left(1-t^4 z^4\right) \left(1-t^2 z^2\right) \left(1-\frac{t^2}{z^2}\right) \left(1-\frac{t^4}{z^4}\right)}\fstop
\end{equation}
This is precisely the Hilbert series of $\text{Sym}^2(\mathbb{C}^2/\ZZ_2)$ as expected.  There is no radical ideal in this case.

\subsection{Open Questions}
As we have seen from the above discussions, the presence of nilpotent operators and radical ideals is {\it generic} when we consider the quotient rings that are invariant under the $\BZ_k$ action \eref{actionphi12}. Note also that the Hilbert series of the radical ideals are not visible upon computing the Coulomb and Higgs branch limits of the index of the corresponding ABJ(M) theory. There are a couple of possibilities that are worth studying in the future. The first one is that the moduli space of the twisted compactification of 4d $\CN=4$ SYM is precisely captured by the ABJ(M) theory and nothing else, and that the moduli space of the ABJ(M) theory in question contains the radical ideal, but this is ``blind'' to the limit of the index. The second possibility, which we consider to be more likely, is that the ABJ(M) theory captures only the component of the moduli space with the largest dimensions, namely $\Sym^n(\BC^2/\BZ_k)$, but there is also the radical ideal, which is not captured by the moduli space of such a theory. If the second possibility is true, it would be nice to find a physical interpretation of the radical ideal, \eg  in terms of motions of branes in the string theoretic configuration.

\acknowledgments 

The authors thank Federico Carta for the useful discussions and previous collaborations. We also thank Guillermo Arias-Tamargo, Craig Lawrie, and Thekla Lepper for helpful discussions on the computation of wreathed quivers, and Julius Grimminger for useful discussions on non-simply laced quivers and twisted compactifications. N. M. is grateful to Alessandro Tomasiello for various insightful discussions on twisted compactifications. The work of S. G. is supported by the INFN grant ``Per attività di formazione per sostenere progetti di ricerca'' (GRANT 73/STRONGQFT). W. H. thanks the Galileo Galilei Institute for Theoretical Physics for the hospitality and the INFN for partial support during the completion of this work. A. M. is supported in part by Deutsche Forschungsgemeinschaft under Germany's Excellence Strategy EXC 2121 Quantum Universe 390833306 and Deutsche Forschungsgemeinschaft through a German-Israeli Project Cooperation (DIP) grant ``Holography and the Swampland''.

\appendix

\section{Wreathing Quiver \texorpdfstring{\eqref{eq:USp2n-4U1}}{} with \texorpdfstring{$n=2$}{n=2}} 
\label{sec:Casen2USp4}

In this appendix, we consider quiver \eref{eq:USp2n-4U1} with $n=2$, in which case it becomes
\begin{align}
    \begin{tikzpicture}[scale=0.86,font=\footnotesize,baseline=0cm]
\node (A1) at (0,0) {$\USp(4)$};
\node (A2) at (1.2,-1.2) {$\U(1)$};
\node (A3) at (-1.2,1.2) {$\U(1)$};
\node (A4) at (1.2,1.2) {$\U(1)$};
\node (A5) at (-1.2,-1.2) {$\U(1)$}; 
\draw (A2)--(A1) (A3)--(A1) (A4)--(A1) (A5)--(A1); 
\path (A1) edge [out=30,in=-30,looseness=6] node[midway,right]{$\mathrm{AS}'$}  (A1);
\end{tikzpicture}
\quad\text{\footnotesize$/\BZ_2$}
\end{align}
where, as in Section \ref{sec:disgauging-affD4}, the notation $/\BZ_2$ means that the $(\BZ_2^{[1]})_C$ 1-form symmetry is gauged. In the following, we also consider the case in which $(\BZ_2^{[1]})_C$ is not gauged, namely
\begin{align}\label{eq:USp4-4U1}
    \begin{tikzpicture}[scale=0.86,font=\footnotesize,baseline=0cm]
\node (A1) at (0,0) {$\USp(4)$};
\node (A2) at (1.2,-1.2) {$\U(1)$};
\node (A3) at (-1.2,1.2) {$\U(1)$};
\node (A4) at (1.2,1.2) {$\U(1)$};
\node (A5) at (-1.2,-1.2) {$\U(1)$}; 
\draw (A2)--(A1) (A3)--(A1) (A4)--(A1) (A5)--(A1); 
\path (A1) edge [out=30,in=-30,looseness=6] node[midway,right]{$\mathrm{AS}'$}  (A1);
\end{tikzpicture}
\end{align}
To be explicit, we denote by \eqref{eq:USp4-4U1}$/\BZ_2$ the option in which $(\BZ_2^{[1]})_C$ is gauged.

The Coulomb branch Hilbert series of theory \eref{eq:USp4-4U1}$/\BZ_2$ is
\begin{equation} \label{CBUSp44U1}
    \begin{split}
        & \mathrm{HS}\left[\text{CB of \eref{eq:USp4-4U1}}/\BZ_2\right]({\vec z}|\omega;t) \\
        & = \sum_{\epsilon=0}^1 \omega^{\epsilon} \sum_{u_1 \in \BZ+\frac{\epsilon}{2}} \, \sum_{u_2 \in \BZ+\frac{\epsilon}{2}} \, \sum_{u_3 \in \BZ+\frac{\epsilon}{2}} \, \sum_{u_4 \in \BZ+\frac{\epsilon}{2}} \, \sum_{m_1 \geq m_2 \geq \frac{\epsilon}{2}} t^{2\Delta} \left( \prod_{i=1}^4 z_i^{u_i} \right)  \frac{P_{\USp(4)}(m_1, m_2;t)}{(1-t^2)^4}~,
    \end{split}
\end{equation}
where $\omega$ (with $\omega^2=1$) is the fugacity for the $\BZ_2^{[0]}$ 0-form symmetry that arises from gauging the 1-form symmetry of \eref{eq:USp4-4U1}, $z_{1,2,3,4}$ are the fugacities for the four $\U(1)$ topological symmetries, which we shall denote by $\U(1)_i^T$, and 
\bes{
2 \Delta &= \sum_{i=1}^2 \sum_{j=1}^4 \sum_{s= \pm 1}|s m_i -u_j| + \sum_{s_1, s_2 = \pm 1} |s_1 m_1 +s_2 m_2|\\
& \quad -2 |m_1-m_2| -2|m_1+m_2| -4|m_1|-4|m_2|~.
}
In the dressing factor $\frac{P_{\USp(4)}(m_1, m_2;t)}{(1-t^2)^4}$, the denominator is the contribution of each $\U(1)$ gauge group, and the numerator, which is the contribution of the $\USp(4)$ gauge group, is given by
\bes{
P_{\USp(4)}(m_1, m_2;t) = 
\begin{cases}
(1-t^2)^{-2} & m_1 >m_2>0 \\
(1-t^2)^{-1}(1-t^4)^{-1} & m_1 >m_2=0~\text{or}~m_1=m_2>0 \\
(1-t^4)^{-1}(1-t^8)^{-1} & m_1=m_2=0
\end{cases}~.
}
The Coulomb branch Hilbert series of theory \eref{eq:USp4-4U1} can then be obtained by gauging the 0-form symmetry associated with $\omega$:
\bes{
\mathrm{HS}[\text{CB of \eref{eq:USp4-4U1}}]({\vec z};t) = \frac{1}{2} \sum_{\omega=\pm 1} \mathrm{HS}[\text{CB of \eref{eq:USp4-4U1}}/\BZ_2]({\vec z}|\omega;t)~.
}

Let us present the result explicitly:
\bes{ \label{CBHSUSp4-4U1}
\scalebox{0.99}{$
\begin{split}
&\mathrm{HS}[\text{CB of \eref{eq:USp4-4U1}}]({\vec z};t) \\&= \PE\left[4 t^2+\left\{\sum_{i=1}^4 \left(z_i + z_i^{-1}\right)+2\right\} t^4+\left\{\sum_{i=1}^4 \left(z_i + z_i^{-1}\right)+1\right\} t^6\right.\\
        &\qquad \quad +\left\{\sum_{i=1}^4 z_i +2 \sum_{1\leq i <j \leq 4} z_i z_j +2 \sum_{1\leq i <j \leq 4} \frac{z_i}{z_j}+ \sum_{\overset{1\leq i <j \leq 4}{k \neq i, \, j}} \frac{z_i z_j}{z_k} + \frac{z_1 z_2}{z_3 z_4} + \frac{z_1 z_3}{z_2 z_4}+ \frac{z_1 z_4}{z_2 z_3}\right. \\
        & \qquad \quad +\left.\left.\sum_{1\leq i <j<k \leq 4} z_i z_j z_k +\sum_{\overset{1\leq i <j<k \leq 4}{l \neq i,\, j, \, k}} \frac{z_i z_j z_k}{z_l} + z_1z_2z_3z_4 + \left(z_i \leftrightarrow z_i^{-1}\right) -1 \right\} t^8 +\mathcal{O}\left(t^{10}\right)\right]~.
        \end{split}$}
        }
Setting $z_i=1$, we obtain the unrefined Hilbert series:
\begin{equation}
    \mathrm{HS}\left[\text{CB of \eref{eq:USp4-4U1}}\right](t) =\PE\left[4 t^2+10 t^4+9 t^6+103 t^8+82 t^{10}-484 t^{12}+\mathcal{O}(t^{14})\right]~.
\end{equation}
On the other hand, the Coulomb branch Hilbert series of \eref{eq:USp4-4U1}$/\BZ_2$ can be written as
\bes{ \label{CBHSUSp4-4U1modZ2}
\scalebox{0.9}{$\displaystyle
\begin{aligned}
&\mathrm{HS}\left[\text{CB of \eref{eq:USp4-4U1}}/\BZ_2\right]({\vec z}| \omega;t) \\ &= \PE\left[4 t^2+\left\{\sum_{i=1}^4 \left(z_i + z_i^{-1}\right) + 2 + \omega \sum_{s_1, \ldots, s_4 = \pm 1} z_1^{\frac{1}{2} s_1} z_2^{\frac{1}{2} s_2} z_3^{\frac{1}{2} s_3} z_4^{\frac{1}{2} s_4}  \right\} t^4 +\left\{\sum_{i=1}^4 \left(z_i + z_i^{-1}\right)+1\right\} t^6 \right.\\
&\left. \qquad \quad  - \left(3\sum_{i=1}^4 (z_i + z_i^{-1}) + 9 + 5\omega \sum_{s_1, \ldots, s_4 = \pm 1} z_1^{\frac{1}{2} s_1} z_2^{\frac{1}{2} s_2} z_3^{\frac{1}{2} s_3} z_4^{\frac{1}{2} s_4}  \right) t^8 +\mathcal{O}\left(t^{10}\right) \right]~,
\end{aligned}$}
}
whose unrefinement, with $z_i = 1$ and $\omega=1$, reads
\begin{equation}
    \mathrm{HS}[\text{CB of \eref{eq:USp4-4U1}}/\BZ_2](t) =\PE\left[4 t^2+26 t^4+9 t^6-113 t^8+\mathcal{O}(t^{10})\right]~.
\end{equation}
Note that the terms at order $t^8$ in  \eref{CBHSUSp4-4U1} receive a cancellation from the terms with $\omega^2=1$ and become those shown in \eref{CBHSUSp4-4U1modZ2}.

We see that gauging the $(\BZ_2^{[1]})_C$ symmetry of theory \eref{eq:USp4-4U1} leads to half-odd-integer powers of the fugacities $z_i$ of the topological symmetries $\U(1)_i^T$. We remark that these half-odd-integer powers are not present in \eref{CBHSUSp4-4U1}. In other words, upon gauging $(\BZ_2^{[1]})_C$, the quantization of the $\U(1)_i^T$ charges changes from integers to half-odd-integers.
This implies that there is a mixed anomaly between the $\U(1)_i^T$ topological symmetries and the $(\BZ_2^{[1]})_C$ 1-form symmetry \cite{Hanany:2010qu, Mekareeya:2022spm, Bhardwaj:2023zix, Sacchi:2023omn}, where the anomaly theory is given by
\bes{ \label{mixedanom}
i \pi \int_{M_4} B^{(2)} \cup \left[c_1(\U(1)^T_i) \,\, \mod \, 2\right]~, \quad \text{with $i=1,2,3,4$}~,
}
where $B^{(2)}$ is the 2-form background gauge field associated with $(\BZ_2^{[1]})_C$ and $c_1(\U(1)^T_i)$ is the first Chern class of the bundle associated with the $\U(1)^T_i$ topological symmetry. 

We can confirm the 1-form symmetry of theory \eqref{SU4w4SU2} at the level of the BPS quiver. For definiteness, we focus on the case $n=2$, although the argument can be easily generalized. For the model at hand, the BPS quiver can be derived exploiting the methods developed in \cite{Alim:2011kw, Cecotti:2012jx, Cecotti:2013lda}:\footnote{In particular, this quiver can be obtained by superimposing four copies of the quiver depicted in \cite[(B.2)]{Cecotti:2012jx}; see also \cite[Section 3.3]{Cecotti:2012jx} and  \cite[Section 4.9]{Alim:2011kw} for further discussions.}
\bes{ \label{BPSSU4}
\begin{tikzpicture}[baseline,font=\footnotesize,scale=1]
\node[circle,draw=black,inner sep=5pt] (A) at (0,0) {}; 
\node[circle,draw=black,inner sep=5pt] (B) at (0,2) {};
\node[circle,draw=black,inner sep=5pt] (C) at (2,2) {};
\node[circle,draw=black,inner sep=5pt] (D) at (2,0) {};
\node[circle,draw=black,inner sep=5pt] (E) at (4,0) {};
\node[circle,draw=black,inner sep=5pt] (F) at (4,2) {};
\node[circle,draw=black,inner sep=5pt] (a1) at (5,2.5) {}; 
\node[circle,draw=black,inner sep=5pt] (b1) at (6,1.5) {};
\node[circle,draw=black,inner sep=5pt] (c1) at (6,3.5) {};
\node[circle,draw=black,inner sep=5pt] (a2) at (5,-0.5) {}; 
\node[circle,draw=black,inner sep=5pt] (b2) at (6,-1.5) {};
\node[circle,draw=black,inner sep=5pt] (c2) at (6,0.5) {};
\node[circle,draw=black,inner sep=5pt] (a3) at (-1,2.5) {}; 
\node[circle,draw=black,inner sep=5pt] (b3) at (-2,1.5) {};
\node[circle,draw=black,inner sep=5pt] (c3) at (-2,3.5) {};
\node[circle,draw=black,inner sep=5pt] (a4) at (-1,-0.5) {}; 
\node[circle,draw=black,inner sep=5pt] (b4) at (-2,-1.5) {};
\node[circle,draw=black,inner sep=5pt] (c4) at (-2,0.5) {};
\draw [->] (B) to (C);
\draw [->] (D) to (A);
\draw [->] (D) to (E);
\draw [->] (F) to (C);
\draw [->] (F) to (a1);
\draw [->] (a1) to (E);
\draw [->] (a1) to (b1);
\draw [->] (c1) to (a1);
\draw [->] (F) to (a2);
\draw [->] (a2) to (E);
\draw [->] (a2) to (b2);
\draw [->] (c2) to (a2);
\draw [->] (B) to (a3);
\draw [->] (a3) to (A);
\draw [->] (a3) to (b3);
\draw [->] (c3) to (a3);
\draw [->] (B) to (a4);
\draw [->] (a4) to (A);
\draw [->] (a4) to (b4);
\draw [->] (c4) to (a4);
\draw [->] (-0.1,0.25) to (-0.1,1.75);
\draw [->] (0.1,0.25) to (0.1,1.75);
\draw [->] (3.9,0.25) to (3.9,1.75);
\draw [->] (4.1,0.25) to (4.1,1.75);
\draw [->] (1.9,1.75) to (1.9,0.25);
\draw [->] (2.1,1.75) to (2.1,0.25);
\draw [->] (5.9,1.75) to (5.9,3.25);
\draw [->] (6.1,1.75) to (6.1,3.25);
\draw [->] (5.9,-1.25) to (5.9,0.25);
\draw [->] (6.1,-1.25) to (6.1,0.25);
\draw [->] (-1.9,1.75) to (-1.9,3.25);
\draw [->] (-2.1,1.75) to (-2.1,3.25);
\draw [->] (-1.9,-1.25) to (-1.9,0.25);
\draw [->] (-2.1,-1.25) to (-2.1,0.25);
\end{tikzpicture}
}
As explained in \cite{DelZotto:2022ras}, the 1-form symmetry of the theory is encoded in the adjacency matrix of the quiver $B_{ij}$, defined as 
$$B_{ij}=\text{number of arrows from node $i$ to node $j$}\,,$$ 
where $i$ and $j$ label the nodes of the quiver and we put a minus sign if the arrows point from $j$ to $i$. For the quiver \eqref{BPSSU4} the adjacency matrix is $18\times 18$ and its Smith normal form turns out to be 
\be\label{Smith} \diag(1,1,1,1,1,1,1,1,1,1,1,1,2,2,0,0,0,0)\,.\ee 
From \eqref{Smith}, we see that the rank of the global symmetry of the theory (which is the dimension of the kernel of $B_{ij}$) is 4 as expected, and the fact that we have two eigenvalues equal to 2, while the others are equal to 1, tells us that the defect group of the theory is $\BZ_2\times \BZ_2$ \cite{DelZotto:2022ras}.

Let us now study discrete gauging of theory \eref{SU4w4SU2} with $n=2$ by performing the $\ZZ_2$, $\ZZ_3$, $S_3$, $\ZZ_4$ and $S_4$ wreathing of \eqref{eq:USp4-4U1} with respect to the four $\U(1)$ gauge nodes. We explicitly indicate how the range of the summations of the gauge fluxes is restricted and how the dressing factors in \eref{CBUSp44U1} get modified. Importantly, we will study how the mixed anomaly \eref{mixedanom} changes upon discrete gauging. 

\subsection*{\texorpdfstring{$\ZZ_2$}{Z2} Wreathing}

Let us perform the $\BZ_2$ wreathing on the $\U(1)$ gauge nodes corresponding to the magnetic fluxes $u_1$ and $u_2$. The gauge fluxes get restricted, and the dressing factor is modified as follows.
\begin{equation}
\renewcommand*{\arraystretch}{1.4}
    \begin{array}{c|c}
        \text{Restriction} & \text{Dressing factor} \\
        \hline
        u_1<u_2 & (1-t^2)^{-4}P_{\USp(4)}(m_1,m_2;t)\\
        u_1=u_2 & (1-t^2)^{-3}(1-t^4)^{-1}P_{\USp(4)}(m_1,m_2;t)
    \end{array}
\end{equation}
We also set $z_1=z_2 \equiv x$. The topological symmetry $\U(1)^T_1 \times \U(1)^T_2$ of the original quiver \eref{eq:USp4-4U1} is broken into a diagonal subgroup $\U(1)^T_{(12)}$, while $\U(1)^T_3$ and $\U(1)^T_4$ are left untouched. After wreathing, the Coulomb branch Hilbert series is the same as that of the following theory:
\bes{ \label{USp4wrZ2}
\begin{tikzpicture}[baseline,font=\footnotesize,scale=1]
\node (U2l) at (0,0) {$\USp(4)$}; 
\node (U2r) at (2,0) {$\U(2)$}; 
\node (U1a) at (0,1) {$\U(1)$}; 
\node (U1b) at (0,-1) {$\U(1)$}; 
\path (U2r) edge [out=45,in=-45,looseness=8] node[midway,right]{$\mathrm{Adj}$} (U2r);
\path (U2l) edge [out=135,in=-135,looseness=8] node[midway,left]{$\mathrm{AS}'$}  (U2l);
\draw (U2l)--(U2r);
\draw (U2l) -- (U1a);
\draw (U2l) -- (U1b);
\end{tikzpicture}
\quad\text{\footnotesize$(/\BZ_2)$}
}
where $(/\BZ_2)$ denotes the option whether the $(\BZ_2^{[1]})_C$ 1-form symmetry is gauged or not.  Explicitly, if we denote the topological fugacity of the $\U(2)$ gauge node of \eref{USp4wrZ2} by $x$, and those of the two $\U(1)$ gauge nodes by $z_3$ and $z_4$, the Coulomb branch Hilbert series without gauging $(\BZ_2^{[1]})_C$ is
\begin{equation}
\begin{split}
    & \mathrm{HS}\left[\text{CB of \eref{eq:USp4-4U1}}_{\text{wr} \, \BZ_2}\right] (x, z_3, z_4; t) = \mathrm{HS}\left[\text{CB of \eref{USp4wrZ2}}\right] (x, z_3, z_4; t)\\ &
    = \PE \left[3 t^2+\left(x+\frac{1}{x}+z_3+\frac{1}{z_3}+z_4+\frac{1}{z_4}+3\right) t^4\right. \\
    & \left.\qquad \quad+\left(2 x+\frac{2}{x}+z_3+\frac{1}{z_3}+z_4+\frac{1}{z_4}+1\right) t^6 +\mathcal{O}(t^{8})\right]~.
\end{split}
\end{equation}
Setting $x=z_3=z_4=1$, we have
\begin{equation}
\begin{split}
    &\mathrm{HS}\left[\text{CB of  \eref{eq:USp4-4U1}}_{\text{wr} \, \BZ_2}\right] (t) = \mathrm{HS}\left[\text{CB of \eref{USp4wrZ2}}\right] (t) \\&=\PE\left[3 t^2+9 t^4+9 t^6+73 t^8+95 t^{10}-175 t^{12}+\mathcal{O}(t^{14})\right]~.
    \end{split}
\end{equation}
Gauging the $(\BZ_2^{[1]})_C$ 1-form symmetry, we obtain
\bes{
& \mathrm{HS}\left[\text{CB of [\eref{eq:USp4-4U1}}/\BZ_2]_{\text{wr} \, \BZ_2}\right] (x, z_3, z_4| \omega; t) = \mathrm{HS}\left[\text{CB of \eref{USp4wrZ2}}/\BZ_2\right] (x, z_3, z_4| \omega; t)\\
&= \mathrm{HS}\left[\text{CB of \eref{eq:USp4-4U1}}_{\text{wr} \, \BZ_2}\right] (x, z_3, z_4; t) \times \PE \left[\omega \sum_{s_1, s_2 = \pm 1} z_3^{\frac{1}{2} s_1} z_4^{\frac{1}{2} s_2} \left( x+1+x^{-1} \right) t^4 \right. \\
& \left.\qquad \, \, \, + \omega \sum_{s_1, s_2 = \pm 1} z_3^{\frac{1}{2} s_1} z_4^{\frac{1}{2} s_2} \left( x+2+x^{-1} \right) t^6 +\mathcal{O}(t^{8}) \right]~.
}
Observe that the quantization of the charges of the $\U(1)^T_{(12)}$ topological symmetry associated with the fugacity $x$ does not change upon gauging $(\BZ_2^{[1]})_C$, whereas that of the $\U(1)^T_{3,4}$ symmetries associated with the fugacities $z_{3,4}$ gets modified. We conclude that, in the discretely gauged theory, there is no mixed anomaly between $(\BZ_2^{[1]})_C$ and $\U(1)^T_{(12)}$. However, there is one involving $(\BZ_2^{[1]})_C$ and $\U(1)^T_{3,4}$, described by the following anomaly theory:
\bes{
i \pi \int_{M_4} B^{(2)} \cup \left[c_1(\U(1)^T_i) \,\, \mod \, 2\right]~, \qquad  \text{with $i=3, 4$}~.
}
This also follows from the fact that the anomaly theory \eref{mixedanom} implies that there is no mixed anomaly between $(\BZ_2^{[1]})_C$ and any diagonal subgroup of $\U(1)^T_i \times \U(1)^T_j$ for $i\neq j$.

\subsection*{\texorpdfstring{$\ZZ_3$}{Z3} Wreathing}

Let us now perform the $\BZ_3$ wreathing on the $\U(1)$ gauge nodes corresponding to the magnetic fluxes $u_{1,2,3}$. The gauge fluxes get restricted, and the dressing factor is modified as follows.
\begin{equation}
\renewcommand*{\arraystretch}{1.4}
    \begin{array}{c|c}
        \text{Restriction} & \text{Dressing factor} \\
        \hline
        u_1<u_2\geq u_3 & (1-t^2)^{-4}P_{\USp(4)}(m_1,m_2;t)\\
        u_1=u_2=u_3 & (1-t^2+t^4)\left(1-t^2\right)^{-3} \left(1-t^6\right)^{-1}P_{\USp(4)}(m_1,m_2;t)
    \end{array}
\end{equation}
We also set $z_1=z_2=z_3 \equiv x$. The topological symmetry $\prod_{i=1}^3 \U(1)^T_{i}$ of the original quiver \eref{eq:USp4-4U1} is thus broken to its diagonal subgroup $\U(1)^T_{(123)}$, and $\U(1)^T_{4}$ is left untouched.
Explicitly, the Coulomb branch Hilbert series is
\bes{
    & \mathrm{HS}\left[\text{CB of \eref{eq:USp4-4U1}}_{\text{wr} \, \BZ_3}\right] (x, z_4; t) \\& = \PE\left[2 t^2+\left(x+\frac{1}{x}+z_4+\frac{1}{z_4}+3\right) t^4+\left(3 x+\frac{3}{x}+z_4+\frac{1}{z_4}+3\right) t^6 +\mathcal{O}(t^{8})\right]
}
and, setting $z_1=z_4=1$, we obtain the unrefined Hilbert series
\begin{equation}
\scalebox{0.95}{$
    \mathrm{HS}\left[\text{CB of \eref{eq:USp4-4U1}}_{\text{wr} \, \BZ_3}\right] (t) =\PE\left[2 t^2+7 t^4+11 t^6+53 t^8+114 t^{10}+83 t^{12}-300 t^{14}+\mathcal{O}(t^{16})\right]~.
    $}
\end{equation}
Gauging the $(\BZ_2^{[1]})_C$ 1-form symmetry, the Coulomb branch Hilbert series gets modified as follows:
\bes{
&\mathrm{HS}\left[\text{CB of [\eref{eq:USp4-4U1}$/\BZ_2$]}_{\text{wr} \, \BZ_3}\right] (x, z_4| \omega; t) \\ &= \mathrm{HS}\left[\text{CB of \eref{eq:USp4-4U1}}_{\text{wr} \, \BZ_3}\right] (x, z_4; t) \times \PE \left[\omega \left(z_4^{\frac{1}{2}}+z_4^{-\frac{1}{2}} \right) \left(x^{\frac{3}{2}}+x^{\frac{1}{2}}+x^{-\frac{1}{2}}+x^{-\frac{3}{2}}   \right) t^4 \right.\\
&\left. \qquad \, \, \, + \omega \left(z_4^{\frac{1}{2}}+z_4^{-\frac{1}{2}} \right) \left(x^{\frac{3}{2}}+3x^{\frac{1}{2}}+3x^{-\frac{1}{2}}+x^{-\frac{3}{2}}   \right) t^6 +\mathcal{O}(t^{8})\right]~.
}
Since the charge quantization of $\U(1)^T_{(123)}$ and $\U(1)^T_4$ changes upon gauging $(\BZ_2^{[1]})_C$, we conclude that the mixed anomalies between each of them and $(\BZ_2^{[1]})_C$ are given by
\bes{ \label{mixedanomZ3wreathing}
i \pi \int_{M_4} B^{(2)} \cup \left[c_1(\U(1)^T_i) \,\, \mod \, 2\right]~, \qquad  \text{with $i=(123), 4$}~.
}

\subsection*{\texorpdfstring{$S_3$}{S3} Wreathing}

In the case of an $S_3$ wreathing, we modify the Coulomb branch formula \eref{CBUSp44U1} as follows:
\begin{equation}
\renewcommand*{\arraystretch}{1.4}
    \begin{array}{c|c}
        \text{Restriction} & \text{Dressing factor} \\
        \hline
        u_1<u_2<u_3 & (1-t^2)^{-4}P_{\USp(4)}(m_1,m_2;t)\\
        u_1=u_2<u_3 & (1-t^2)^{-3}(1-t^4)^{-1}P_{\USp(4)}(m_1,m_2;t) \\
        u_1<u_2=u_3 & (1-t^2)^{-3}(1-t^4)^{-1}P_{\USp(4)}(m_1,m_2;t) \\
        u_1=u_2=u_3 & (1-t^2)^{-2}(1-t^4)^{-1}(1-t^6)^{-1}P_{\USp(4)}(m_1,m_2;t) \\
    \end{array}
\end{equation}
Also in this case, we set $z_1=z_2=z_3=x$. As before, the symmetry $\prod_{i=1}^3 U(1)^T_i$ is broken to its diagonal subgroup $\U(1)^T_{(123)}$ and only $\U(1)_4^T$ is left untouched. In this case, the Coulomb branch Hilbert series of the wreathed quiver is equal to that of the following quiver:
\bes{ \label{USp4wrS3}
\begin{tikzpicture}[baseline,font=\footnotesize,scale=1]
\node (U2l) at (0,0) {$\USp(4)$}; 
\node (U2r) at (1.6,0) {$\U(3)$}; 
\node (U1a) at (0,1.2) {$\U(1)$}; 
\path (U2r) edge [out=45,in=-45,looseness=8] node[midway,right]{$\mathrm{Adj}$} (U2r);
\path (U2l) edge [out=135,in=-135,looseness=8] node[midway,left]{$\mathrm{AS}'$}  (U2l);
\draw (U2l)--(U2r);
\draw (U2l) -- (U1a);
\end{tikzpicture}
\quad\text{\footnotesize$(/\BZ_2)$}
}
Explicitly, it reads
\bes{
        &\mathrm{HS}\left[\text{CB of \eref{eq:USp4-4U1}}_{\text{wr} \, S_3}\right] (x, z_4; t) = \mathrm{HS}\left[\text{CB of \eref{USp4wrS3}}\right] (x, z_4; t) \\ &= \PE\left[2 t^2+\left(x+\frac{1}{x}+z_4+\frac{1}{z_4}+3\right) t^4+\left(2 x+\frac{2}{x}+z_4+\frac{1}{z_4}+2\right) t^6 +\mathcal{O}(t^{8})\right]~,
}
with the unrefinement
\bes{
    &\mathrm{HS}\left[\text{CB of \eref{eq:USp4-4U1}}_{\text{wr} \, S_3}\right] (t) = \mathrm{HS}\left[\text{CB of \eref{USp4wrS3}}\right] (t) \\ &=\PE\left[2 t^2+7 t^4+8 t^6+44 t^8+72 t^{10}+18 t^{12}-203 t^{14}+\mathcal{O}(t^{16})\right]~.
}
Gauging the $(\BZ_2^{[1]})_C$ 1-form symmetry, we obtain
\bes{
&\mathrm{HS}\left[\text{CB of [\eref{eq:USp4-4U1}}/\BZ_2]_{\text{wr} \, S_3}\right] (x, z_4| \omega; t) = \mathrm{HS}\left[\text{CB of \eref{USp4wrS3}}/\BZ_2\right] (x, z_4| \omega; t) \\ &= \mathrm{HS}\left[\text{CB of \eref{eq:USp4-4U1}}_{\text{wr} \, S_3}\right] (x, z_4; t) \times \PE \left[\omega \left(z_4^{\frac{1}{2}}+z_4^{-\frac{1}{2}} \right) \left(x^{\frac{3}{2}}+x^{\frac{1}{2}}+x^{-\frac{1}{2}}+x^{-\frac{3}{2}}   \right) t^4\right. \\
& \left.\qquad \, \, \, + \omega \left(z_4^{\frac{1}{2}}+z_4^{-\frac{1}{2}} \right) \left(x^{\frac{3}{2}}+2x^{\frac{1}{2}}+2x^{-\frac{1}{2}}+x^{-\frac{3}{2}}   \right) t^6 +\mathcal{O}(t^{8})\right]~.
}
By the same reasoning as in the case of the $\BZ_3$ wreathing, the mixed anomalies between $(\BZ_2^{[1]})_C$ and $\U(1)^T_{(123)}$ and $\U(1)^T_4$ are given by \eref{mixedanomZ3wreathing}.

\subsection*{\texorpdfstring{$\ZZ_4$}{Z4} Wreathing}

For the $\BZ_4$ wreathing on the $\U(1)$ gauge nodes of quiver \eqref{eq:USp4-4U1}, we modify the gauge fluxes and dressing factor in the Coulomb branch formula \eref{CBUSp44U1} as follows:
\begin{equation}
\renewcommand*{\arraystretch}{1.4}
    \begin{array}{c|c}
       \text{Restriction} & \text{Dressing factor} \\
        \hline
        u_1<u_2 \land u_3\geq u_4 & (1-t^2)^{-4}P_{\USp(4)}(m_1,m_2;t)\\
        u_1=u_2<u_3=u_4 & (1+t^4)\left(1-t^2\right)^{-2} \left(1-t^4\right)^{-2}P_{\USp(4)}(m_1,m_2;t)\\
        u_1=u_2=u_3=u_4 & (1-t^2+t^4+t^6)\left(1-t^2\right)^{-2} \left(1-t^4\right)^{-1} \left(1-t^8\right)^{-1}P_{\USp(4)}(m_1,m_2;t)
    \end{array}
\end{equation}
This time, we take $z_1=z_2=z_3=z_4 \equiv x$, and the whole topological symmetry $\prod_{i=1}^4 \U(1)_{i}^T$ is broken to the diagonal subgroup $\U(1)_{(1234)}^T$. The Coulomb branch Hilbert series is
\begin{equation}
    \begin{split}
        &\mathrm{HS}\left[\text{CB of \eref{eq:USp4-4U1}}_{\text{wr} \, \BZ_4}\right] (x; t) \\& =\PE\left[t^2+\left(x+4+\frac{1}{x}\right) t^4+\left(4 x+3+\frac{4}{x}\right) t^6\right.\\
    & \qquad \quad \left. +\left(x^4+x^3+7 x^2+11 x+15+\frac{11}{x}+\frac{7}{x^2}+\frac{1}{x^3}+\frac{1}{x^4}\right) t^8+\mathcal{O}(t^{10})\right]~,
    \end{split}
\end{equation}
from which, setting $x=1$, we obtain the unrefined Hilbert series as
\bes{
&\mathrm{HS}\left[\text{CB of \eref{eq:USp4-4U1}}_{\text{wr} \, \BZ_4}\right] (t) \\ &=\PE\left[t^2+6 t^4+11 t^6+55 t^8+135 t^{10}+214 t^{12}+11 t^{14}-1626 t^{16}+\mathcal{O}(t^{18})\right]~.
}
Gauging the $(\BZ_2^{[1]})_C$ 1-form symmetry, the Coulomb branch Hilbert series reads
\bes{
        & \mathrm{HS}\left[\text{CB of [\eref{eq:USp4-4U1}$/\BZ_2$]}_{\text{wr} \, \BZ_4}\right] (x| \omega; t) \\&= \mathrm{HS}\left[\text{CB of \eref{eq:USp4-4U1}}_{\text{wr} \, \BZ_4}\right] (x; t) \times \PE\left[\omega \left(x^2+x+2+\frac{1}{x}+\frac{1}{x^2} \right) t^4 \right. \\& \left. \qquad \, \, \,+ \omega \left(x^2+4 x+6+\frac{4}{x}+\frac{1}{x^2} \right) t^6
        + \left \{ \omega \left(3 x^2+8 x+12+\frac{8}{x}+\frac{3}{x^2} \right) \right. \right. \\& \left. \left. \qquad \, \, \,-\left( x^4+x^3+3 x^2+3 x+5+\frac{3}{x}+\frac{3}{x^2}+\frac{1}{x^3}+\frac{1}{x^4} \right) \right \} t^8 +\mathcal{O}(t^{10})
        \right]~.
}
Setting $x=1$ and $\omega=1$, we obtain the unrefined Hilbert series as
\bes{
    &\mathrm{HS}\left[\text{CB of [\eref{eq:USp4-4U1}$/\BZ_2$]}_{\text{wr} \, \BZ_4}\right] (t) \\ &=\PE\left[t^2+12 t^4+27 t^6+68 t^8+75 t^{10}-236 t^{12}-1489 t^{14} -4054 t^{16}+\mathcal{O}(t^{18})\right]~.
}
Observe that the quantization of the charges of the $\U(1)_{(1234)}^T$ symmetry does not change upon gauging the $(\BZ_2^{[1]})_C$ 1-form symmetry. Therefore, we conclude that there is no mixed anomaly between $(\BZ_2^{[1]})_C$ and $\U(1)_{(1234)}^T$.

\subsection*{\texorpdfstring{$S_4$}{S4} Wreathing}

Finally, in the case of the $S_4$ wreathing, we modify the Coulomb branch formula \eref{CBUSp44U1} as follows:
\begin{equation}
\renewcommand*{\arraystretch}{1.4}
    \begin{array}{c|c}
       \text{Restriction} & \text{Dressing factor} \\
        \hline
        u_1<u_2<u_3<u_4 & (1-t^2)^{-4}P_{\USp(4)}(m_1,m_2;t)\\
        u_1=u_2<u_3<u_4 & (1-t^2)^{-3}(1-t^4)^{-1}P_{\USp(4)}(m_1,m_2;t) \\
        u_1<u_2=u_3<u_4 & (1-t^2)^{-3}(1-t^4)^{-1}P_{\USp(4)}(m_1,m_2;t) \\
        u_1<u_2<u_3=u_4 & (1-t^2)^{-3}(1-t^4)^{-1}P_{\USp(4)}(m_1,m_2;t) \\
        u_1=u_2=u_3<u_4 & (1-t^2)^{-2}(1-t^4)^{-1}(1-t^6)^{-1}P_{\USp(4)}(m_1,m_2;t) \\
        u_1<u_2=u_3=u_4 & (1-t^2)^{-2}(1-t^4)^{-1}(1-t^6)^{-1}P_{\USp(4)}(m_1,m_2;t) \\
        u_1=u_2<u_3=u_4 & (1-t^2)^{-2}(1-t^4)^{-2}P_{\USp(4)}(m_1,m_2;t) \\
        u_1=u_2=u_3=u_4 & (1-t^2)^{-1}(1-t^4)^{-1}(1-t^6)^{-1}(1-t^8)^{-1}P_{\USp(4)}(m_1,m_2;t) \\
    \end{array}
\end{equation}
As in the case of the $\BZ_4$ wreathing, we also set $z_1=z_2=z_3=z_4 \equiv x$, and the whole topological symmetry $\prod_{i=1}^4 \U(1)_{i}^T$ is broken to the diagonal subgroup $\U(1)_{(1234)}^T$. The resulting Coulomb branch Hilbert series is equal to the one of the following quiver:
\bes{ \label{USp4wrS4}
\begin{tikzpicture}[baseline,font=\footnotesize,scale=1]
\node (U2l) at (0,0) {$\USp(4)$}; 
\node (U2r) at (2,0) {$\U(4)$}; 
\path (U2r) edge [out=45,in=-45,looseness=8] node[midway,right]{$\mathrm{Adj}$} (U2r);
\path (U2l) edge [out=135,in=-135,looseness=8] node[midway,left]{$\mathrm{AS}'$}  (U2l);
\draw (U2l)--(U2r);
\end{tikzpicture}
\quad\text{\footnotesize$(/\BZ_2)$}
}
Explicitly, the Coulomb branch Hilbert series reads
\begin{equation}
    \begin{split}
        &\mathrm{HS}\left[\text{CB of \eref{eq:USp4-4U1}}_{\text{wr} \, S_4}\right] (x; t) = \mathrm{HS}\left[\text{CB of \eref{USp4wrS4}}\right] (x; t) \\ &=  \PE \left[t^2+\left(x+3+\frac{1}{x}\right) t^4+\left(2 x+2+\frac{2}{x}\right) t^6\right.\\
    &\left. \qquad \, \, \, +\left(x^4+x^3+4 x^2+4 x+7+\frac{4}{x}+\frac{4}{x^2}+\frac{1}{x^3}+\frac{1}{x^4}\right) t^8+\mathcal{O}(t^{10})\right]~,
    \end{split}
\end{equation}
with the unrefinement
\bes{
   &\mathrm{HS}\left[\text{CB of \eref{eq:USp4-4U1}}_{\text{wr} \, S_4}\right] (t) = \mathrm{HS}\left[\text{CB of \eref{USp4wrS4}}\right] (t) \\ &=\PE\left[t^2+5 t^4+6 t^6+27 t^8+46 t^{10}+61 t^{12}+26 t^{14}-223 t^{16}+\mathcal{O}(t^{18})\right]~.
}
Let us also gauge the $(\BZ_2^{[1]})_C$ 1-form symmetry. In which case, the Coulomb branch Hilbert series becomes
\bes{
    &\mathrm{HS}\left[\text{CB of [\eref{eq:USp4-4U1}$/\BZ_2$]}_{\text{wr} \, S_4}\right] (x| \omega; t) = \mathrm{HS}\left[\text{CB of \eref{USp4wrS4}}/\BZ_2\right] (x| \omega; t) \\&
    = \mathrm{HS}\left[\text{CB of \eref{eq:USp4-4U1}}_{\text{wr} \, S_4}\right] (x; t) \times \PE\left[\omega \left(x^2+x+1+\frac{1}{x}+\frac{1}{x^2}\right) t^4 \right. \\ & \left. \qquad \, \, \, + \omega \left( x^2+2 x+2+\frac{2}{x}+\frac{1}{x^2} \right) t^6+ \left \{ \omega \left( x^2+2 x+4+\frac{2}{x}+\frac{1}{x^2} \right) \right. \right. \\
    & \left. \left. \qquad \, \, \, - \left(x^4+x^3+2 x^2+2 x+3+\frac{2}{x}+\frac{2}{x^2}+\frac{1}{x^3}+\frac{1}{x^4} \right)\right \} t^8 +\mathcal{O}(t^{10})\right]~.
}
Unrefining it by setting $x=1$ and $\omega=1$, we obtain
\begin{equation}
\begin{split}
    &\mathrm{HS}\left[\text{CB of [\eref{eq:USp4-4U1}$/\BZ_2$]}_{\text{wr} \, S_4}\right] (t) = \mathrm{HS}\left[\text{CB of \eref{USp4wrS4}}/\BZ_2\right] (t) \\ &=\PE\left[t^2+10 t^4+14 t^6+22 t^8+13 t^{10}-50 t^{12}-177 t^{14} -309 t^{16}+\mathcal{O}(t^{18})\right]~.
\end{split}
\end{equation}
Again, as for the $\BZ_4$ case, the quantization of the charges of the $\U(1)_{(1234)}^T$ symmetry does not change upon gauging the $(\BZ_2^{[1]})_C$ 1-form symmetry. Therefore, this leads to the conclusion that there is no mixed anomaly between $(\BZ_2^{[1]})_C$ and $\U(1)_{(1234)}^T$.

\section{Discrete Gauging of the \texorpdfstring{$(A_{N},A_{N})$}{(A,A)} Theories}
\label{sec:discretegaugingAA}

In this appendix, we consider discrete gauging of the Argyres--Douglas $(A_N, A_N)$ theories. One of the features of such theories is that their 3d mirror is described by a complete graph with edge multiplicity 1 of $N+1$ $\U(1)$ gauge nodes \cite{Xie:2012hs, Benvenuti:2017bpg, Giacomelli:2020ryy}, where, upon decoupling an overall $\U(1)$, we arrive at a complete graph of $N$ $\U(1)$ gauge nodes such that each of them has one hypermultiplet of charge 1 transforming under it. The latter description allows us to wreath such a quiver with any subgroup of $S_N$. Therefore, the Higgs branch of the discretely gauged $(A_N, A_N)$ theory can be determined by studying the Coulomb branch of the wreathed magnetic quiver.
We remark that discrete gauging of Argyres--Douglas theories whose 3d mirror theories are complete graphs with even edge multiplicity has been studied in
\cite{Hanany:2023uzn}. The results in this appendix are complementary to those of \cite{Hanany:2023uzn}.

\subsection{\texorpdfstring{$(A_3,A_3)$}{(A3,A3)}}

Let us consider the 3d mirror theory of $(A_3,A_3)$, which is depicted by the following quiver.
\begin{align} \label{A3A3}
    \begin{tikzpicture}[baseline=10pt,font=\footnotesize,scale=1.2]
        \node (U1n) at (0,1) {$\U(1)$};
        \node (U1e) at (1,0) {$\U(1)$};
        \node (U1w) at (-1,0) {$\U(1)$};
        \node (Fn) at (0,1.8) {[1]};
        \node (Fe) at (2,0) {[1]};
        \node (Fw) at (-2,0) {[1]};
        \draw (U1n) -- (U1e) --  (U1w) -- (U1n);
        \draw (Fn) -- (U1n);
        \draw (Fw) -- (U1w);
        \draw (Fe) -- (U1e);
    \end{tikzpicture}
\end{align}
The Coulomb branch Hilbert series is given by
\bes{ \label{CBHSA3A3}
\mathrm{HS}\left[\text{CB of \eqref{A3A3}}\right](\vec{z}; t) = (1-t^2)^{-3} \sum_{u_1, u_2, u_3 \in \BZ} t^{2 \Delta} z_1^{u_1}z_2^{u_2}z_3^{u_3} ~,
}
where $z_{1,2,3}$ are the fugacities associated with each $\U(1)$ topological symmetry, $(1-t^2)^{-3}$ is the dressing factor and
\bes{
2 \Delta = \sum_{i=1}^3 |u_i| +\sum_{1\leq j < k \leq 3} |u_j-u_k|~.
}
Explicitly, it reads
\begin{equation}
\scalebox{0.92}{$\displaystyle
\begin{aligned}
\mathrm{HS}\left[\text{CB of \eqref{A3A3}}\right](\vec{z}; t) = \PE &\left[3 t^2+\left(z_1 + \frac{1}{z_1}+z_2+\frac{1}{z_2}+z_3+\frac{1}{z_3}+z_1 z_2 z_3+\frac{1}{z_1 z_2 z_3}\right) t^3\right.\\
    &  +\left.\left(z_1 z_2+\frac{1}{z_1 z_2}+z_1 z_3+\frac{1}{z_1 z_3}+z_2 z_3+\frac{1}{z_2 z_3}\right) t^4\right.\\
    &  -\left.\left(2 z_1 z_2+\frac{2}{z_1 z_2}+2 z_1 z_3+\frac{2}{z_1 z_3}+2 z_2 z_3+\frac{2}{z_2 z_3}+4\right) t^6+\mathcal{O}\left(t^{7}\right)\right]\fstop
\end{aligned}$}
\end{equation}
From the coefficient at order $t^2$, we observe that the Coulomb branch symmetry is indeed $\U(1)^3$.  Setting $z_i=1$, the unrefined Hilbert series admits the following closed form (see \cite[(4.8)]{Beratto:2020wmn}):
\begin{equation}
\begin{split}
    \mathrm{HS}\left[\text{CB of \eqref{A3A3}}\right](t) &= \frac{1-2 t+3 t^2+2 t^3-2 t^4+2 t^5+3 t^6-2 t^7+t^8}{(1-t)^6 (1+t)^2 \left(1+t^2\right) \left(1+t+t^2\right)^2}
   \fstop
\end{split}
\end{equation}

In the following subsections, we consider the $\ZZ_2$, $\ZZ_3$ and $S_3$ wreathing of this theory. The gauge fluxes get restricted and the dressing factors are consequently modified in a similar manner as in the previous examples. 

\subsubsection{\texorpdfstring{$\ZZ_2$}{Z2} Wreathing}

Let us perform the $\ZZ_2$ wreathing on two of the $\U(1)$ gauge nodes, e.g. those associated to the magnetic fluxes $u_1$ and $u_3$. The gauge fluxes restrictions and the corresponding dressing factors are as follows. 
\begin{equation}
\renewcommand*{\arraystretch}{1.4}
    \begin{array}{c|c}
        \text{Restriction} & \text{Dressing factor} \\
        \hline
        u_1<u_3 & (1-t^2)^{-3}\\
        u_1=u_3 & (1-t^2)^{-2}(1-t^4)^{-1}
    \end{array}
\end{equation}
Setting $z_1=z_3 \equiv x$, we obtain the following Coulomb branch Hilbert series:
\begin{equation}
\begin{split}
\mathrm{HS}\left[\text{CB of \eqref{A3A3}}_{\text{wr} \, \BZ_2}\right](x,z_2;t) = \PE &\left[2 t^2+\left(x+\frac{1}{x}+z_2+\frac{1}{z_2}+x^2 z_2+\frac{1}{x^2 z_2}\right) t^3 \right. \\& \left. +\left(x^2+\frac{1}{x^2}+x z_2+\frac{1}{x z_2}+1\right) t^4+\left(x+\frac{1}{x}\right) t^5\right.\\
    & \left.-\left(x^2+\frac{1}{x^2}+x z_2+\frac{1}{x z_2}+2\right) t^6+\mathcal{O}\left(t^7\right)\right]\fstop
\end{split}
\end{equation}
The unrefined Hilbert series is
\bes{
    \mathrm{HS}\left[\text{CB of \eqref{A3A3}}_{\text{wr} \, \BZ_2}\right](t) = \frac{1-2 t+3 t^2+2 t^5+3 t^8-2 t^9+t^{10}}{(1-t)^6 (1+t)^2 \left(1+t^2\right)^2 \left(1+t+t^2\right)^2} 
    ~,
}
from which it is clear that the dimension of the CB is not changed by the discrete gauging.

\subsubsection{\texorpdfstring{$\ZZ_3$}{Z3} Wreathing}

Let us now perform the $\BZ_3$ wreathing on all the $\U(1)$ gauge nodes. The gauge fluxes get restricted and the dressing factor is modified as follows.
\begin{equation}
\renewcommand*{\arraystretch}{1.4}
    \begin{array}{c|c}
        \text{Restriction} & \text{Dressing factor} \\
        \hline
        u_1<u_2\geq u_3 & (1-t^2)^{-3}\\
        u_1=u_2=u_3 & (1-t^2+t^4)\left(1-t^2\right)^{-2} \left(1-t^6\right)^{-1}
    \end{array}
\end{equation}
Setting $z_1=z_2=z_3 \equiv x$, the Coulomb branch Hilbert series is 
\begin{equation}
\scalebox{0.92}{$\displaystyle
\begin{aligned}
\mathrm{HS}\left[\text{CB of \eqref{A3A3}}_{\text{wr} \, \BZ_3}\right](x;t) = \PE &\left[t^2+\left(x^3+x+\frac{1}{x}+\frac{1}{x^3}\right) t^3+\left(x^2+1+\frac{1}{x^2}\right) t^4\right.\\
    & \left. +\left(2 x+\frac{2}{x}\right) t^5+\left(x^2+2+\frac{1}{x^2}\right) t^6+\left(x^3 + x +\frac{1}{x}+\frac{1}{x^3}\right) t^7\right.\\
    &\left.-t^8-\left(x^3 + 3 x +\frac{3}{x} + \frac{1}{x^3}\right) t^9+\mathcal{O}\left(t^{10}\right)\right]~,
\end{aligned}$ }
\end{equation}
whose unrefinement leads to the following closed expression:
\begin{equation}
    \mathrm{HS}\left[\text{CB of \eqref{A3A3}}_{\text{wr} \, \BZ_3}\right](t) = \frac{1-2 t+t^2+2 t^3-2 t^4+2 t^5+t^6-2 t^7+t^8}{(1-t)^6 (1+t)^2 \left(1+t^2\right) \left(1+t+t^2\right)^2} 
    \fstop
\end{equation}

\subsubsection{\texorpdfstring{$S_3$}{S3} Wreathing}

Finally, we can consider the $S_3$ wreathing of quiver \eref{A3A3}. The restrictions and the corresponding dressing factors are listed in the table below.
\begin{equation}
\renewcommand*{\arraystretch}{1.4}
    \begin{array}{c|c}
        \text{Restriction} & \text{Dressing factor} \\
        \hline
        u_1<u_2<u_3 & (1-t^2)^{-3}\\
        u_1=u_2<u_3 & (1-t^2)^{-2}(1-t^4)^{-1} \\
        u_1<u_2=u_3 & (1-t^2)^{-2}(1-t^4)^{-1} \\
        u_1=u_2=u_3 & (1-t^2)^{-1}(1-t^4)^{-1}(1-t^6)^{-1}
    \end{array}
\end{equation}
Also in this case, we set $z_1=z_2=z_3 \equiv x$, obtaining the following Coulomb branch Hilbert series
\begin{equation}
\begin{split}
\mathrm{HS}\left[\text{CB of \eqref{A3A3}}_{\text{wr} \, S_3}\right](x;t) =  \PE &\left[t^2+\left(x^3 + x + \frac{1}{x}+\frac{1}{x^3}\right) t^3+\left(x^2+1+\frac{1}{x^2}\right) t^4\right.\\
    & \left.+\left(x+\frac{1}{x}\right) t^5-t^8-\left(x+\frac{1}{x}\right) t^9+\mathcal{O}\left(t^{10}\right)\right]\fstop
\end{split}
\end{equation}
Once unrefined, this admits the following closed formula:
\begin{equation}
    \mathrm{HS}\left[\text{CB of \eqref{A3A3}}_{\text{wr} \, S_3}\right](t) = \frac{1 - 2 t + 2 t^2 - t^4 + 2 t^5 - t^6 + 2 t^8 - 2 t^9 + t^{10}}{(1-t)^6 (1+t)^2 \left(1+t^2\right)^2 \left(1+t+t^2\right)^2} 
    \fstop
\end{equation}

We remark that the pole in the denominator of the unrefined Hilbert series is the same for all types of wreathing we are considering.  This phenomenon is very similar to that of \cite[Figure 9]{Bourget:2020bxh}, in agreement with the fact that the dimension of the Coulomb branch of magnetic quivers is preserved upon wreathing.

\subsection{\texorpdfstring{$(A_4,A_4)$}{(A4,A4)}}

The 3d mirror theory of $(A_4, A_4)$ is given by the following quiver
\begin{align}\label{A4A4quiver3d}
    \begin{tikzpicture}[baseline=0pt,font=\footnotesize,scale=1.2]
        \node (U1n) at (0,1) {$\U(1)$};
        \node (U1e) at (1,0) {$\U(1)$};
        \node (U1s) at (0,-1) {$\U(1)$};
        \node (U1w) at (-1,0) {$\U(1)$};
        \node (Fn) at (0,2) {[1]};
        \node (Fe) at (2,0) {[1]};
        \node (Fs) at (0,-2) {[1]};
        \node (Fw) at (-2,0) {[1]};
        \draw (U1n) -- (U1e) -- (U1s) -- (U1w) -- (U1n);
        \draw (U1n)  -- (U1s);
        \draw (U1e) -- (U1w);
        \draw (Fn) -- (U1n);
        \draw (Fs) -- (U1s);
        \draw (Fw) -- (U1w);
        \draw (Fe) -- (U1e);
    \end{tikzpicture}
\end{align}
Similarly to \eref{CBHSA3A3}, the Coulomb branch Hilbert series without wreathing is given by
\begin{equation}
     \mathrm{HS}\left[\text{CB of \eqref{A4A4quiver3d}}\right](t) =\frac{1+8 t^4+18 t^6+11 t^8+44 t^{10}+11 t^{12}+18 t^{14}+8 t^{16}+t^{20}}{\left(1-t\right)^8 \left(1+t\right)^8 \left(1+t^2\right)^2 \left(1-t+t^2\right)^2 \left(1+t+t^2\right)^2}\fstop
\end{equation}

Let us now consider the $\BZ_2$, $\BZ_3$, $S_3$, $\BZ_4$ and $S_4$ wreathings of \eref{A4A4quiver3d}. The procedure is similar to the one described in Section \ref{sec:disgauging-hooft} and Appendix \ref{sec:Casen2USp4}, but without the $P_{\USp(2n)}$ factor. In the following, we simply report the closed forms of the unrefined Coulomb branch Hilbert series of \eref{A4A4quiver3d}.

\bes{
\begin{tabular}{c|l}
Wreathing & \multicolumn{1}{c}{Hilbert series}  \\  
\hline
\\ [-4mm]
$\BZ_2$ & $\frac{1-t^2+7 t^4+7 t^6+5 t^8+22 t^{10}+5 t^{12}+7 t^{14}+7
t^{16}-t^{18}+t^{20}}{\left(1-t\right)^8 \left(1+t\right)^8 \left(1+t^2\right)^2 \left(1-t+t^2\right)^2 \left(1+t+t^2\right)^2}$ \\[2mm]
\hline
\\ [-4mm]
$\BZ_3$ & $ \frac{1-2 t^2+6 t^4+2 t^6+5 t^8+16 t^{10}+5 t^{12}+2 t^{14}+6 t^{16}-2 t^{18}+t^{20}}{\left(1-t\right)^8 \left(1+t\right)^8 \left(1+t^2\right)^2 \left(1-t+t^2\right)^2 \left(1+t+t^2\right)^2}$ \\[2mm]
\hline
\\ [-4mm]
$S_3$ & $ \frac{1-2 t^2+6 t^4-t^6+2 t^8+8 t^{10}+2 t^{12}-t^{14}+6 t^{16}-2 t^{18}+t^{20}}{\left(1-t\right)^8 \left(1+t\right)^8 \left(1+t^2\right)^2 \left(1-t+t^2\right)^2 \left(1+t+t^2\right)^2}$ \\[2mm]
\hline
\\ [-4mm]
$\BZ_4$ & $ \frac{1-3t^2+7t^4-t^6+9t^8+4t^{10}+9t^{12}-t^{14}+7t^{16}-3t^{18}+t^{20}}{\left(1-t\right)^8 \left(1+t\right)^8 \left(1+t^2\right)^2 \left(1-t+t^2\right)^2 \left(1+t+t^2\right)^2}$ \\[2mm]
\hline
\\ [-4mm]
$S_4$ & $ \frac{1-3 t^2+6 t^4-5 t^6+4 t^8-t^{10}+4 t^{12}-5 t^{14}+6 t^{16}-3 t^{18}+t^{20}}{\left(1-t\right)^8 \left(1+t\right)^8 \left(1+t^2\right)^2 \left(1-t+t^2\right)^2 \left(1+t+t^2\right)^2}$ 
\end{tabular}
\label{eq:A4A4wreathings}
}

\section{The Hilbert Series of \texorpdfstring{$\BC^{n}/G(k, p, n)$}{} and \texorpdfstring{$\BC^{2 n}/G(k, p, n)$}{}} 
\label{app:complexreflectiongroup}

In this appendix, we present a systematic method of computing the Hilbert series of $\BC^{n}/G(k, p, n)$ and $\BC^{2n}/G(k, p, n)$, where $G(k, p, n)$ is a complex reflection group in the notation of \cite{ShephardTodd} such that $p$ divides $k$. For a detailed exposition on the subject, we refer the reader to the works \cite{Aharony:2016kai, Argyres:2019ngz, Tachikawa:2019dvq, Kaidi:2022lyo} available in the physics literature, as well as the works \cite{GeckMalle, LehrerTaylor} available in the mathematics literature.  The space $\BC^{2n}/G(k, p, n)$ is of particular significance, as it is isomorphic to both the Higgs and Coulomb branches of the $[\U(n + x)_k \times \U(n)_{-k}]$ ABJ(M) theory \cite{Aharony:2008ug, Aharony:2008gk}, as well as its $\BZ_p$ quotient \cite{Tachikawa:2019dvq, Bergman:2020ifi}, viewed as a 3d $\CN=4$ SCFT. As pointed out in \cite{Kaidi:2022uux}, this arises from twisted compactification of the 4d $\CN=4$ SYM theory with a classical gauge algebra.  In the following, we start by discussing the Hilbert series of $\BC^{n}/G(k, p, n)$ and then generalize this result to $\BC^{2n}/G(k, p, n)$.

\subsection{Complex Reflection Groups \texorpdfstring{$G(k, p, n)$}{} Acting on \texorpdfstring{$\BC^n$}{}}

If we take the coordinates on $\BC^n$ to be $(z_1, \ldots, z_n)$, then the complex reflection groups $G(k, p, n)$ acting on $\BC^n$ are generated by permutations of $z_1, \ldots, z_n$ together with
\bes{ \label{zzGkpnCn}
(z_i, z_j) \to (e^{2 \pi i/k} z_i, e^{-2 \pi i/k} z_j)~, \,\ \text{other $z_l$ fixed}~,
}
\bes{ \label{zGkpnCn}
z_i \to e^{2 \pi i p/k} z_i~, \,\ \text{other $z_j$ fixed}~.
}
Some special cases are the following:
\bi
\item $G(k, p, 1)$ is $\BZ_{k/p}$~,
\item $G(1, 1, n)$ is the $n$-fold symmetric product of $\BC$, \ie $\Sym^n(\BC) \cong \BC^n/S_n$~,
\item $G(k, k, 2)$ is the dihedral group $\BD_{2 n}$ of order $2 n$~.
\ei
The Hilbert series of $\BC^{n}/G(k, p, n)$ can be derived starting from the Molien formula
\bes{ \label{MolienCn}
\mathrm{HS}\left[{\BC^{n}/G(k, p, n)}\right] (t) = \frac{1}{|G(k, p, n)|} \sum_{g \in G(k, p, n)} \frac{1}{\det{(\ID_{n} - t g)}}~,
}
where $|G(k, p, n)| = k^n n!/p$ is the order of the group, $\ID_{n}$ is the $n \times n$ identity matrix and the summation is over $n$-dimensional representations of the group elements. An $n$-dimensional representation of the generators consists of the permutation matrices $\SM_i$ of $n$ objects,\footnote{In general, the permutation group $S_n$ of $n$ objects is generated by the adjacent transpositions, namely $\langle (12), (23), \ldots, (n-1,n) \rangle$. As an example, the matrix representation of $(12)$ is given by
\bes{ 
        \left(\begin{array}{c;{2pt/2pt}c}
        X & \ZM \\ \hdashline[2pt/2pt]
        \ZM & \ID \end{array}\right)~,~ \text{with} ~ X= \begin{pmatrix} 0 & 1 \\ 1 & 0 \end{pmatrix}.
        \nonumber
}
Several explicit examples will be given below.} with $i = 1, \ldots n!$, together with the $n \times n$ matrices
\bes{ \label{Agen}
        \MA_a &= \left\{ 
        %\scalebox{1}{$
        \begin{pmatrix}
        e^{2 \pi i/k} & & & & & \\
        & e^{-2 \pi i/k} & & & & \\ 
        & & 1 & & & \\ 
        & & & \ddots & & \\
        & & & & 1 & \\
        & & & & & 1 \end{pmatrix}~, \ldots,
        \begin{pmatrix}
        1 & & & & & \\
        & e^{2 \pi i/k} & & & & \\ 
        & & e^{-2 \pi i/k} & & & \\ 
        & & & 1 & & \\
        & & & & \ddots & \\
        & & & & & 1 \end{pmatrix}
        %$}
        ~, \right. \\
         & \, \ldots, \left.
        %\scalebox{1}{$
        \begin{pmatrix}
        1 & & & & & \\
        & \ddots & & & & \\ 
        & & e^{2 \pi i/k} & & & \\ 
        & & & e^{2 \pi i/k} & & \\
        & & & & \ddots & \\
        & & & & & 1 \end{pmatrix}~, \ldots,
        \begin{pmatrix}
        1 & & & & & \\
        & 1 & & & & \\ 
        & & \ddots & & & \\ 
        & & & 1 & & \\
        & & & & e^{2 \pi i/k} & \\
        & & & & & e^{-2 \pi i/k} \end{pmatrix} 
        %$}
        \right\}~,
}
where ${a = 1, \ldots, n (n - 1)/2}$, and
\bes{ \label{Bgen}
        \BM_b &= \left\{ \begin{pmatrix}
        e^{2 \pi i p/k} & & & & \\
        & 1 & & & \\ 
        & & \ddots & & \\
        & & & 1 & \\
        & & & & 1 \end{pmatrix}~, \ldots,
        \begin{pmatrix}
        1 & & & & \\
        & e^{2 \pi i p/k} & & & \\ 
        & & 1 & & \\ 
        & & & \ddots & \\
        & & & & 1 \end{pmatrix}~, \right. \\
        & \, \, \ldots, \left. \begin{pmatrix}
        1 & & & & \\
        & \ddots & & & \\ 
        & & 1 & & \\ 
        & & & e^{2 \pi i p/k} & \\
        & & & & 1 \end{pmatrix}~, \ldots,
        \begin{pmatrix}
        1 & & & & \\
        & 1 & & & \\ 
        & & \ddots & & \\ 
        & & & 1 & \\
        & & & & e^{2 \pi i p/k} \end{pmatrix} \right\}~,
}
with ${b = 1, \ldots, n}$, implementing \eref{zzGkpnCn} and \eref{zGkpnCn}, respectively. It turns out that
\bes{ \label{HSCnModGkpn}
\mathrm{HS}\left[{\BC^{n}/G(k, p, n)}\right] (t) = \prod_{j = 1}^n \frac{1}{1 - t^{d_j}}~,
}
where $d_j = \{k, 2 k, \ldots, (n - 1) k, k n/p\}$ are the degrees of the basic invariants of the group (see \cite[Table 1]{GeckMalle}). Note that, for $p=1$, this is actually equal to the dressing factor $P_{\U(n)}(m,m,\ldots,m;t^{k/2})$ used throughout this paper (see \cite[(A.12)]{Cremonesi:2013lqa});\footnote{We remark that the conventions of $t$ in \cite{Cremonesi:2013lqa} and in our paper are different: $t_{\text{ours}}= t^2_{\text{\cite{Cremonesi:2013lqa}}}$.} the examples for $n=2$ and $n=3$ are given explicitly in \eref{PU2U3}. 

Before discussing the action of the groups $G(k, p, n)$ on $\BC^{2 n}$, we can verify the validity of equation \eref{HSCnModGkpn} by performing the explicit computation of the Hilbert series in the simple cases $n = 1, 2 ,3$.

\subsubsection*{The Case of $n=1$}
The generator is simply $\BM = \exp{(2 \pi i p/k)}$ and the Molien formula \eref{MolienCn} coincides with the Hilbert series of $\BC/\BZ_{k/p}$: 
\bes{
\mathrm{HS}\left[{\BC/G(k, p, 1)}\right] (t) = \mathrm{HS}\left[{\BC/\BZ_{k/p}}\right] (t) = \frac{1}{k/p} \sum_{j = 0}^{k/p - 1} \frac{1}{1 - t \BM^j} = \frac{1}{1 - t^{k/p}} = {\PE} \left[t^{k/p}\right]~.
}
This is consistent with the fact that $G(k, p, 1)$ is $\BZ_{k/p}$.

\subsubsection*{The Case of $n=2$}

The generators are
\bes{ \label{C2Gkp2rep1}
\SM_1 = \begin{pmatrix} 
	1 & 0 \\
	0 & 1 \end{pmatrix}~, \,\ 
\SM_2  = \begin{pmatrix} 
	0 & 1 \\
	1 & 0 \end{pmatrix}~,
}
\bes{ \label{C2Gkp2rep2} 
\MA =  \begin{pmatrix} 
	e^{2 \pi i/k} & 0 \\
	0 & e^{-2 \pi i/k} \end{pmatrix}~,
}
\bes{ \label{C2Gkp2rep3}
\BM_1 = \begin{pmatrix} 
	e^{2 \pi i p/k} & 0 \\
	0 & 1 \end{pmatrix}~, \,\ 
\BM_2 = \begin{pmatrix} 
	1 & 0 \\
	0 & e^{2 \pi i p/k} \end{pmatrix}~.
}
The Molien formula \eref{MolienCn} reads
\bes{
\mathrm{HS}\left[{\BC^{2}/G(k, p, 2)}\right] (t) &= \frac{1}{2} \frac{1}{2 k^2/p} \sum_{i, b = 1}^2 \sum_{l = 0}^{k - 1} \sum_{j = 0}^{k/p - 1} \frac{1}{\det{(\ID_2 - t \SM_i \MA^l \BM_b^j )}} \\
&= \frac{1}{(1 - t^k) (1 - t^{2 k/p})} = {\PE} \left[t^k + t^{2 k/p}\right]~,
}
in agreement with the results of \cite[Table 2]{Argyres:2019ngz}.

\subsubsection*{The Case of $n=3$}

The generators are
\bes{ 
\SM_1 = \begin{pmatrix} 
	1 & 0 & 0 \\
        0 & 1 & 0 \\
	0 & 0 & 1 \end{pmatrix}~, \,\ 
\SM_2 = \begin{pmatrix} 
        1 & 0 & 0 \\
        0 & 0 & 1 \\
	0 & 1 & 0 \end{pmatrix}~, \,\ 
\SM_3 = \begin{pmatrix} 
        0 & 1 & 0 \\
        1 & 0 & 0 \\
	0 & 0 & 1 \end{pmatrix}~, \\
\SM_4 = \begin{pmatrix} 
	0 & 1 & 0 \\
        0 & 0 & 1 \\
	1 & 0 & 0 \end{pmatrix}~, \,\ 
\SM_5 = \begin{pmatrix} 
        0 & 0 & 1 \\
        1 & 0 & 0 \\
	0 & 1 & 0 \end{pmatrix}~, \,\ 
\SM_6 = \begin{pmatrix} 
        0 & 0 & 1 \\
        0 & 1 & 0 \\
	1 & 0 & 0 \end{pmatrix}~,
}
\bes{ 
\MA_1 = \scalebox{0.94}{$\begin{pmatrix} 
	e^{2 \pi i/k} & 0 & 0 \\
	0 & e^{-2 \pi i/k} & 0 \\
        0 & 0 & 1 \end{pmatrix}~, \,\
\MA_2 = \begin{pmatrix} 
	e^{2 \pi i/k} & 0 & 0 \\
	0 & 1 & 0 \\
        0 & 0 & e^{-2 \pi i/k} \end{pmatrix}~, \,\
\MA_3 = \begin{pmatrix} 
        1 & 0 & 0 \\
	0 & e^{2 \pi i/k} & 0 \\
        0 & 0 & e^{-2 \pi i/k} \end{pmatrix}$}~,
}
\bes{ 
\BM_1 = \begin{pmatrix} 
	e^{2 \pi i p/k} & 0 & 0 \\
        0 & 1 & 0 \\
	0 & 0 & 1 \end{pmatrix}~, \,\ 
\BM_2 = \begin{pmatrix} 
	1 & 0 & 0 \\
	0 & e^{2 \pi i p/k} & 0 \\
        0 & 0 & 1 \end{pmatrix}~, \,\ 
\BM_3 = \begin{pmatrix} 
	1 & 0 & 0 \\
	0 & 1 & 0 \\
        0 & 0 & e^{2 \pi i p/k} \end{pmatrix}~.
}
The Molien formula \eref{MolienCn} reads
\bes{
\mathrm{HS}\left[{\BC^{3}/G(k, p, 3)}\right] (t) = & \,\frac{1}{3} \frac{1}{6 k^3/p} \sum_{i = 1}^6 \sum_{l, m = 0}^{k - 1} \sum_{j = 0}^{k/p - 1} \left[ \frac{1}{\det{(\ID_3 - t \SM_i \MA_1^l \MA_2^m \BM_1^j )}} + \right. \\ & +\left. \frac{1}{\det{(\ID_3 - t \SM_i \MA_1^l \MA_3^m \BM_2^j)}} + \frac{1}{\det{(\ID_3 - t \SM_i \MA_2^l \MA_3^m \BM_3^j)}} \right] \\
=& \,\frac{1}{(1 - t^k) (1 - t^{2 k}) (1 - t^{3 k/p})} = {\PE} \left[t^k + t^{2 k} + t^{3 k/p}\right]~.
}

\subsection{Complex Reflection Groups \texorpdfstring{$G(k, p, n)$}{} Acting on \texorpdfstring{$\BC^{2 n}$}{}}

The Molien formula for the Hilbert series of $\BC^{2 n}/G(k, p, n)$ is
\bes{ \label{MolienC2n}
\mathrm{HS}\left[{\BC^{2 n}/G(k, p, n)}\right] (z;t) = \frac{1}{|G(k, p, n)|} \sum_{g \in G(k, p, n)} \frac{1}{\det{(\ID_{2 n} - \Bt_{2 n} g)}}~,
}
where $\ID_{2 n}$ is the $2 n \times 2 n$ identity matrix, $\Bt_{2 n}$ is the $2 n \times 2 n$ diagonal matrix
\bes{ 
        \Bt_{2 n} = \left(\begin{array}{c;{2pt/2pt}c}
        t_1 \ID_n & \ZM \\ \hdashline[2pt/2pt]
        \ZM & t_2 \ID_n \end{array}\right)~,
}
and the summation is over $2 n$-dimensional representations of the group. The fugacity $z$ corresponding to the $\mathfrak{u}(1)$ or $\su(2)$ symmetry can be introduced by taking $t_1 = t z^{p/k}$ and $t_2 = t z^{-p/k}$, from which we can recover the {\it unrefined} Hilbert series by setting $z = 1$.

We take as generators the $2 n \times 2 n$ block diagonal matrices $\TM_i = \SM_i \oplus \SM_i$, explicitly
\bes{ 
        \TM_i = \left(\begin{array}{c;{2pt/2pt}c}
        \SM_i & \ZM \\ \hdashline[2pt/2pt]
        \ZM & \SM_i \end{array}\right)~,~ i = 1, \ldots, n!~,
}
together with the $2 n \times 2 n$ matrices
\bes{ \label{Pgen}
        \PM_a = \left(\begin{array}{c;{2pt/2pt}c}
        \MA_a & \ZM \\ \hdashline[2pt/2pt]
        \ZM & \MA_a^{-1} \end{array}\right)~,~ a = 1, \ldots, n (n - 1)/2~,
}
and
\bes{ \label{Qgen}
        \QM_b = \left(\begin{array}{c;{2pt/2pt}c}
        \BM_b & \ZM \\ \hdashline[2pt/2pt]
        \ZM & \BM_b^{-1} \end{array}\right)~,~ b = 1, \ldots, n~,
}
corresponding to the $2 n$-dimensional uplift of the $n \times n$ matrices \eref{Agen} and \eref{Bgen} respectively. Observe that \eref{Pgen} and \eref{Qgen} are block diagonal matrices with {\it inverse actions} between the two $n$-dimensional blocks, meaning that the complex reflection group acts with opposite parameters $k$ and $-k$ on the two copies of $\BC^n$.

From the special cases listed at the beginning of this section, it follows that in general
\bes{
\mathrm{HS}\left[{\BC^{2 n}/G(k, 1, n)}\right] (z;t) = \mathrm{HS}\left[{\Sym^n(\BC^2/\BZ_k)}\right] (z;t)~.
}
\subsubsection*{The Case of $n=1$}

The generators
\bes{
\TM = \left(\begin{array}{c;{2pt/2pt}c}
    1 & 0 \\ \hdashline[2pt/2pt]
    0 & 1 \end{array}\right)~, \,\ 
\QM = \left(\begin{array}{c;{2pt/2pt}c}
    e^{2 \pi i p/k} & 0 \\ \hdashline[2pt/2pt]
    0 & e^{-2 \pi i p/k} \end{array}\right)
}
coincide with those of the $A_{k/p - 1}$ singularity \cite[(3.9)]{Benvenuti:2006qr}, meaning that the Molien formula \eref{MolienC2n} coincides with the Hilbert series of $\BC^{2}/\BZ_{k/p}$:
\bes{
\mathrm{HS}\left[{\BC^{2}/G(k, p, 1)}\right] (z;t) = \mathrm{HS}\left[{\BC^{2}/\BZ_{k/p}}\right] (z;t) &= \frac{1 - t^{2 k/p}}{(1 - t^2) (1 - t^{k/p} z) (1 - t^{k/p} z^{-1})}\\ 
&= {\PE} \left[t^2 + (z+z^{-1}) t^{k/p} - t^{2 k/p}\right]~.
}
Again, this is consistent with the fact that $G(k, p, 1)$ is $\BZ_{k/p}$.

\subsubsection*{The Case of $n=2$}

The generators are
\bes{ \label{SgenC4Gkp2}
\TM_1 = \left(\begin{array}{cc;{2pt/2pt}cc}
    1 & 0 & 0 & 0 \\
    0 & 1 & 0 & 0 \\ \hdashline[2pt/2pt]
    0 & 0 & 1 & 0 \\
    0 & 0 & 0 & 1 \end{array}\right)~, \,\
\TM_2 = \left(\begin{array}{cc;{2pt/2pt}cc}
    0 & 1 & 0 & 0 \\
    1 & 0 & 0 & 0 \\ \hdashline[2pt/2pt]
    0 & 0 & 0 & 1 \\
    0 & 0 & 1 & 0 \end{array}\right)~,
}
\bes{ \label{AgenC4Gkp2}
\PM = \left(\begin{array}{cc;{2pt/2pt}cc}
    e^{2 \pi i/k} & 0 & 0 & 0 \\
    0 & e^{-2 \pi i/k} & 0 & 0 \\ \hdashline[2pt/2pt]
    0 & 0 & e^{-2 \pi i/k} & 0 \\
    0 & 0 & 0 & e^{2 \pi i/k} \end{array}\right)~,
}
\bes{ \label{BgenC4Gkp2}
\QM_1 = \left(\begin{array}{cc;{2pt/2pt}cc}
    e^{2 \pi i p/k} & 0 & 0 & 0 \\
    0 & 1 & 0 & 0 \\ \hdashline[2pt/2pt]
    0 & 0 & e^{-2 \pi i p/k} & 0 \\
    0 & 0 & 0 & 1 \end{array}\right)~, \,\
\QM_2 = \left(\begin{array}{cc;{2pt/2pt}cc}
    1 & 0 & 0 & 0 \\
    0 & e^{2 \pi i p/k} & 0 & 0 \\ \hdashline[2pt/2pt]
    0 & 0 & 1 & 0 \\
    0 & 0 & 0 & e^{-2 \pi i p/k} \end{array}\right)~.
}
The Molien formula \eref{MolienC2n} reads
\bes{ \label{MolienC4Gkp2}
\mathrm{HS} \left[\BC^{4}/G(k, p, 2)\right] (z;t) = \frac{1}{2} \frac{1}{2 k^2/p} \sum_{i, b = 1}^2 \sum_{l = 0}^{k - 1} \sum_{j = 0}^{k/p - 1} \frac{1}{\det{(\ID_4 - \Bt_4 \SM_i \MA^l \BM_b^j)}}
}
and we have the following special cases:
\bi
\item $\mathrm{HS}\left[{\BC^{4}/G(k, k, 2)}\right] (z;t) = \mathrm{HS}\left[{\BC^{4}/\BD_{2 k}}\right] (z;t)$~,
\item $\mathrm{HS}\left[{\BC^{4}/G(k, 1, 2)} \right](z;t) = \mathrm{HS}\left[{\Sym^2(\BC^2/\BZ_k)}\right] (z;t)$~,
\item $\mathrm{HS}\left[{\BC^{4}/G(4, 4, 2)} \right](z;t) = \mathrm{HS}\left[{\BC^{4}/\BD_{8}} \right](z;t) = \mathrm{HS}\left[{\Sym^2(\BC^2/\BZ_2)}\right] (z^2;t)$~,
\ei
where the last equality follows from $G(4, 4, 2) \cong G(2, 1, 2)$ (see \cite[Theorem 2.16]{LehrerTaylor}). Observe that, for $p = k$, since the determinant in \eref{MolienC4Gkp2} is multilinear and alternating, the generators \eref{SgenC4Gkp2}, \eref{AgenC4Gkp2}, and \eref{BgenC4Gkp2} produce the same result of the four-dimensional representations of the dihedral group of order $2 k$ $\BD_{2 k}$ reported in \cite[(C.5)]{Hayashi:2022ldo} for $z = 1$.

\subsubsection*{The Case of $n=3$}

The generators are
\bes{ 
\TM_1 = \left(\begin{array}{ccc;{2pt/2pt}ccc}
	1 & 0 & 0 & 0 & 0 & 0 \\
        0 & 1 & 0 & 0 & 0 & 0 \\
	0 & 0 & 1 & 0 & 0 & 0 \\ \hdashline[2pt/2pt]
        0 & 0 & 0 & 1 & 0 & 0 \\
        0 & 0 & 0 & 0 & 1 & 0 \\
        0 & 0 & 0 & 0 & 0 & 1 \end{array}\right)~, \,\ 
\TM_2 =  \left(\begin{array}{ccc;{2pt/2pt}ccc}
        1 & 0 & 0 & 0 & 0 & 0 \\
        0 & 0 & 1 & 0 & 0 & 0 \\
	0 & 1 & 0 & 0 & 0 & 0 \\ \hdashline[2pt/2pt]
        0 & 0 & 0 & 1 & 0 & 0 \\
        0 & 0 & 0 & 0 & 0 & 1 \\
	0 & 0 & 0 & 0 & 1 & 0 \end{array}\right)~, \,\ 
\TM_3 =  \left(\begin{array}{ccc;{2pt/2pt}ccc}
        0 & 1 & 0 & 0 & 0 & 0 \\
        1 & 0 & 0 & 0 & 0 & 0 \\
	0 & 0 & 1 & 0 & 0 & 0 \\ \hdashline[2pt/2pt]
        0 & 0 & 0 & 0 & 1 & 0 \\
        0 & 0 & 0 & 1 & 0 & 0 \\
	0 & 0 & 0 & 0 & 0 & 1 \end{array}\right)~, \\
\TM_4 =  \left(\begin{array}{ccc;{2pt/2pt}ccc}
	0 & 1 & 0 & 0 & 0 & 0 \\
        0 & 0 & 1 & 0 & 0 & 0 \\
	1 & 0 & 0 & 0 & 0 & 0 \\ \hdashline[2pt/2pt]
        0 & 0 & 0 & 0 & 1 & 0 \\
        0 & 0 & 0 & 0 & 0 & 1 \\
	0 & 0 & 0 & 1 & 0 & 0 \end{array}\right)~, \,\ 
\TM_5 = \left(\begin{array}{ccc;{2pt/2pt}ccc}
        0 & 0 & 1 & 0 & 0 & 0 \\
        1 & 0 & 0 & 0 & 0 & 0 \\
	0 & 1 & 0 & 0 & 0 & 0 \\ \hdashline[2pt/2pt]
        0 & 0 & 0 & 0 & 0 & 1 \\
        0 & 0 & 0 & 1 & 0 & 0 \\
	0 & 0 & 0 & 0 & 1 & 0 \end{array}\right)~, \,\ 
\TM_6 = \left(\begin{array}{ccc;{2pt/2pt}ccc}
        0 & 0 & 1 & 0 & 0 & 0 \\
        0 & 1 & 0 & 0 & 0 & 0 \\
	1 & 0 & 0 & 0 & 0 & 0 \\ \hdashline[2pt/2pt]
        0 & 0 & 0 & 0 & 0 & 1 \\
        0 & 0 & 0 & 0 & 1 & 0 \\
	0 & 0 & 0 & 1 & 0 & 0 \end{array}\right)~,
}
\bes{ 
&\scalebox{0.9}{$ \PM_1 = \left(\begin{array}{ccc;{2pt/2pt}ccc} 
	e^{2 \pi i/k} & 0 & 0 & 0 & 0 & 0 \\
	0 & e^{-2 \pi i/k} & 0 & 0 & 0 & 0 \\
        0 & 0 & 1 & 0 & 0 & 0 \\ \hdashline[2pt/2pt]
        0 & 0 & 0  & e^{-2 \pi i/k} & 0 & 0 \\
	0 & 0 & 0  & 0 & e^{2 \pi i/k} & 0 \\
        0 & 0 & 0 & 0 & 0 & 1 \end{array}\right)~, \,\
\PM_2 =   \left(\begin{array}{ccc;{2pt/2pt}ccc}
        e^{2 \pi i/k} & 0 & 0 & 0 & 0 & 0 \\
	0 & 1 & 0 & 0 & 0 & 0 \\
        0 & 0 & e^{-2 \pi i/k} & 0 & 0 & 0 \\ \hdashline[2pt/2pt]
        0 & 0 & 0 & e^{-2 \pi i/k} & 0 & 0 \\
	0 & 0 & 0 & 0 & 1 & 0 \\
        0 & 0 & 0 & 0 & 0 & e^{2 \pi i/k} \end{array}\right)$}~, \\
& \qquad \qquad \qquad \qquad \quad
\scalebox{0.9}{$\PM_3 = \left(\begin{array}{ccc;{2pt/2pt}ccc}
        1 & 0 & 0 & 0 & 0 & 0 \\
	0 & e^{2 \pi i/k} & 0 & 0 & 0 & 0 \\
        0 & 0 & e^{-2 \pi i/k} & 0 & 0 & 0 \\ \hdashline[2pt/2pt]
        0 & 0 & 0 & 1 & 0 & 0 \\
	0 & 0 & 0 & 0 & e^{-2 \pi i/k} & 0 \\
        0 & 0 & 0 & 0 & 0 & e^{2 \pi i/k} \end{array}\right)$}~,
}
\bes{ 
&\QM_1 = \left(\begin{array}{ccc;{2pt/2pt}ccc} 
        e^{2 \pi i p/k} & 0 & 0 & 0 & 0 & 0 \\
	0 & 1 & 0 & 0 & 0 & 0 \\
        0 & 0 & 1 & 0 & 0 & 0 \\ \hdashline[2pt/2pt]
        0 & 0 & 0  & e^{-2 \pi i p/k} & 0 & 0 \\
	0 & 0 & 0  & 0 & 1 & 0 \\
        0 & 0 & 0 & 0 & 0 & 1 \end{array}\right)~, \,\
\QM_2 = \left(\begin{array}{ccc;{2pt/2pt}ccc}
	1 & 0 & 0 & 0 & 0 & 0 \\
	0 & e^{2 \pi i p/k} & 0 & 0 & 0 & 0 \\
        0 & 0 & 1 & 0 & 0 & 0 \\ \hdashline[2pt/2pt]
        0 & 0 & 0 & 1 & 0 & 0 \\
	0 & 0 & 0 & 0 & e^{-2 \pi i p/k} & 0 \\
        0 & 0 & 0 & 0 & 0 & 1 \end{array}\right)~, \\
& \qquad \qquad \qquad \qquad \quad
\QM_3 = \left(\begin{array}{ccc;{2pt/2pt}ccc}
        1 & 0 & 0 & 0 & 0 & 0 \\
	0 & 1 & 0 & 0 & 0 & 0 \\
        0 & 0 & e^{2 \pi i p/k} & 0 & 0 & 0 \\ \hdashline[2pt/2pt]
        0 & 0 & 0 & 1 & 0 & 0 \\
	0 & 0 & 0 & 0 & 1 & 0 \\
        0 & 0 & 0 & 0 & 0 & e^{-2 \pi i p/k} \end{array}\right)~.
}
The Molien formula \eref{MolienC2n} reads
\bes{
\mathrm{HS}\left[{\BC^{6}/G(k, p, 3)}\right] (z;t) = &\, \frac{1}{3} \frac{1}{6 k^3/p} \sum_{i = 1}^6 \sum_{l, m = 0}^{k - 1} \sum_{j = 0}^{k/p - 1} \left[ \frac{1}{\det{(\ID_6 - \Bt_6 \TM_i \PM_1^l \PM_2^m \QM_1^j )}}  \right. \\ & \left. +\frac{1}{\det{(\ID_6 - \Bt_6 \TM_i \PM_1^l \PM_3^m \QM_2^j)}} + \frac{1}{\det{(\ID_6 - \Bt_6 \TM_i \PM_2^l \PM_3^m \QM_3^j)}} \right]~,
}
which, in the special case $p = 1$, coincides with the Hilbert series of $\Sym^3(\BC^2/\BZ_k)$.

\newpage

\bibliographystyle{JHEP}
\bibliography{mybib}

\end{document}